\documentclass[11pt]{article}
\usepackage{amstext,amsfonts}
\usepackage{rotating,graphicx}
\usepackage{verbatim}
\makeatletter
\@addtoreset{equation}{section}
\@addtoreset{table}{section}

\makeatother

\newcommand{\dis}{\displaystyle}

\newcommand{\tA}{\theta^\alpha}

\newcommand{\xM}{x^\mu}
\newcommand{\vfmn}{\varphi_{\mu\nu}}
\newcommand{\vfma}{\varphi_{\mu\alpha}}
\newcommand{\vfab}{\varphi_{\alpha\beta}}
\newcommand{\ZMN}{Z^{\mu\nu}}
\newcommand{\ZMA}{Z^{\mu\alpha}}

\newcommand{\ZAB}{Z^{\alpha\beta}}
\newcommand{\Gmnab}{(C \Gamma_{\mu\nu})_{\alpha\beta}}

\newcommand{\Gmab}{(C \Gamma_\mu)_{\alpha\beta}}
\newcommand{\GMab}{(C \Gamma^\mu)_{\alpha\beta}}
\newcommand{\GMcd}{(C \Gamma^\mu)_{\gamma\delta}}
\newcommand{\Gmcd}{(C \Gamma_\mu)_{\gamma\delta}}
\newcommand{\be}{\begin{equation}}
\newcommand{\bae}{\begin{eqnarray}}
\newcommand{\ee}{\end{equation}}
\newcommand{\eae}{\end{eqnarray}}
\newcommand{\eg}{\emph{e.g.}}
\newcommand{\ie}{\emph{i.e.}}

\begin{document}
\renewcommand{\thefootnote}{\fnsymbol{footnote}}
\thispagestyle{empty}

\vspace*{-1cm}
\begin{flushright}
FTUV--99/20\quad IFIC--99/21
\\
April 10, 1999
\\
REVISED (shortened): July 13. Version to appear in NPB
\\
hep-th/9904137
\\[3cm]
\end{flushright}

\begin{center}
\begin{Large}
\bfseries{The geometry of branes and extended superspaces}
\end{Large}

\vspace*{1.6cm}

\begin{large}
C.~Chryssomalakos$^1$, J.~A.~de~Azc\'arraga$^1$, J.~M. Izquierdo$^2$
and \\
J.~C.~P\'erez Bueno$^1$\footnote{
e-mail addresses: chryss@lie3.ific.uv.es,
azcarrag@lie1.ific.uv.es, izquierd@fta.uva.es,
pbueno@lie.ific.uv.es}
\end{large}

\vspace*{0.6cm}
\begin{it}
$1$ Departamento de F\'{\i}sica Te\'orica, Universidad de Valencia
\\
and IFIC, Centro Mixto Universidad de Valencia--CSIC
\\
E--46100 Burjassot (Valencia), Spain
\\[0.4cm]
$2$ Departamento de F\'{\i}sica Te\'orica, Universidad de Valladolid
\\
E--47011 Valladolid, Spain\\
\end{it}

\end{center}
\vspace*{1.4cm}
\begin{abstract}

We argue that a description of supersymmetric
extended objects from a unified geometric point of view 
requires an enlargement of superspace.
To this aim we study in a systematic way how superspace groups
and algebras arise from Grassmann spinors when
these are assumed to be the only primary entities.
In the process, we recover generalized spacetime
superalgebras and extensions of supersymmetry found earlier.
The enlargement of ordinary superspace with new parameters
gives rise to  extended superspace groups, on which
manifestly supersymmetric actions may be constructed
for various types of $p$-branes, including D-branes (given by
Chevalley-Eilenberg cocycles) with
their Born-Infeld fields. This results in a field/extended 
superspace democracy for superbranes: all brane fields appear as
pull-backs from a suitable target superspace. 
Our approach also clarifies 
some facts concerning the origin of the central charges for 
the different $p$-branes.
\end{abstract}

\setcounter{footnote}{0}
\renewcommand{\thefootnote}{\arabic{footnote}}

\newpage
%%%%%%%%%%%%%%%%%%%%%%%%%%%%%%%%%%%%%%%%%%%%%%%%%%%%%%%%%%%%%%%%%
%%%%%%%%%%%%%%%%%%%%%%%%%%%%%%%%%%%%%%%%%%%%%%%%%%%%%%%%%%%%%%%%%
\section{Introduction}

As is well known, the existence of consistent classical 
actions for extended supersymmetric objects of spatial dimension 
$p$ is restricted to certain dimensions $D$ of spacetime.
This is \eg\ the case of the $p$-branes of the minimal or `old' 
branescan \cite{Ach.Eva.Tow.Wil:87}, which restricts the actions 
to certain values $(D,p)$ for which there exists a  Wess-Zumino (WZ) 
term. This is needed for the $\kappa$-symmetry of the full action
that matches the physical bosonic and fermionic degrees 
of freedom on the worldvolume $W$, and the WZ term is given 
by a closed $(p+2)$-form which can be interpreted \cite{Azc.Tow:89} 
as a Chevalley-Eilenberg \cite{Che.Eil:48} (CE)
$(p+2)$-cocycle on superspace.
The first classification of $p$-branes \cite{Ach.Eva.Tow.Wil:87} 
was restricted to fields forming a scalar supermultiplet on 
$W$, consisting of scalars and spinors  after gauging the 
$\kappa$-symmetry.
The restriction to superspace coordinates $x^\mu,\theta^\alpha$ on
$W$ was later removed with the addition of higher spin fields, vectors 
or antisymmetric tensors, forming vector or antisymmetric tensor 
supermultiplets on $W$. This, together with the Bose-Fermi 
matching conditions led to an enlargement of the possibilities 
(see \cite{Duf.Khu.Lu:95,Duf:96b} and earlier 
references therein) for the classically allowed supermembranes.
Recently, $p$-branes including an abelian vector gauge 
field on $W$ have been interpreted as (Dirichlet) 
$D$-branes \cite{Pol:95} (see \cite{Pol.Cha.Joh:96} for a review).
Their kinetic term is described by a Born-Infeld 
type Lagrangian which replaces the usual Nambu-Goto
one to accommodate the vector potential; 
in similarity with the $p$-branes in \cite{Ach.Eva.Tow.Wil:87}, 
there also exists a $\kappa$-symmetric 
worldvolume action \cite{Aga.Pop.Sch:97} for them.
The introduction of other objects such as $L$-branes (which 
have linear supermultiplets on $W$) \cite{How.Rae.Rud.Sez:98} 
etc., have enlarged the number and types of $p$-branes.
Finally,
the emergence of a web of dualities among the five consistent 
$10$-dimensional string theories, all presumably 
subsumed, together with $D=11$
supergravity, in the eleven dimensional $M$-theory (see, \eg\ 
\cite{Sch:97b})
has led to the `second superstring revolution' and to a 
change of the conventional views of supersymmetry.
One version of the $M$-theory, $M$-atrix theory
\cite{Ban.Fis.She.Sus:97}, even
reinterprets spacetime coordinates as non-commuting matrices.

The existence of various 
extended objects for which there is no unified description 
suggests that, in the same way Minkowski space was enlarged 
to the superspace $\Sigma$ to treat bosons and fermions 
simultaneously, it may be necessary to extend $\Sigma$ further 
to accommodate in a unified point of view a number of 
the physical aspects mentioned above. In particular, 
one might hope to remove the need of defining fields 
{\it directly} on $W$ if an extended superspace ${\tilde \Sigma}$
is introduced, as it will be seen to be the case. This 
extension of $\Sigma$
is tantamount to enlarging the $D$-dimensional superPoincar\'e 
$sP$ to $\widetilde{sP}$ and to defining the extended superspace 
$\tilde{\Sigma}$ by 
the quotient $\widetilde{sP}/Spin(1,D-1)$.
Endowing $\tilde{\Sigma}$ with a supergroup structure means that 
there must exist new superalgebras going beyond the 
ordinary supersymmetry algebra, and several of them have 
been discussed in various contexts
\cite{Azc.Gau.Izq.Tow:89,Gre:89,Ber.Sez:89,
Azc.Izq.Tow:91, Ber.Sez:95, Bar:96,
Sor.Tow:97,Sez:97,Der.Gal:97,Bar:97,Sak:98}.  
Our point of view, however, will be to assume that fermions 
(in the form of odd abelian spinor translations) 
are the only basic ({\it i.e.}, initial) entities. 
We shall then look for the most 
general superspace groups that are allowed by group 
extension theory and discuss their consequences for 
a unified picture of superbranes.
We find this path rather natural, but it is 
not the only one. Another possibility is to take
the worldvolume supersymmetry of the $p$-branes into account
by elevating the target superspace coordinates to worldvolume
superfields \cite{Ban.Pas.Sor.Ton.Vol:95,How.Sez:97} 
(see also \cite{How.Rae.Rud.Sez:98}
and references therein), but we shall not follow this 
superembedding or `double supersymmetry' approach.

Stated as above, the problem is first a mathematical one.
Much in the same way that rigid superspace is itself a group 
extension, and hence supersymmetry is the result of
the non-trivial cohomology of a certain odd superstranslation 
group $\text{sTr}_D$ \cite{Ald.Azc:85,Azc.Izq:95}, it is worth
looking for all the possible group extensions of 
the various $\text{sTr}_D$ ({\it i.e.}, $\text{sTr}_D$= 
\{$N$=1, $\text{sTr}_{11}$, $\text{sTr}_{10}$, IIA, IIB, etc\}) 
to explore their r\^ole in more general theories.
At the algebra level, the possible supersymmetry algebras were already 
investigated in \cite{Haa.Lop.Soh:75} and, allowing tensor 
`central' charges, in \cite{Hol.Pro:82} (tensorial charges 
were also considered in \cite{DAu.Fre:82,Ziz:84b,Ziz:84}).
But there is also a physical reason behind the mathematical
extension problem.
It is known that the quasi-invariance of a Lagrangian under a symmetry 
indicates that the (second) cohomology group is non-trivial,
and that the symmetry group may hence be extended.\footnote{
For a detailed account of quasi-invariance, Noether currents, 
cohomology and 
extensions, see \cite{Azc.Izq:95,Azc:92}.}
This was exploited in \cite{Azc.Gau.Izq.Tow:89} to extend the supersymmetry 
algebra for the supersymmetric extended objects by topological charges.
For Lagrangians containing a quasi-invariant piece $\phi^*(b)$ constructed
from a form $b$ on a group 
$\Sigma$ by pulling it back to a manifold $W$ by $\phi^*$, where 
$\phi:W\to\Sigma$, the extension may allow us, as it is the case 
with the WZ terms $b$ of supersymmetric objects, to obtain 
manifestly invariant terms $\tilde b$
\footnote{This is not always possible.
When the group $G$ is simple, the extension appears only at the 
loop algebra (of charge densities) level,
as for the $su(2)$ Kac-Moody algebra for a WZW model, and
disappears for the 
algebra of charges, as required by Whitehead's lemma.}
by defining them on an \emph{extended} group manifold $\tilde \Sigma$.
In fact, it was shown in \cite{Ber.Sez:95} 
that to every free differential algebra in
\cite{Azc.Tow:89} corresponds a new spacetime superalgebra, from which 
invariant forms can be found to define new WZ terms
\footnote{Although it may be argued that these invariant
terms should no longer be called WZ terms, we shall retain this
name for them.}. We shall take the analysis of 
\cite{Ber.Sez:95, Azc.Tow:89,Sez:97} further 
(see also \cite{Sor.Tow:97}) by considering various
superbrane types and by emphasizing the supergroup manifold 
point of view. Thus, we shall look for and introduce extended 
superspace groups $\tilde \Sigma$ in a systematic way 
(we restrict our attention here to rigid superspaces). The 
additional variables in these will determine symmetries to which 
(topological) charges may correspond via the standard Noether 
theorem. For the branes of the old branescan, these new 
variables will appear only  in the WZ term and as a total 
differential. This will be different for the D-branes, 
for which we will obtain, nevertheless, that it is also possible 
to find an action defined on an extended superspace (thus
removing the necessity of introducing directly worldvolume
fields) with a WZ term given by a CE $(p+2)$-cocycle. 
By showing that all these structures and extended superspaces
${\tilde \Sigma}$ follow from a basic odd translation group 
$\text{sTr}_D$ 
defined by the Grassmann spinors of the specific theory, 
we may conclude that the $\tilde \Sigma$'s (and 
the corresponding extended superPoincar\'e groups $\widetilde{sP}$)
are in a way as fundamental as the standard one, and 
necessary for a proper description of the physics involved around 
$M$-theory and its six weak coupling limit corners. 
The new variables may be relevant in the search for superbrane
actions, in the description of 
dualities or in the quantisation process.

This paper is organised as follows. Sec. 2 contains 
all central extensions of $\text{sTr}_D$, including ordinary
superspace, for various dimensions, and its
results are summarised in a table. Sec. \ref{NCESA} considers in 
general the inclusion of additional non-central generators.
Sec. \ref{Nns} is devoted to the structure of 
the new superspaces ${\tilde \Sigma}$ and provides a compact 
expression for the contribution to the Noether 
charges coming from the WZ terms of the various possible actions, 
once they are formulated on ${\tilde \Sigma}$. Sec. 
\ref{p12} shows how the simplest $D=10,11$ extended 
superspaces are relevant to construct a manifestly invariant 
WZ term, both for the Green-Schwarz superstring  \cite{Sie:94,Gre:89}
(which we will complete with an additional contribution),
and for the supermembrane. We shall recover there the results of 
\cite{Ber.Sez:95} and compute the topological charges which, in
our approach, correspond to the new group variables. 
The question of the linearity of the group action is seen in Sec.
\ref{css} to be associated with a coboundary election. 
Sections \ref{casedbr} and \ref{m5b}  show how the case of the IIA 
D$p$-branes and $M$5-brane may also be treated within the same 
framework \emph{i.e.}, how branes containing vector and tensor
fields on $W$ may be defined directly on suitably extended 
superspaces. We shall argue,
in fact, that the picture is general and that suitable 
target superspaces exist on which to define all the fields
appearing in the $p$-brane actions, including the various
vector (see also~\cite{Sak:98}), tensor, etc. worldvolume fields. 
This is tantamount to establishing a general fields/extended 
superspace democracy in which the worldvolume fields and the 
extended superspace 
variables are on the same footing, as it was already the
case for the minimal branescan. Indeed this correspondence
between coordinates and fields has occasionally been 
discussed in the past in other contexts 
(see \cite{Ber:79,Sch:80,Gay.Rom.Sch:81,Ald.Azc:83}).
Sec. \ref{2bn} contains a brief discussion of the origin 
of the contributions to the Noether charges in the D-branes 
case \cite{Ham:98} in our approach. Finally, an Appendix 
complements the general theory of non-central extensions 
of superspace in Sec. \ref{NCESA} and gives the proof of some
needed $\Gamma$-matrix identities.

%%%%%%%%%%%%%%%%%%%%%%%%%%%%%%%%%%%%%%%%%%%%%%%%%%%%%%%%%%%%%%%%%
%%%%%%%%%%%%%%%%%%%%%%%%%%%%%%%%%%%%%%%%%%%%%%%%%%%%%%%%%%%%%%%%%
\section{Central extensions and their superspaces}
\label{CESA}
%%%%%%%%%%%%%%%%%%%%%%%%%%%%%%%%%%%%%%%%%%%%%%%%%%%%%%%%%%%%%%%%%
\subsection{Standard superspace as a central extension}
\label{Ssce}
Let $\theta$ be an arbitrary Grassmann spinor
in a $D$-dimensional spacetime. Its components 
$\theta^\alpha$ ($2^{[D/2]}$ where $[D/2]$ denotes the 
integer part of $D/2$, or $2^{(D/2)-1}$ in the Weyl case) 
determine an abelian group of supertranslations, generically 
denoted  $\text{sTr}_D$, with group composition law
\begin{equation}
\theta''^\alpha = \theta'^\alpha + \theta^\alpha
\quad.
\label{st}
\end{equation}
When the Lorentz part is considered explicitly, there is an action
$\rho$ of $Spin(1,D-1)$ on $\text{sTr}_D$ and the relevant
group becomes $\text{sTr}_D \circ Spin(1,D-1)$, where $\circ$
indicates semidirect product. Then~(\ref{st}) is replaced by
\begin{equation}
\theta'' = \theta' + \rho(A) \theta \quad, \quad \quad
A'' = A'A \quad,
\label{stlor}
\end{equation}
where $A \in Spin(1,D-1)$ and $\rho(A)$ is the appropriate 
spin representation.
The spinor $\theta$ is
often restricted to be of some specific type, usually minimal
(\eg, Majorana (M), Weyl (W) or Majorana-Weyl (MW), when possible); 
it may carry an 
additional index $i=1,\dots,N$ if there is more
than one supersymmetry. Associated with~(\ref{st}) is the abelian Lie
superalgebra $\{D_\alpha,D_\beta\}=0$~\footnote{ 
Since we shall be considering left-invariant (LI) generators and forms, 
we shall use here $D$'s (rather than $Q$'s) to denote the
generators of the right translations (the  $Q$'s being realized 
as the right-invariant (RI) generators 
of the left translations). This distinction is of course 
irrelevant for an abelian group such as 
(\ref{st}) but it is not so when non-abelian parts are added
(nevertheless, the corresponding structure constants differ only
in a sign). Furthermore, LI and RI generators commute, $\{Q,D\}=0$.
We may look at the $D$'s as covariant 
derivatives and at the $Q$'s as the generators of the (left) 
supersymmetry transformations; see Sec. \ref{Nns}.}, 
which can also be described in terms of the left-invariant 
(LI) one-forms $\Pi^\alpha=d\theta^\alpha$ and the trivial 
Maurer-Cartan (MC) equation
\begin{equation}
d \Pi^\alpha = 0 \quad.
\label{2.2}
\end{equation}

\noindent Extending $\text{sTr}_D$ by the Minkowski
translations $x^\mu$ , $\mu=0,\dots , D-1$  leads to standard
(rigid) superspace~\cite{Ald.Azc:85,Azc.Izq:95}.
Let us adopt the free differential algebra
(FDA)\footnote{The term refers to an algebra generated 
by differential forms which is closed under the action of
$d$~\cite{Sul:77}.  For early physical applications of
FDA in supersymmetry see \cite{DAu.Fre.Reg:80,PvN:83} and 
references therein.} point of view to discuss the extension problem, 
since forms are especially convenient in the construction of 
actions for extended objects.

Let $\theta^\alpha$ be Majorana, and
consider the two-form
$(C \Gamma^\mu)_{\alpha\beta} \Pi^\alpha\wedge \Pi^\beta$ on 
$\text{sTr}_D$. It defines a 
non trivial CE two-cocycle on the superalgebra of 
$\text{sTr}_D$, \ie \ it is left-invariant (LI), 
closed and not given by the differential of a LI one-form. 
Since by construction
the two-cocycle transforms as a Lorentz
vector, it is consistent to extend the FDA~(\ref{2.2}) by a one-form
$\Pi^\mu$ such that
\begin{equation}
d \Pi^\mu = {1\over 2} (C \Gamma^\mu)_{\alpha\beta}
            \Pi^\alpha \Pi^\beta
\label{2.3}
\end{equation}
(we omit the wedge product henceforth).
The above extension immediately implies
$\{D_\alpha,D_\beta\}=(C \Gamma^\mu)_{\alpha\beta} X_\mu$, with
$X_\mu$ central.
One still has to relate the newly introduced one-form to the
coordinate $x^\mu$. We define
\begin{equation}
\Pi^\mu = dx^\mu + {1\over 2} (C \Gamma^\mu)_{\alpha \beta}
          \theta^\alpha d\theta^\beta
\quad
\label{2.3b}
\end{equation}
and choose the transformation law for $x^\mu$ so that
$\Pi^\mu$ is LI
\begin{equation}
x''^\mu = x'^\mu +x^\mu -{1 \over 2} (C \Gamma^\mu)_{\alpha\beta}
            \theta'^\alpha \theta^\beta
\quad.
\label{xgcl}
\end{equation}
 This gives rigid 
superspace $\Sigma$, parametrized by $(\theta^\alpha,x^\mu)$ and 
with group law given by (\ref{st}) and (\ref{xgcl}).

The above simple example
exhibits already the key features of the extension algorithm.
Given a particular FDA to be extended, one identifies in general
a non-trivial two-cocycle of a desired Lorentz covariant nature and
introduces a new LI one-form, the differential of which is given by
the cocycle. The new form (here, (\ref{2.3b})) together with the 
MC equations (here, eqns. (\ref{2.2}) and (\ref{2.3})) 
automatically define by duality an extended Lie algebra.
The new LI one-form is given by the sum of the
differential of the new group parameter and the potential one-form
of the CE two-cocycle on $\text{sTr}_D$, which is {\it not} LI.
Finally, the transformation properties of the
new coordinate are fixed so as to guarantee the left
invariance of the new one-form, while those of the original
manifold are unmodified. The additional one-form can be made LI 
only if it is defined on 
the \emph{extended} superspace manifold ${\tilde \Sigma}$.
The new (central) generator, associated with translations
along the new coordinates, modifies the r.h.s. of the
original commutators of the algebra.
Since adding (\ref{2.3}) to (\ref{2.2}) involves a \emph{central}
extension, we could have introduced a dimensionful constant\footnote{
The value of this constant determines the specific element in 
cohomology space that characterizes the central extension.} as a
factor in the r.h.s. of (\ref{2.3}).
By not doing so, the dimensions of $\Pi^\mu$ are fixed to be
$[\Pi^\alpha]^2=[\theta^\alpha]^2$.
We shall, as usual, take
$[\theta^\alpha]=L^{1\over 2}$ so that $[\Pi^\mu]=L$.

If we add the Lorentz group, the result must reflect the action 
$\sigma$ of $Spin(1,D-1)$ on the extension 
cocyle (see, \eg\ \cite{Azc.Izq:95},
Sec. 5.3), but we shall not consider explicitly the effect of 
the simple part of the algebra which, apart from 
extracting the various tensor-valued second cohomology groups from 
that of the trivial ($\sigma$=0) action $H_0^2(\text{sTr}_D)$ 
(see below), plays no essential r\^ole in our discussion once only
Lorentz covariant objects are used. Thus, \emph{central} means, 
where appropriate, central up to Lorentz transformations. 

The extension procedure described above can be applied more than
once ---there are two
basic patterns one may follow in this case. One can start in
each step with the same original manifold $\text{sTr}_D$, 
and keep adding two-cocycles and
central generators or, in each step, one can consider 
the result of the previous extension as the starting manifold.
In the first case (Sec.~\ref{Mces}) all new
generators remain central and appear only at the
r.h.s. of the original $\{D,D\}$ anticommutator. In the second case,
a richer structure emerges since the generators introduced
at each step can, in principle, modify all previous commutators.
We shall give the details of this second construction in
Sec.~\ref{NCESA}.

%%%%%%%%%%%%%%%%%%%%%%%%%%%%%%%%%%%%%%%%%%%%%%%%%%%%%%%%%%%%%%%%%
\subsection{Maximal central extensions of superspace}
\label{Mces}

Let $\theta^\alpha$ be Majorana.
We may obtain additional Lorentz tensors, leading to new central
charges, by considering
\begin{equation}
d\Pi^{\mu_1\dots \mu_p} \equiv {1\over 2}
(C \Gamma^{\mu_1\dots \mu_p})_{\alpha\beta} \Pi^\alpha \Pi^\beta
\, ,\qquad (\Gamma^{\mu_1\dots \mu_p} = 
\Gamma^{[\mu_1} \Gamma^{\mu_2} \cdots
 \Gamma^{\mu_p]}
\equiv {1 \over p!} \epsilon^{\mu_1 \dots \mu_p}_{\nu_1 \dots
\nu_p} \Gamma^{\nu_1} \dots \Gamma^{\nu_p}) \, ,
\label{2.4}
\end{equation}
where\footnote{%
We adopt $C\equiv C_{-}$ for simplicity.
By not considering $C_{+} \Gamma^{\mu} C_{+}^{-1} = {\Gamma^{\mu}}^T$
we rule out, \eg, the pseudoMajorana spinors that exist for $D=8,9$
$[\text{mod}\, 8]$ (see \cite{Nie:84}).%
}
$C \Gamma^{\mu} C^{-1} = - {\Gamma^{\mu}}^T$.
The antisymmetry in the Lorentz indices is needed to rule out trivial 
dependences coming from the fact that 
$\{\Gamma^\mu,\Gamma^\nu\}=2\eta^{\mu \nu}$.
The left invariance of the new forms in~(\ref{2.4})
\emph{requires} new group parameters
$\varphi^{\mu_1 \dots \mu_p}$ so that (\emph{cf.} (\ref{2.3b}))
\begin{equation}
\Pi^{\mu_1\dots \mu_p} =
d \varphi^{\mu_1 \dots \mu_p}  +
{1\over 2}(C \Gamma^{\mu_1\dots \mu_p})_{\alpha \beta}
\theta^\alpha \Pi^\beta
\quad.
\label{2.8}
\end{equation}
The superalgebra generator (LI vector field) $Z_{\mu_1\dots \mu_p}$,
corresponding to $\Pi^{\mu_1\dots \mu_p}$, is realized 
by $Z_{\mu_1\dots \mu_p}= \partial / \partial
\varphi^{\mu_1\dots \mu_p}$ on the extended group manifold.

At this stage there are no restrictions coming from the 
Jacobi identity, equivalent to
$d(d \Pi^{\mu_1\dots \mu_p})= 0$, which follows
trivially from $d \Pi^\alpha =0$. This is an 
alternative way of stating that the $p$-tensor-valued 
mapping on $\text{sTr}_D \otimes \text{sTr}_D$,
\begin{equation}
\xi^{\mu_1\dots \mu_p} (\theta',\theta)=
\theta'^\alpha (C \Gamma^{\mu_1\dots \mu_p})_{\alpha \beta} 
\theta^\beta\quad,
\label{jcocycle}
\end{equation}
satisfies trivially the two-cocycle condition
\begin{equation}
\xi(\theta,\theta') +
\xi(\theta + \theta' ,\theta'') =
\xi(\theta , \theta' + \theta'') +
\xi(\theta' ,\theta'')
\, .
\label{cocycle}
\end{equation}
The symmetry of $(C\Gamma ^{\mu_1 \dots \mu_p})_{\alpha \beta}$ 
is needed to prevent 
the two-cocycle (\ref{jcocycle}) from being trivial (\ie, a 
two-coboundary), since $\eta(\theta)$=
$\theta^\alpha (C \Gamma^{\mu_1\dots 
\mu_p})_{\alpha \beta} \theta^\beta$ on $\text{sTr}_D$, which might 
generate $\xi$ through $\xi_{cob}(\theta',
\theta)\equiv \eta(\theta'+\theta) - \eta(\theta') 
-\eta(\theta)$, is identically zero.
Thus, (\ref{jcocycle}) defines a non-trivial extension.
For a given spacetime dimension $D$, the symmetry condition 
restricts the rank of the tensors that are allowed in (\ref{2.4}).
Hence, the problem of finding all \emph{central}
extensions of the algebra $\{D_\alpha,D_\beta\}=0$
(or of the Lie FDA (\ref{2.2})) is reduced
to finding a basis of the symmetric space $\Pi^{(\alpha}\otimes
\Pi^{\beta)}$ in terms of tensors 
$(C \Gamma^{\mu_1\dots \mu_p})_{\alpha\beta}$ symmetric in 
$\alpha,\beta$; they define the Lie algebra CE two-cocycles
$\Pi^\alpha (C \Gamma^{\mu_1\dots \mu_p})_{\alpha \beta}\Pi^\beta$.

When $D$ is \emph{even}, the space of matrices with
indices $(\alpha \beta)$ is $2^D$-dimensional. Since 
$2^{D}= \sum_{p=0}^D {D \choose p}$, a basis for this 
space is provided by the $(2^D-1)$ matrices given by the Lorentz tensors
$\Gamma^{\mu_1\dots \mu_p}$ of rank $1\le p \le D$ plus the unit matrix.
For $D$ \emph{odd}, the spinors have dimension $2^{(D-1)/2}$ and, since
$2^{D-1} =  \sum_{p=0}^{(D-1)/2} {D \choose p}$, a basis is
provided by the $(2^{D-1} -1)$ matrices given by the
tensors $\Gamma^{\mu_1\dots \mu_p}$
of rank $1\le p \le (D-1)/2$ plus the unit matrix.
The difference is a consequence of the fact that, for any $D$,
\begin{equation}
\Gamma^{\mu_1\dots \mu_p} \Gamma^{D+1} \propto
\epsilon^{\mu_1\dots \mu_D} \Gamma_{\mu_{p+1}\dots \mu_D}
\quad,
\label{2.6}
\end{equation}
where $\Gamma^{D+1}$ is the chirality matrix.
For $D$ odd, $\Gamma^{D+1}\propto 1$,
and only the tensors of rank $0\le p \le (D-1)/2$ 
are linearly independent.

For $D$ \emph{even}, $C \Gamma^{\mu_1\dots \mu_p}$
satisfies (see \eg, \cite{Wes:98})
\begin{equation}
\begin{array}{c}
\displaystyle
(C \Gamma^{\mu_1\dots \mu_p}) = \epsilon (-1)^{(p-1)(p-2)/2}
(C\Gamma^{\mu_1\dots \mu_p})^T \quad,\quad
\mu=0,1,\dots,D-1
\\[0.3cm]
\displaystyle
\epsilon = -\sqrt{2} \cos {\pi\over 4} (D+1)
\quad.
\end{array}
\label{2.9}
\end{equation}
Thus, $\epsilon$=$1$ $(-1)$ for $D$=$2,4\ (6,8)\ [\text{mod}\,8]$
so that $(C \Gamma^{\mu_1\dots \mu_p})_{\alpha\beta}$ is 
symmetric for $p$= $1,2\ [\text{mod}\,4]$ 
if $D=2,4\ [\text{mod}\,8]$ and for $p=3,4$ if 
$D=6,8\ [\text{mod}\,8]$.
For $D$ \emph{odd}, it turns out that the same condition,
$\epsilon (-1)^{(p-1)(p-2)/2}$=$1$, holds for
$D$=$3\ [\text{mod}\, 4]$ with $\epsilon$ = 
$\displaystyle -\sqrt{2} \cos {\pi\over 4}D$.
We have excluded here (somewhat arbitrarily) the $D$=$5, 9$ cases 
because in these dimensions no $C$ such that
$C \Gamma^{\mu} C^{-1} = - {\Gamma^{\mu}}^T$ exists.

The number of cohomology spaces 
$H_{\sigma}^2(\text{sTr}_D \circ Spin(1,D-1))$
for various $\text{sTr}_D$ groups is given in the table. 
As a two-form, the various CE two--cocycles 
are given by $d\theta (C\Gamma^{\mu_1 \dots \mu_p}) d\theta$. 
The corresponding new generators $Z_{\mu_1 \dots \mu_p}$ are
all central, as is $X_{\mu}$ itself. They are on the same footing
and may thought of as generalised momenta. Each of 
the resulting extensions {\it defines} an extended superspace 
group; we will denote them generically by $\tilde{\Sigma}$.

The table also includes the cases in which the spinor is 
Majorana-Weyl or complex (Dirac and Weyl).
If the spinor is complex the independent tensors $\Gamma^0 
\Gamma^{\mu_1\dots \mu_p}$ may appear. The effect of 
considering Weyl spinors is taken into account by introducing 
a chiral projector ($\mathcal{P}_+$, say).

The different extended supersymmetry algebras can be easily
found from the results in the table. We shall only give below
two examples which contain formulae that will be explicitly 
used later on. 
To avoid cumbersome 
factorials, we use a normalization of the generators which is
tantamount to defining the duality relations by
$\Pi^{\mu_1\dots\mu_p}(Z_{\nu_1\dots\nu_p})=
\frac{1}{p!}\epsilon^{\mu_1\dots\mu_p}_{\nu_1\dots\nu_p}$ so that
$\Pi^{\mu_1\dots\mu_p}(C\Gamma^{\nu_1\dots\nu_p}Z_{\nu_1\dots\nu_p})=
C\Gamma^{\mu_1\dots\mu_p}$.

\begin{figure}

\begin{turn}{90}
\begin{minipage}{\textheight}
\scalebox{.9}
{
\begin{tabular}{|c|c|c|lcl|c|c|c|c|cc|}
\hline
\bf{1}
&
\bf{2}
&
\bf{3}
&
\,
&
\bf{4}
&
\,
&
\bf{5}
&
\bf{6}
&
\bf{7}
&
\bf{8}
&
\multicolumn{2}{c|}
{\bf{9}}
\\
\hline
$D$ & 
$n$=$2^{[D/2]}$ & ${n(n+1)\over 2}$ &
\multicolumn{3}{l|}{
\begin{minipage}{0.1\textheight} real $n'$ [complex] dim. of spinor
\end{minipage}
}
&
\begin{minipage}{0.15\textheight}
rank $p$ of symmetric $(C \Gamma^{\mu_1\dots \mu_p})_{\alpha\beta}$
$\{(C \Gamma^{\mu_1\dots \mu_p}\mathcal{P}_+)_{\alpha\beta}\}$
\end{minipage}
&
\multicolumn{1}{l|}
{\begin{minipage}{0.18\textheight}
$\dim (C \Gamma^{\mu_1\dots \mu_p})$ 
$\{\dim (C \Gamma^{\mu_1\dots \mu_p}\mathcal{P}_+)\}$\, ;
total real [complex] dimension
\end{minipage}}
&
\multicolumn{1}{l|}{\begin{minipage}{0.1\textheight}
rank of 
$(\Gamma^0 \Gamma^{\mu_1\dots \mu_p})$
$\{(\Gamma^0 \Gamma^{\mu_1\dots \mu_p}\mathcal{P}_+)\}$
\end{minipage}}
&
\begin{minipage}{0.13\textheight}
$\dim(\Gamma^0 \Gamma^{\mu_1\dots \mu_p})$
$\{\dim(\Gamma^0 \Gamma^{\mu_1\dots \mu_p}\mathcal{P}_+)\}$
\end{minipage}
&
\multicolumn{2}{l|}{
\begin{minipage}{0.14\textheight}
real $H^2_\sigma(\text{sTr}_D\circ Spin(1,D-1))$ spaces
$(\dim H^2_0(\text{sTr}_D))$
\end{minipage}
}
\\
\hline
2 & 2 & 3 & 1 & & {\bfseries MW} & \{1\} & \{$\frac 12$ 2 ; 1 \} &
\multicolumn{2}{c|}{} & 1 & (1)
\\
2 & 2 & 3 & 2 & &  M & 1,2 & 2,1; 3 &
\multicolumn{2}{c|}{do not contribute further} &
2 & (3)
\\
3 & 2 & 3 & 2 & & {\bfseries M} & 1 & 3; 3 &
\multicolumn{2}{c|}{in the real case}
& 1 & (3)
\\
4 & 4 & 10 & 4 & & {\bfseries M} & 1,2 & 4,6; 10 &
\multicolumn{2}{c|}{} & 2 & (10)
\\
\hline
6 & 8 & 36 & 8 & [4] & {\bfseries W} & \{3\} & \{$\frac 12$ 20 ; 20 [10]\}
& \{1,3,5\} & \{6, $\frac 12$ 20 ; 16\} & 3 & (36)
\\
6 & 8 & 36 & 16 & [8] &  D  & 0,3,4 & 1,20,15 ; 72 [36] &
0,1,2,3,4,5,6 & 1,6,15,20,15,6,1 ; 64 & 10 & (136)
\\
7 & 8 & 36 & 16 & [8] & {\bfseries D} & 0,3 & 1,35 ; 72 [36] &
0,1,2,3, & 1,7,21,35 ; 64 & 6 & (136)
\\
8 & 16 & 136 & 16 & [8] & {\bfseries W} & \{0,4,8\} &
\{1, $\frac 12$ 70 ; 72 [36]\} & \{1,3,5,7\} &
\{8,56 ; 64\} & 4 & (136)
\\
8 & 16 & 136 & 32 & [16] &  D  & 0,3,4,7,8 &
1,56,70,8,1; 272 [136] & 0,1,2,3,4,5,6,7,8 & 1,8,28,56,70,56,28,8,1 ; 256 &
14 & (528)
\\
\hline
10 & 16 & 136 & 16 & & {\bfseries MW} & \{1,5,9\} 
& \{10, $\frac 12$ 252; 136\} &
\multicolumn{2}{c|}{} & 2 & (136)
\\
10 & 32 & 528 & 32 & &  M  & 1,2,5,6,9,10 & 10,45,252,210,10,1; 528 &
\multicolumn{2}{c|}{do not contribute further} &
6 & (528)
\\
11 & 32 & 528 & 32 & & {\bfseries M} & 1,2,5 & 11,55,462; 528 &
\multicolumn{2}{c|}{in the real case}
& 3 & (528)
\\
12 & 64 & 2080 & 64 & & {\bfseries M} 
& 1,2,5,6,9,10 & 12,66,792,924,220,66; 2080 &
\multicolumn{2}{c|}{} & 6 & (2080)
\\
\hline
\end{tabular}
}
\vspace*{0.5cm}
\newline
\begin{small}
\emph{Some Lie algebra second cohomology groups for 
$\text{sTr}_D$ (minimal spinors are in boldface)}.
$n$ is the complex dimension of a Dirac spinor, 
equal to the real dimension of Majorana spinors for 
$D=2,3,4\ [\text{mod}\,8]$.
The fourth column gives the dimension of the spinor indicated.
The fifth and sixth column give the ranks for which
$(C \Gamma^{\mu_1\dots \mu_p})_{\alpha\beta}$
(or $(C \Gamma^{\mu_1\dots \mu_p} \mathcal{P}_+)_{\alpha\beta}$)
are symmetric (as deduced from
(\ref{2.9})) and the dimension of these Lorentz tensors; $C$ itself is 
symmetric in $D=6,7,8\ [\text{mod}\,8]$.
The seventh and eigth columns do the same for the additional tensors
$(\Gamma^0 \Gamma^{\mu_1\dots \mu_p})$
$(\Gamma^0 \Gamma^{\mu_1\dots \mu_p}\mathcal{P}_+)$ appearing in 
the complex spinor case. These hermitian (adding $i$ when needed)
tensors are limited by duality (eqn.~(\ref{2.6})) in the odd 
($D=7$) case and to odd rank by the presence of $\mathcal{P}_+$ 
in the Weyl case. The  $\frac 12$ indicates 
halving due to self-duality.
The number of real cohomology groups
$H_\sigma^2(\text{sTr}_D\circ Spin(1,D-1))$
is given by the first number in the last column.
These spaces are the relevant (\ie tensorial) ones,
once the Lorentz symmetry is considered since in this case
$\text{sTr}_D\circ Spin(1,D-1)$ 
(rather than $\text{sTr}_D$)
is the group to be extended.
The action $\sigma$ of $Spin(1,D-1)$ on the extension cocycles 
is automatically taken 
into account by using only Lorentz covariant objects for them.
The bracketed number in the last column ignores the Lorentz
part and, as a result,
$\dim H^2_0(\text{sTr}_D) ={n' \choose 2} + n'$
since the elements of $\text{sTr}_D$ are odd
(for an ordinary $n'$-dimensional  abelian group $\dim H_0^2 = 
{n' \choose 2}$).
The number ${n' \choose 2} + n'$ is given nevertheless since it 
serves as a check on the degrees of freedom: it is equal to the sum of the 
total real dimensions in the sixth and eigth columns.
%\label{table2.1}
\end{small}
\end{minipage}
\end{turn}
\end{figure}
%%%%%%%%%%%%%%%%%%%%%%%%%%%%%%%%%%%%%%%%%%%%%%%%%%%%%%%%%%%%%%%%%%
%%%%%%%%%%%%%%%%%%%%%%%%%%%%%%%%%%%%%%%%%%%%%%%%%%%%%%%%%%%%%%%%%%
\subsection{Applications}
%%%%%%%%%%%%%%%%%%%%%%%%%%%%%%%%%%%%%%%%%%%%%%%%%%%%%%%%%%%%%%%%%%
%%%%%%%%%%%%%%%%%%%%%%%%%%%%%%%%%%%%%%%%%%%%%%%%%%%%%%%%%%%%%%%%%%%%%%
\subsubsection{N=1 theory extended superspace}
\label{sectene1}
For $D$ even, the basic spinors in (\ref{st}) may be reduced to 
$2^{D/2-1}$-dimensional Weyl spinors, 
and the discussion of the possible 
$H_\sigma^2(\text{sTr}_{D}\circ Spin(1,D-1))$ spaces must take 
this into account.
Let $D=2\ [\text{mod}\, 8]$ and let $\theta^\alpha$ be
MW. The symmetry of 
$(C\Gamma^{\mu_1\dots \mu_p}\mathcal{P}_\pm)_{\alpha\beta}$
is now achieved if \emph{both}
$C\Gamma^{\mu_1\dots \mu_p}$ and $C\Gamma^{\mu_1\dots 
\mu_p}\Gamma^{D+1}$
(or, on account of (\ref{2.6}), $C\Gamma^{\mu_1\dots \mu_{D-p}}$)
are symmetric .
Hence, there are central charges for $D=2,\ p=1$
and $D=10,\ p=1,5,9$ (\ie, $p$=1 $[\text{mod}\, 4]$).
As a result, the \emph{$D$=10,$\,$N=1 extended superspace algebra} 
has the form
\begin{equation}
\{D_\alpha^+,D_\beta^+\}=
(C\Gamma^\mu \mathcal{P}_+)_{\alpha\beta} X_\mu
+ (C\Gamma^{\mu_1\dots \mu_5} \mathcal{P}_+)_{\alpha\beta} 
Z_{\mu_1\dots \mu_5}
+ (C\Gamma^{\mu_1\dots \mu_9} \mathcal{P}_+)_{\alpha\beta} 
Z_{\mu_1\dots \mu_9} 
\quad.
\label{ene1}
\end{equation}
Due to $\mathcal{P}_+$ and to
(\ref{2.6}), the first and last term in the r.h.s 
may be grouped into a single one,
$(C\Gamma^\mu \mathcal{P}_+)(X_\mu+Z_\mu)$;
classically, $Z_\mu$ may be absorbed by redefining $X_\mu$ and
the previous analysis shows that the vector--valued cohomology
space is one--dimensional. The second term may be rewritten as
$(C\Gamma^{\mu_1\dots \mu_5}) Z^+_{\mu_1\dots \mu_5}$ where 
$Z^+_{\mu_1\dots \mu_5}$ is a self-dual 5-tensor \cite{Hol.Pro:82},
$\displaystyle Z^+_{\mu_1\dots \mu_5} = 
{1\over 2}(Z_{\mu_1\dots \mu_5} +
{\epsilon_{\mu\dots\mu_5}}^{\mu_6\dots \mu_{10}} 
Z_{\mu_6\dots \mu_{10}}$),
with half the number of components of $Z_{\mu_1\dots \mu_5}$.
As a result, the degrees of freedom in eqn. (\ref{ene1}) 
match:
${16\choose 2} + 16 = 136 = 10 + \frac{1}{2} 
{10\choose 5}$ (see table). But in general $Z_\mu$ 
cannot be reabsorbed, since the Green-Schwarz action for the 
heterotic superstring produces such a contribution to the 
algebra \cite{Azc.Gau.Izq.Tow:89}, of an origin different
from that of $X_\mu$. Mathematically, this corresponds to 
the fact that the group parameters are different for $X_\mu$ 
(translations $x^\mu$) and $Z_\mu$ ($\varphi^\mu$);
they are locally equivalent, much in the same way $\mathbb{R}\sim S^1$
locally, but they are different globally.
We may, however, achieve the symmetry under the exchange 
of $X_9$ 
and $Z_9$ (say) when the 9-direction is a circle of radius $R$.
Then the spectra of $X_9$ and $Z_9$ are isomorphic under the 
T-duality exchange $R\to 1/R$ (see \cite{Tow:97,Giv.Por.Rab:94}).

The FDA form of the $D$=10, $N$=1 superalgebra (\ref{ene1}) is  
given by the MC relations
\begin{equation}
d\Pi^\alpha =0\, , \qquad
d\Pi^\mu = {1\over 2} (C\Gamma^\mu)_{\alpha\beta} \Pi^\alpha \Pi^\beta
\, ,
\qquad d\Pi^{\mu_1\dots \mu_5} = 
{1\over 2} (C\Gamma^{\mu_1\dots \mu_5})_{\alpha\beta} \Pi^\alpha \Pi^\beta
\, ,
\label{MCene1}
\end{equation}
where $\Pi^\alpha$, $\Pi^\mu$ and $\Pi^{\mu_1\dots \mu_5}$ are 
defined as ($\alpha=1,\dots,32$)
\begin{equation}
\Pi^\alpha = \mathcal{P}_+ d \theta^\alpha 
\, , \qquad
\Pi^\mu = dx^\mu + {1\over2} (C\Gamma^\mu)_{\alpha\beta} \theta^\alpha 
\Pi^\beta
\, , \qquad
\Pi^{\mu_1\dots \mu_5} = d \varphi^{\mu_1\dots \mu_5} +
{1\over2} (C\Gamma^{\mu_1\dots \mu_5})_{\alpha\beta} \theta^\alpha
\Pi^\beta
\, .
\end{equation}
If $Z_\mu$ is included separately, this 
introduces a further extension which requires adding a new LI 
form associated with it, $\Pi_\mu^{(\varphi)}$,
$d\Pi_\mu^{(\varphi)}={1 \over 2} (C \Gamma_{\mu}
\mathcal{P}_+)_{\alpha \beta} d\theta^\alpha
d\theta^\beta$ (we have written the index down for consistency 
with later notation, as in Sec. \ref{css}).
At the group level this means that the MW 
translations generate two types of transformations
\ie, one has to distinguish between the translations $x^\mu$ and 
the $\varphi_\mu$, some of which may be compact, in which case
the corresponding group law expression should be understood locally.
%%%%%%%%%%%%%%%%%%%%%%%%%%%%%%%%%%%%%%%%%%%%%%%%%%%%%%%%%%%%%%%%%%%%%
\subsubsection{IIA theory centrally extended superspace}
%%%%%%%%%%%%%%%%%%%%%%%%%%%%%%%%%%%%%%%%%%%%%%%%%%%%%%%%%%%%%%%%%%%%%
Let us consider now the 
$H_\sigma^2(\text{sTr}_{10}\circ Spin(1,9),\text{(IIA)})$ spaces.
The IIA superalgebra is the $D=10$ algebra associated with
two $16$-dimensional spinors of opposite chiralities which may be
combined into a Majorana spinor. Then (see table),
the \emph{IIA theory maximally extended algebra} \cite{Hol.Pro:82}
is found to be
\begin{eqnarray}
\{D_\alpha,D_\beta\} & = &
(C \Gamma^{\mu})_{\alpha\beta} X_\mu
+
(C \Gamma^{\mu_1 \mu_2})_{\alpha\beta} Z_{\mu_1 \mu_2}
+
(C \Gamma^{\mu_1 \dots \mu_5})_{\alpha\beta} Z_{\mu_1  \dots \mu_5}
\nonumber \\
 & & \mbox{} + 
(C \Gamma^{11})_{\alpha \beta} Z
+
(C \Gamma^{\mu}\Gamma^{11})_{\alpha \beta}Z_{\mu}
+
(C \Gamma^{\mu_1 \dots \mu_4} \Gamma^{11} )_{\alpha \beta}
Z_{\mu_1 \dots \mu_4}
\quad,
\label{2.14}
\end{eqnarray}
since the tensor spaces
\{$\Gamma^{\mu_1\dots \mu_p}$\} and \{$\Gamma^{\mu_1\dots
\mu_{D-p}}\Gamma^{11}$\} are isomorphic by eqn.~(\ref{2.6}).
Notice that $X_\mu$ and $Z_\mu$ belong to different cohomology classes
(their corresponding two-cocycles are not cohomologous
due to the presence of $\Gamma^{11}$ in the $Z$ term).
The associated IIA Lie FDA, involving the LI one-forms dual to
the generators in (\ref{2.14}), is given by
\begin{equation}
\addtolength{\arraycolsep}{-1ex}
\begin{array}{rclcrclcrcl}
d \Pi^\alpha & = & 0 
\, , & &
d \Pi^\mu & = & {1\over 2}
(C \Gamma^{\mu})_{\alpha\beta} \Pi^\alpha \Pi^\beta
\, , &  &
d \Pi^\mu_{(z)} & = & {1\over 2} (C \Gamma^{\mu}\Gamma^{11})_{\alpha 
\beta} \Pi^\alpha\Pi^\beta 
\\
d \Pi^{\mu_1 \mu_2} & = & {1\over 2} (C \Gamma^{\mu_1 \mu_2})_{\alpha
\beta} \Pi^\alpha \Pi^\beta
\, , &  &
d \Pi^{\mu_1 \dots \mu_5} & = & {1\over 2}
(C \Gamma^{\mu_1 \dots \mu_5})_{\alpha\beta} \Pi^\alpha \Pi^\beta
\, , &  &
 & & 
\\
d \Pi^{\mu_1 \dots \mu_4} & = & {1\over 2}
(C \Gamma^{\mu_1 \dots \mu_4} \Gamma^{11})_{\alpha \beta}
\Pi^\alpha \Pi^\beta
\, , &  &
d \Pi & = & {1\over 2} (C \Gamma^{11})_{\alpha \beta} \Pi^\alpha 
\Pi^\beta
\, .
\end{array}
\label{2.15}
\end{equation}
The new group parameters define the \emph{IIA theory centrally extended
superspace}, parametrized by the coordinates
$(\theta^\alpha,x^\mu,\varphi^\mu,\varphi^{\mu_1 \mu_2},
\varphi^{\mu_1 \dots \mu_5},
\varphi^{\mu_1 \dots \mu_4},\varphi)$. 

The IIB case with $\Pi^{\alpha\,i} \equiv \mathcal{P}_+ 
d\theta^{\alpha\,i}$
$(i=1,2)$ is treated similarly by noticing that the presence of 
$\epsilon_{ij}$ allows for $C\Gamma^{\mu_1\mu_2\mu_3}\mathcal{P}_+$,
which is skew-symmetric.
%%%%%%%%%%%%%%%%%%%%%%%%%%%%%%%%%%%%%%%%%%%%%%%%%%%%%%%%%%%%%%%%%%%%%%
%%%%%%%%%%%%%%%%%%%%%%%%%%%%%%%%%%%%%%%%%%%%%%%%%%%%%%%%%%%%%%%%%
%%%%%%%%%%%%%%%%%%%%%%%%%%%%%%%%%%%%%%%%%%%%%%%%%%%%%%%%%%%%%%%%%
\section{Non-central extensions and their superspaces}
\label{NCESA}
%%%%%%%%%%%%%%%%%%%%%%%%%%%%%%%%%%%%%%%%%%%%%%%%%%%%%%%%%%%%%%%%%
We start now from standard rigid superspace, eqns.~(\ref{2.2}),
(\ref{2.3}) for real, odd translations.
To keep the discussion as general as possible, we rescale 
$\Pi^\mu$, $\Pi_{\mu_1\dots \mu_p}$ by an arbitrary dimensionless 
constant $a_s$, so
that~(\ref{2.3}), (\ref{2.4}) become
\begin{equation}
d \Pi^\mu = a_s (C \Gamma^\mu)_{\alpha\beta}
            \Pi^\alpha \Pi^\beta
\quad,\quad
d\Pi_{\mu_1\dots \mu_p} \equiv a_0
(C \Gamma_{\mu_1\dots \mu_p})_{\alpha\beta} \Pi^\alpha \Pi^\beta
\quad.
\label{dPimu}
\end{equation}

  Let us fix $p$ and consider the
resulting extended superspace, parametrized by
$(\theta^\alpha, x^\mu, \varphi_{\mu_1 \dots \mu_p})$, as our
starting group manifold. We look for a non-trivial CE two-cocycle 
with $p$ indices on the above extended superspace.
This may now involve any of the LI forms available,
$\Pi^\mu, \Pi^\alpha$ or $\Pi_{\mu_1 \dots \mu_p}$. Inspection of the
possible Lorentz tensors shows that the external Lorentz
indices of this two-cocycle have to be of the type 
$(\mu_1 \dots \mu_{p-1}
\alpha_1)$ and, hence, the only available LI two-forms are
\begin{equation}
\rho^{(1)}_{\mu_1 \dots \mu_{p-1} \alpha_1}=(C \Gamma_{\nu \mu_1 \dots
\mu_{p-1}})_{\beta \alpha_1} \Pi^\nu \Pi^\beta
\quad, \quad \quad
\rho^{(2)}_{\mu_1 \dots \mu_{p-1} \alpha_1}= (C \Gamma^\nu)_{\beta
\alpha_1} \Pi_{\nu \mu_1 \dots \mu_{p-1}} \Pi^\beta
\label{cand1}
\quad .
\end{equation}
For $p=1$, both are closed. For $p \geq 2$, 
$d (\rho^{(1)} + \lambda_2 \rho^{(2)})=0$ gives
$\displaystyle \lambda_2={a_s \over a_0}$ 
provided\footnote{
Primed indices are understood to be symmetrised (with unit weight).}
\begin{equation}
(C \Gamma^\nu)_{\alpha' \beta'} (C \Gamma_{\nu \mu_1 \dots
\mu_{p-1}})_{\gamma' \delta'}=0
\quad,
\label{Gammaprop}
\end{equation}
which holds only for certain values of 
$(D,p)$~\cite{Ach.Eva.Tow.Wil:87} (for $p=1$, $D=3,4,10$ and, with 
the appropriate modifications for complex spinors, $D=6$). The 
existence of such a constraint, on both $D$ and $p$, is a new 
feature---the non-triviality of (\ref{jcocycle})
only restricted $p$. We introduce now a new one--form 
$\Pi_{\mu_1 \dots \mu_{p-1} \alpha_1}$ with
\begin{equation}
d\Pi_{\mu_1 \dots \mu_{p-1} \alpha_1} = a_1 \left((C \Gamma_{\nu
\mu_1 \dots \mu_{p-1}})_{\beta \alpha_1} \Pi^\nu \Pi^\beta
+ {a_s \over a_0} (C\Gamma^\nu)_{\beta \alpha_1} \Pi_{\nu \mu_1
\dots \mu_{p-1}} \Pi^\beta)\right)
\label{dPim1a1}
\end{equation}
(for $p=1$ the coefficient of the second term can be arbitrary,
see Sec.~\ref{css}).\footnote{
For $p=1$, the one-form in the l.h.s. of~(\ref{dPim1a1}) becomes
$\Pi_\alpha$ -- notice that this is unrelated to $\Pi^\alpha$ (so
that \eg \ $d\Pi_\alpha$ is non-zero). In general, we will not
raise or lower the Lorentz indices of forms, their position being
used to distinguish between different types of them as in
\cite{Sez:97}.}
The above MC equation implies that both $\left[D,X\right]$
and $\left[D,Z^{\mu_1 \dots \mu_p}\right]$  are 
modified by a
term proportional to $Z^{\mu_1 \dots \mu_{p-1} \alpha_1}$, the
latter being the only central generator at this stage ($Z^{\mu_1
\dots \mu_{p-1} \alpha_1}$ is central because,
by construction,  $\Pi_{\mu_1 \dots\mu_{p-1} \alpha_1}$ 
cannot appear at the r.h.s. of a MC equation
expressing the differential of a LI form). 
This is a general feature of the extension scheme in this section: 
at any stage in the chain of extensions, the only central
generator present is the last one introduced. Thus, each extension
is central, but the resulting algebra/group is not a central 
extension of superspace: all generators but the last one 
have non-zero commutators as a consequence of the 
subsequent extensions. A second feature here is that 
successive extensions substitute one spinorial index for a vectorial 
one, preserving  the total number of indices. The chain
ends with the introduction of a generator 
with $p$ spinorial indices.

Repeating the above procedure, one finds that the next three
extensions are in some sense exceptional (see~(\ref{dPim1a24})
below), while the one introducing five spinorial indices and 
all others after it follow a pattern which can be used to 
derive a recursion formula. Skipping the somewhat involved 
algebra (see Appendix A), we list first the results for the next 
three extensions

\begin{eqnarray}
d \Pi_{\mu_1 \dots \mu_{p-2} \alpha_1 \alpha_2} & = &
a_2 \left( (C \Gamma_{\nu \rho \mu_1 \dots \mu_{p-2}})_{ \alpha_1
\alpha_2} \Pi^\nu \Pi^\rho
+ {a_s \over a_0} (C \Gamma^\nu)_{\alpha_1 \alpha_2}
  \Pi_{\nu \rho \mu_1 \dots \mu_{p-2}} \Pi^\rho \right. \nonumber \\
& & - {a_s \over a_1}  \left. (C \Gamma^\nu)_{\alpha_1 \alpha_2}
   \Pi_{\nu \mu_1 \dots \mu_{p-2} \beta} \Pi^\beta
   -8 {a_s \over a_1}  (C \Gamma^\nu)_{\alpha'_1 \beta}
   \Pi_{\nu \mu_1 \dots \mu_{p-2} \alpha'_2} \Pi^\beta \right)
\quad,
\nonumber \\
d \Pi_{\mu_1 \dots \mu_{p-3} \alpha_1 \alpha_2 \alpha_3} & = &
a_3 \left( (C \Gamma^\nu)_{\alpha'_1 \alpha'_2}
  \Pi_{\nu \rho \mu_1 \dots \mu_{p-3} \alpha'_3} \Pi^\rho \right.
+ {5a_1 \over 4a_2}  (C \Gamma^\nu)_{\alpha'_1 \beta}
   \Pi_{\nu \mu_1 \dots \mu_{p-3} \alpha'_2 \alpha'_3} \Pi^\beta
\nonumber \\
& & + {a_1 \over 4a_2}  \left. (C \Gamma^\nu)_{\alpha'_1 \alpha'_2}
   \Pi_{\nu \mu_1 \dots \mu_{p-3} \beta \alpha'_3} \Pi^\beta \right)
\quad,
\nonumber \\
d \Pi_{\mu_1 \dots \mu_{p-4} \alpha_1 \alpha_2 \alpha_3
        \alpha_4} & = &
a_4 \left( (C \Gamma^\nu)_{\alpha'_1 \alpha'_2}
  \Pi_{\nu \rho \mu_1 \dots \mu_{p-4} \alpha'_3 \alpha'_4} \Pi^\rho
\right.
 - {48a_sa_2 \over 5a_1 a_3}  (C \Gamma^\nu)_{\alpha'_1 \beta}
   \Pi_{\nu \mu_1 \dots \mu_{p-4} \alpha'_2 \alpha'_3 \alpha'_4}
   \Pi^\beta \nonumber \\
& & - {12a_s a_2\over 5a_1 a_3}  \left. 
(C \Gamma^\nu)_{\alpha'_1 \alpha'_2}
   \Pi_{\nu \mu_1 \dots \mu_{p-4} \beta \alpha'_3 \alpha'_4} \Pi^\beta
\right)
\label{dPim1a24}
\end{eqnarray}
(the $a_k$'s in the r.h.s. normalise the $\Pi$'s with $k$
spinorial indices). For the remaining extensions, which introduce 
one-forms with five or more spinorial indices, one
establishes the following recursion formula
\begin{eqnarray}
d \Pi_{\mu_1 \dots \mu_{p-(k+2)} \alpha_1 \dots \alpha_{k+2}} & = &
a_{k+2} \left\{ (C \Gamma^\nu)_{\alpha'_1 \alpha'_2}
  \Pi_{\nu \rho \mu_1 \dots \mu_{p-(k+2)} \alpha'_3 \dots
   \alpha'_{k+2}} \Pi^\rho \right. \nonumber \\
& &
    + \lambda^{(k+2)}_2 (C \Gamma^\nu)_{\alpha'_1 \beta}
   \Pi_{\nu \mu_1 \dots \mu_{p-(k+2)} \alpha'_2 \dots \alpha'_{k+2}}
   \Pi^\beta \nonumber \\
& & 
   \left. + \lambda^{(k+2)}_3 (C \Gamma^\nu)_{\alpha'_1 \alpha'_2}
   \Pi_{\nu \mu_1 \dots \mu_{p-(k+2)} \beta \alpha'_3 \dots
   \alpha'_{k+2}}
   \Pi^\beta \right\}
\quad,
\label{dPim1ak2}
\end{eqnarray}
where
\begin{equation}
\lambda^{(k+2)}_2  =  -{a_s \over a_{k+1}}\left( {2 \over
\lambda^{(k+1)}_2} + {k \over \lambda^{(k+1)}_3}\right) \quad ,\quad
\lambda^{(k+2)}_3  =  -{a_s \over a_{k+1}}{k+1 \over
\lambda^{(k+1)}_2}
\quad.
\label{la23}
\end{equation}
Notice that the above recursion starts at $k=3$, which implies $p
\geq 5$. On the other hand, the maximum value of $p$ (of 
interest to us) for 
which~(\ref{Gammaprop}) holds true is $p=5$,
\ie~(\ref{dPim1ak2}) is relevant here only for $k=3$, $p=5$. 
It is easily checked that $\left[ \Pi_{\mu_1 \dots \mu_{p-l} \alpha_1
\dots \alpha_l} \right]=L^{1+{l \over 2}}$.
We give related explicit results, for $p=1,2$, in Sec.~\ref{p12}. 

%%%%%%%%%%%%%%%%%%%%%%%%%%%%%%%%%%%%%%%%%%%%%%%%%%%%%%%%%%%%%%%%%
%%%%%%%%%%%%%%%%%%%%%%%%%%%%%%%%%%%%%%%%%%%%%%%%%%%%%%%%%%%%%%%%%
%%%%%%%%%%%%%%%%%%%%%%%%%%%%%%%%%%%%%%%%%%%%%%%%%%%%%%%%%%%%%%%%%
\section{Structure of the new superspaces and Noether currents}
\label{Nns}
%%%%%%%%%%%%%%%%%%%%%%%%%%%%%%%%%%%%%%%%%%%%%%%%%%%%%%%%%%%%%%%%%

\subsection{Fibre bundle structure}
\label{fbs}

  All extended superspaces have a natural  
bundle structure, in which the basis is the group to be extended
and the fibre is the group by which we extend. For instance,
superspace $\Sigma$ itself and the various extensions
$\tilde \Sigma$ in Sec. \ref{Mces} may be considered as the 
total spaces of principal bundles over the $\text{sTr}_D$'s
of the specific theory. The two-forms which define the 
extensions are curvatures of invariant 
connections valued on the central algebras by 
which $\text{sTr}_D$ is extended. The $D$'s are then the 
horizontal lifts of the vector fields 
${\partial \over\partial \theta^\alpha}$ on the specific 
base manifold $\text{sTr}_D$; this justifies the `covariant
derivative' name which may be given to the $D$'s in
the algebra of the $\tilde\Sigma$'s.
 Similar considerations apply at any step in the chain of extensions 
in Sec. \ref{NCESA}. From this point of view, after the last
step, one has a 
bundle structure with the last coordinate in the fibre and all
the rest in the base. As we show below, there is also
another relevant 
 bundle structure with $\Sigma$ in the
base and all new coordinates in the fibre.

 Let us now discuss the general case treated in
Sec.~\ref{NCESA}. $\tilde{\Sigma}$ is parametrised by the
coordinates $(\theta^\alpha, x^\mu, \varphi_{\mu_1 \dots
\mu_p},\varphi_{\mu_1 \dots\mu_{p-1}\alpha_1},\dots,
\varphi_{\alpha_1 \dots \alpha_p})$. We will denote
them collectively by the row vector $\phi = (z^a, \varphi_A)$ where $z$
parametrizes the base (superspace $\Sigma$) and $\varphi$ the fibre (the
space of all new coordinates). Referring to 
this block form, we will say that the superspace part is
`two--dimensional' while the fibre part has `dimension' $p+1$. 
The LI
one--forms that reduce to the differentials of these coordinates
at the identity will be denoted by $\Pi=(H^a \, , \, \, \Theta_A)$ 
and the
dual LI vector fields by the column vector $Z=(D_a \, ,  \, \, Y^A)^t$ 
($t$
denotes matrix transposition). The corresponding RI objects will
carry an additional hat. 

Under a right group transformation, $g
\rightarrow gg'$, $Z$ transforms like $Z \rightarrow T'^t Z$,
$T'$
being a matrix of (primed) functions on the group, called the
adjoint representation. It holds 
\begin{equation} 
T''=T'T \quad,
\quad \quad \quad
Z \cdot T \vert_e = \rho_{\hbox{\small adj}}(Z)
\quad,
\label{Tprop}
\end{equation} 
$\rho_{\mbox{\small adj}}(Z)$ being the adjoint representation of $Z$,
given by the structure constants. Inspection of the MC equations
then reveals that $T$ is a lower triangular matrix with units along
the diagonal. We put accordingly 
\begin{equation}
T^t =\left( \begin{array}{cc} A & C \\ 0 & B \end{array} \right) 
\quad,
\quad \quad \quad
(T^t)^{-1} =\left( \begin{array}{cc} A^{-1} & -A^{-1} C B^{-1} \\ 
0 & B^{-1} \end{array} \right) 
\label{Tblock}
\end{equation}
with $A$, $B$ upper triangular matrices. The dimensions of $A$,
$B$, $C$ (in block form) are $2 \times 2$, $(p+1) \times (p+1)$,
$2 \times (p+1)$ respectively. (\ref{Tblock}) shows that the
fibre is a subgroup, with adjoint representation given by $B^t$. 
In this notation, the LI vector fields transform like
\begin{equation}
\left( \begin{array}{c} D \\Y \end{array} \right)
\rightarrow 
\left( \begin{array}{c}  A' \, D + C' \, Y \\
B' \,Y 
 \end{array} \right)
\quad .
\label{Ztrans}
\end{equation}
The LI forms similarly transform
according to $\Pi \rightarrow \Pi (T'^t)^{-1}$, \ie 
\begin{equation}
( \begin{array}{cc} H \, , & \Theta \end{array} )
\rightarrow 
( \begin{array}{cc} H \, A'^{-1} \, , & -H \, A'^{-1} C' B'^{-1} +
\Theta \, B'^{-1} \end{array} )
\quad .
\label{Pitrans}
\end{equation}
For the RI objects it holds $\hat{Z}=(T^t)^{-1} \, Z$, 
$\hat{\Pi}= \Pi \, T^t$ \ie
\begin{equation}
\left( \begin{array}{c} \hat{D} \\ \hat{Y} \end{array} \right)
= 
\left( \begin{array}{c} A^{-1} \, D - A^{-1}CB^{-1} \, Y \\ 
B^{-1} \, Y \end{array} \right)
\quad, \quad \quad \quad
( \begin{array}{cc} \hat{H} \, , &  \hat{\Theta} \end{array} )
=
( \begin{array}{cc} H \, A  \, , & H \, C + \Theta \, B \end{array})
\quad.
\label{ZPihat}
\end{equation}
The Lie algebra valued one--form $\omega = \Theta \, Y$ serves as a
connection in the bundle. Indeed, one easily verifies that $\omega$ is
invariant under~(\ref{Ztrans}),~(\ref{Pitrans}) when $T$ is
restricted to the subgroup of the fibre ($A=I$, $C=0$). The horizontal
subspace is spanned by the kernel of $\omega$, \ie \ by
the components of $D$, the latter being the horizontal lifts of the  
standard superspace generators $D^{(s)}_\alpha ={\partial \over
\partial \theta^\alpha}$, $X^{(s)}_\mu = {\partial \over \partial
x^\mu} + {1 \over 2} \Gmab \theta^\beta {\partial \over \partial
\theta^\alpha}$.

In later applications, in section~\ref{p12}, the
explicit form of the matrix $B^{-1}$ is needed -- we present here
a few remarks that facilitate its computation. Inspection of the
r.h.s. of the MC equations for the new one--forms, in
section~\ref{NCESA}, shows that they always contain one new
one--form, multiplied by a $\Pi^\alpha$ or $\Pi^\mu$. For the
dual Lie algebra this implies that the new generators commute
among themselves and only have, in general, non--zero commutators  
with the superspace generators $D_\alpha$, $X_\mu$. In other
words, the group by which we extend $\Sigma$ is {\em abelian} 
(to begin with) and its generators
acquire, as a result of the extension, non--zero commutators only
with the superspace generators. 
The structure of the resulting Lie algebra is, in symbolic form, 
\be
[D,D] \sim D+Y
\quad, \quad \quad \quad
[D,\, Y] \sim Y
\quad, \quad \quad \quad
[Y,\, Y] = 0
\quad,
\label{LAsymb}
\ee
in the notation introduced earlier. For $T$, the expression 
\be
T = e^{\phi^A \, \rho_{adj}(Z_A)} \equiv e^{z^a \, \rho_{adj}(D_a) 
                                    + \varphi_A \, \rho_{adj}(Y^A)}
\label{Texpr}
\ee
is well known.  From the third of~(\ref{LAsymb}) though, we infer that 
the restriction of $\rho_{adj} \, (Y)$ to the fibre (\ie \
to the sub-block corresponding to $B^t$) is zero.
Denoting this sub-block by $\rho^{(f)}_{adj} \, (Y)$ we find for
$B$
\be
B = e^{z^a \, \,  \rho^{(f)}_{adj} \, (D_a)^t} \equiv e^{\theta^\alpha
\, \, \rho^{(f)}_{adj} \, (D_\alpha)^t + x^\mu \, \,
\rho^{(f)}_{adj} \, (X_\mu)^t}
\quad,
\label{Bexpr}
\ee
where the matrices $\rho^{(f)}_{adj} \, (D_a)$ are given by 
the structure constants that appear in (the explicit form of) the 
second of~(\ref{LAsymb}). The interesting point here is that $B$
depends on $(\theta,x)$ only -- the new variables enter in $T$
only through $A$, $C$. 

%%%%%%%%%%%%%%%%%%%%%%%%%%%%%%%%%%%%%%%%%%%%%%%%%%%%%%%%%%%%%%%%%
\subsection{Invariant actions for the minimal branescan}
\label{seWZt}
As already mentioned, part of
the motivation for studying superspace extensions comes from
their relevance in the construction of manifestly invariant
$p$--brane actions. For the branes of the old branescan,
WZ terms on $\Sigma$ have the form

\begin{equation}
S_{WZ} = \int_W d^{p+1}\xi \, {\cal L}_{WZ} = \lambda \int_W 
\phi^{*} (b)
\quad,
\label{3.1}
\end{equation}
where $b$ is defined\footnote{For an explicit form of the 
quasi-invariant $b$ on $\Sigma$ see \cite{Eva:88}.}
as the potential of the closed $(p+2)$--form $h$ on superspace
\begin{equation}
h = (C \Gamma_{\mu_1\dots \mu_p})_{\alpha\beta} \Pi^{\mu_1}\dots
\Pi^{\mu_p}
\Pi^\alpha \Pi^\beta \quad,\quad db=h
\quad.
\label{3.2}
\end{equation}
$W$ in~(\ref{3.1}) is the $(p+1)$--dimensional worldvolume
swept out by the $p$--brane, parametrized by
$\{\xi^i\} =(\tau, \sigma^1, \dots ,\sigma^p), i=0,1\dots p$
and $\phi^{*}$ is the pullback of the embedding $\phi: W 
\rightarrow \Sigma$. 
The constant $\lambda$ is fixed by the requirement of
$\kappa$--invariance of the total action \cite{Ach.Eva.Tow.Wil:87} 
(we will ignore $\lambda$ henceforth).
As is well known \cite{Ach.Eva.Tow.Wil:87} (see also~\cite{Azc.Tow:89}),
the closure of $h$ is equivalent to the condition (\ref{Gammaprop})
which we have seen to guarantee the existence of the non--central 
extensions of Sec.~\ref{NCESA}. Using the new LI one--forms available
we may obtain a LI potential $\tilde{b}$ for $h$ on $\tilde{\Sigma}$.
Its general form is
\begin{equation}
\tilde{b} = \sum_{k=0}^p  \Pi_{\mu_1\dots \mu_{p-k} \alpha_1 \dots
\alpha_k} ( b_{k}
\Pi^{\mu_1}\dots \Pi^{\mu _{p-k}} \Pi^{\alpha_1} \dots \Pi^{\alpha_k}
) \equiv \Theta_C \Lambda^C
\quad,
\label{bdef}
\end{equation}
where the last equation uses the notation of the previous section and
defines $\Lambda^C$. 
The $b_k$'s are numerical constants, determined by the second eqn. 
in ~(\ref{3.2}). We check that $[ \tilde{b}]=
L^{1+{k \over 2}} L^{p-k} L^{k \over 2} = L^{p+1}$.

We compute now the explicit form for an invariant ${\cal L}_{WZ}$ 
in~(\ref{3.1}) (\ie, with $\tilde{b}$ instead of $b$). For a general 
one--form $\Pi$ we put $\phi^{*}(\Pi) 
\equiv \Pi_i d\xi^i$, so that
\begin{equation}
\Pi^\alpha_i = \partial_i{\theta^\alpha}, \quad \quad \quad
\Pi^\mu_i = \partial_i{x^\mu} + a_s (C \Gamma^\mu)_{\alpha \beta} 
\theta^\alpha \partial_i{\theta^\beta}
\quad.
\label{Pialmui}
\end{equation}
Eqns.~(\ref{bdef}) and~(\ref{3.1}) give for the Lagrangian
\begin{equation}
{\cal L}_{WZ} =   \Theta_{Ci} \Lambda^{Ci}
\quad,
\label{Lexpr2}
\end{equation}
where, expanding multi-indices and using the above notation,
\begin{equation}
\Lambda^{\mu_1 \dots \mu_{p-k} \alpha_1 \dots \alpha_k i} \equiv 
b_k \epsilon^{ij_1 \dots j_p}
   \Pi^{\mu_1}_{j_1} \dots \Pi^{\mu_{p-k}}_{j_{p-k}}
      \Pi^{\alpha_1}_{j_{p-k+1}} \dots \Pi^{\alpha_k}_{j_p}
\quad.
\label{LCidef}
\end{equation}
%%%%%%%%%%%%%%%%%%%%%%%%%%%%%%%%%%%%%%%%%%%%%%%%%%%%%%%%%%%%%%%%%
\subsection{Noether currents for the new symmetries}
\label{noecur}

The invariance under translations of the 
action of the supersymmetric objects implies 
the existence of conserved currents. The integrals of the charge 
densities over a spacelike section of the worldvolume give 
constants of the motion for the
$p$--brane. When the action contains the standard WZ term $b$, the 
Noether current includes a term $\Delta$ coming from the 
quasi-invariance of $b$, $\delta b=d \Delta$. This was used 
in~\cite{Azc.Gau.Izq.Tow:89} to
find topological extensions of the supersymmetry algebra. When $b$ is
replaced by the invariant $\tilde{b}$ in~(\ref{bdef}), $\Delta$ is no
longer present. However, the Noether current receives now a
contribution from the additional fields $\phi^{*} (\varphi_{A})$, 
which leads to the same result.  

One can derive general expressions for the currents 
associated with the new generators.
In the present case, 
where the relevant part of the Lagrangian is obtained
by pulling back to $W$ forms initially defined on
$\tilde{\Sigma}$, it is convenient to work 
on the extended superspace, where quantities have 
a direct geometrical interpretation, and to pull the 
result back to $W$ at the end. 
To keep the discussion general, consider a manifold
$M$ which can serve as worldvolume  and a target space $N$, 
of dimensions $m$, $n$ respectively ($m < n$) and an
embedding $\phi$ of $M$ into $N$, $\phi: x \mapsto y(x)$ where
$\{x^i\}$ ($\{y^j\}$) are local coordinates on $M$ ($N$). Consider 
furthermore an action $S$  given by 
\begin{equation}
S = \int_M \phi^*(\alpha)
\quad,
\label{Sdef}
\end{equation}
where $\alpha$ is a $k$--form on $N$ and $\phi^*$ is the pullback
map associated with the embedding. We assume that the
submanifolds $x^0=x^0_{init}$, $x^0=x^0_{fin}$ of $M$ form its 
boundary
$\partial M$ -- their embeddings in $N$ are the initial and final
configuration respectively. The equations of motion are
\begin{equation}
\delta_Y S= \int_M \phi^*( L_Y \alpha)=0
\quad,
\label{deltaY}
\end{equation}
where $Y$ is an arbitrary vector field on $N$
which vanishes on $\phi(\partial M)$ -- we denote their solutions
generically by $\phi_{cl}$.
Proceeding along the lines of the standard derivation, with inner
derivations taking up the role of the partials
$\frac{\partial}{\partial \phi_{,i}}$, one finds the equations of
motion in the form
\begin{equation}
\phi^*(i_Y d \alpha)=0 \quad ,\label{eom3}
\end{equation}
where now $Y$ is an arbitrary vector field, not necessarily
vanishing on $\partial M$.
For a symmetry generated by $Y_0$ one obtains
\begin{equation}
d \left( \phi^*_{cl} (J_{(Y_0)}) \right) = 0 
\, , \qquad \qquad 
J_{(Y_0)} \equiv i_{Y_0} \alpha
\, ,
\label{Jcons}
\end{equation}
which is the current conservation equation. 
For a quasi-invariant Lagrangian, $\phi^*(L_{Y_0}
\alpha)=\phi^*(d\Delta)$, the conserved current picks up 
a term in
$\Delta$, $\phi^*(J_{(Y_0)})=\phi^*(i_{Y_0} \alpha -
\Delta)$. 

In the present case, $(M,N,\alpha)$ correspond to
$(W,\tilde{\Sigma},{\tilde b})$. The variation of the total action 
from that of
the new coordinates comes only from ${\cal L}_{WZ}$ so
that~(\ref{eom3}) gives
\begin{equation}
0= \phi^*(i_{Y}d \tilde{b}) = \phi^*(i_{Y} h)
\quad,
\label{eomS1}
\end{equation}
where $Y$ is an arbitrary vector field along the fibre. Since $h$
is horizontal, the above equations of motion obtained from
variations of the $\varphi_A$'s, are satisfied trivially,
consistent with the appearence of the 
$\varphi_A$'s in the Lagrangian through exact differentials. For
the Noether currents associated to translations along the new
coordinates we have $Y_0 \rightarrow \hat{Y}^A$ and~(\ref{Jcons})
gives
\begin{equation}
d \left( \phi^*_{cl} (J^A) \right) = 0
\quad, \quad \quad \quad
J^A = i_{\hat{Y}^A} \tilde{b}
\quad.
\label{JconsS}
\end{equation}
With $\tilde{b}$ as in~(\ref{bdef}) and $\hat{Y}^A={(B^{-1})^A}_C
Y^C$ (see~(\ref{ZPihat})),
 the second of~(\ref{JconsS}) gives for $J^A$
\begin{equation}
J^A = {(B^{-1})^A}_D i_{Y^D} \Pi_C \Lambda^C
 = {(B^{-1})^A}_C  \Lambda^C
\quad,
\label{JAS}
\end{equation}
since $i_{Y^D} \Pi_C= {\delta^D}_C$. Notice that $J^A$ is, in
this case, a form on $\Sigma$ (rather than $\tilde{\Sigma}$).
Effecting explicitly the pullback in the first of~(\ref{JconsS}) we
find\footnote{Eqn. (\ref{jAi}) for $j^{Ai}$ also follows from
the standard expression for the current associated with an
`internal' symmetry of a Lagrangian ${\cal L}$,
$j^{Ai}=\delta^{A}\varphi(\xi) {\partial{\cal L} \over \partial_i\varphi
(\xi)}$. However, for the currents considered here the relevant
part of ${\cal L}$ is just ${\cal L}_{WZ}$. Since
${\cal L}_{WZ}$ is obtained from a form on ${\tilde \Sigma}$,
the above derivation allows us to exploit the geometry
of ${\tilde\Sigma}$. The above expression for the current also
makes clear that, within the canonical formalism, the integrated
charge operators will reproduce the original symmetry algebra.}
\begin{equation}
\partial_i j^{Ai}=0
\quad,
\quad \quad \quad 
j^{Ai}(\xi) \equiv {(B^{-1})^A}_C(\xi) \, \Lambda^{Ci}(\xi)
\quad.
\label{jAi}
\end{equation}
Finally, the conserved charges  
$Q^A$ are given by (expanding multi-indices)
\begin{equation}
Q^{\mu_1 \dots \mu_{p-k} \alpha_1 \dots \alpha_k} =
       \int_{W_\tau} d \sigma^1 \dots d \sigma^p\sum^p_{m=0}
       (B^{-1})^{\mu_1 \dots \mu_{p-k} \alpha_1 \dots \alpha_k}_
          {\, \, \, \nu_1  \dots \nu_{p-m} \beta_1 \dots \beta_m}
      b_m \epsilon^{0j_1 \dots j_p} \Pi^{\nu_1}_{j_1} \dots
       \Pi^{\nu_{p-m}}_{j_{p-m}}
       \Pi^{\beta_1}_{j_{p-m+1}} \dots
       \Pi^{\beta_m}_{j_p}
\quad,
\label{jAexpr}
\ee
where $W_\tau$ is a hypersurface of constant $\tau$. Notice that
since $B=B(\theta,\, x)$, the integrand above involves only
superspace variables.
%%%%%%%%%%%%%%%%%%%%%%%%%%%%%%%%%%%%%%%%%%%%%%%%%%%%%%%%%%%%%%
%%%%%%%%%%%%%%%%%%%%%%%%%%%%%%%%%%%%%%%%%%%%%%%%%%%%%%%%%%%%%%
%%%%%%%%%%%%%%%%%%%%%%%%%%%%%%%%%%%%%%%%%%%%%%%%%%%%%%%%%%%%%%
\section{Applications: $p$=1,2}
\label{p12}
%%%%%%%%%%%%%%%%%%%%%%%%%%%%%%%%%%%%%%%%%%%%%%%%%%%%%%%%%%%%%%%%%
\subsection{$D$=10, N=1 and the Green--Schwarz superstring}
\label{css}

The case of the superstring is somewhat special, from the
point of view of the extension algorithm of Sec.~\ref{NCESA}:
the first additional generator to be introduced,
$Z^\mu$, is a vector, as $X_\mu$.
We shall keep it here separate and denote by $\varphi_\mu$
the associated parameter. Fixing 
$(a_s,a_0,a_1)=({1 \over 2},{1 \over 2},1)$
in~(\ref{dPimu}),(\ref{dPim1a1}), 
we find for the FDA\footnote{In Sec. \ref{css} all 
spinors are Majorana-Weyl ($\theta^\alpha\equiv 
(\mathcal{P}_+\theta)^\alpha$, \em{etc.}).}
\be
\begin{array}{rclcrcl}
d\Pi^\alpha & = & 0 
\, ,
& \qquad &
d \Pi^\mu  & = &  {1 \over 2} (C \Gamma^\mu)_{\alpha\beta} \Pi^\alpha
\Pi^\beta 
\, , 
\\
d \Pi_\alpha & = &  (C \Gamma_\mu)_{\alpha\beta}
\Pi^\mu \Pi^\beta
+  (C \Gamma^\mu)_{\alpha\beta}\Pi_\mu \Pi^\beta
\, ,
& \qquad &
d \Pi_\mu^{(\varphi)} & = & {1 \over 2} (C \Gamma_\mu)_{\alpha\beta}
\Pi^\alpha \Pi^\beta
\, ,
\label{ssFDA}
\end{array}
\ee
where $\mu=0, \dots , 9$.
Notice that $d(d\Pi_\alpha)=0$ is implied by  
(\ref{Gammaprop}) for $p=1$, $(C \Gamma^\mu)_{\alpha'\beta'} 
(C \Gamma_\mu)_{\gamma'\delta'}=0$.
$\Pi_\mu^{(\varphi)}$ in the above equation is the one obtained
from the second of
(\ref{dPimu}) for $p=1$ -- we have added a superscript to
avoid any confusion with $\Pi^\mu$, since they have similar
differentials; recall also that $\Pi_\alpha$ and $\Pi^\alpha$ are
unrelated. 
As mentioned in Sec.~\ref{NCESA}, the two terms in the r.h.s.
of the last of~(\ref{ssFDA}) are individually closed and hence
their relative normalization cannot be fixed by requiring
$d(d\Pi_\alpha)=0$. We have nevertheless chosen the above
symmetric  normalization
for convenience -- the results that follow, and in
particular~(\ref{ssJAS}) that involves cancellations, do not
depend essentially on this choice.

The corresponding Lie algebra is given by
\be
\{ D_\alpha, D_\beta\} = \GMab X_\mu + \Gmab Z^\mu
\, , \qquad
 \left[D_\alpha, X_\mu\right] =  \Gmab Z^\beta
\, , \qquad
\left[D_\alpha, Z^\mu\right] = \GMab Z^\beta
\, , \qquad
\label{ssLa}
\ee
which reduces to the Green algebra~\cite{Gre:89} if one omits
$Z^\mu$. 
The associated group manifold (extended superspace)
$\tilde{\Sigma}$ is parametrized by 
$(\theta^\alpha,x^\mu,\varphi_\mu,\varphi_\alpha)$ via
\begin{equation}
g(\theta^\alpha,x^\mu,\varphi_\mu,\varphi_\alpha)
 = e^{\theta^\alpha D_\alpha + x^\mu X_\mu +
   \varphi_\mu Z^\mu + \varphi_\alpha Z^\alpha}
\quad.
\label{ssge}
\end{equation}
Making use of the Baker-Cambell-Hausdorff (BCH) formula, where,
for the algebra~(\ref{ssLa}),
terms of order four and higher vanish,
we find the $\tilde{\Sigma}$ group law
\be
\addtolength{\arraycolsep}{-.6ex}
\begin{array}{rclcrcl}
\dis 
\theta''^\alpha &=& \dis \theta'^\alpha + \theta^\alpha 
\, , 
& \qquad &
\dis
x''^\mu & = & \dis x'^\mu + x^\mu - {1 \over 2} \GMab \theta'^\alpha
\theta^\beta 
\\[1.5ex]
\dis 
\varphi''_\alpha &=& \dis \varphi'_\alpha +\varphi_\alpha
           +{1 \over 2}  \Gmab \theta'^\beta x^\mu
             - {1 \over 2} \Gmab x'^\mu \theta^\beta 
& \qquad &
\dis 
\varphi''_\mu &=& \dis \varphi'_\mu + \varphi_\mu - {1 \over 2} 
\Gmab \theta'^\alpha \theta^\beta 
\, .
\\[1.5ex]
 & & \dis  \mbox{}+{1 \over 2}  \GMab \theta'^\beta \varphi_\mu
      -{1 \over 2} \GMab \varphi'_\mu \theta^\beta
& \qquad &
& & 
\\[1.5ex]
 & & \dis  \mbox{}+{1 \over 6} \Gmab (C \Gamma^\mu)_{\gamma\delta}
       (\theta'^\gamma \theta'^\beta \theta^\delta
      +\theta'^\delta \theta^\gamma \theta^\beta)
\, , 
& \qquad &
& & 
\label{ssgcl}
\end{array}
\ee
The bilinear terms in the expression for
$\varphi''_\alpha$ are the ones that give rise to the fourth of the MC
equations~(\ref{ssFDA})---the trilinear terms are required by 
the associativity of the group law. Their sum gives 
the spinor valued two-cocycle $\xi_\alpha$ associated with the 
central extension of ${\tilde \Sigma} (\theta^\alpha,
x^\mu, \varphi_\mu)$ by $\varphi_\alpha$. 

One can now relate the LI one-forms
to the coordinate differentials
\be
\addtolength{\arraycolsep}{-.5ex}
\begin{array}{rclcrcl}
\dis
\Pi^\alpha &=& \dis d \theta^\alpha
\, , 
& \quad &
\dis
\Pi^\mu & = & \dis d x^\mu +{1 \over 2} \GMab \theta^\alpha d
\theta^\beta
\\[1.5ex]
\dis
\Pi_\alpha &=& \dis d \varphi_\alpha
-{1 \over 2} \Gmab \theta^\beta
dx^\mu 
-{1 \over 2} \GMab \theta^\beta d\varphi_\mu
& \quad &
\dis
\Pi_\mu^{(\varphi)} &=& \dis d \varphi_\mu +{1 \over 2} \Gmab
\theta^\alpha 
\, .
\\[1.5ex]
 & & 
\dis
\mbox{} +{1 \over 2} \Gmab x^\mu d \theta^\beta
+{1 \over 2} \GMab \varphi_\mu d \theta^\beta
& \quad &
 & &
\\[1.5ex]
 & &
\dis
\mbox{} +{1 \over 3} \Gmab (C \Gamma^\mu)_{\gamma\delta} \theta^\gamma
\, , & \quad &
 & &
\label{ssLIf}
\end{array}
\ee
(see also~\cite{Der.Gal:97} although, omitting 
$\Pi_\mu^{(\varphi)}$, we disagree with the corresponding
expressions 
there). One may check that the LI forms in  (\ref{ssLIf})
satisfy the FDA (\ref{ssFDA}). From the group composition law, one 
can compute the LI vector fields dual to (\ref{ssLIf})
satisfying (\ref{ssLa})
\begin{eqnarray}
D_\alpha &=&  \partial_\alpha  
+ {1 \over 2} \GMab \theta^\beta
\partial_\mu 
+{1 \over 2} \Gmab \theta^\beta \partial^\mu 
\nonumber \\
 & & 
-{1 \over 2} \Gmab x^\mu \partial^\beta
-{1 \over 2} \GMab \varphi_\mu \partial^\beta
+{1 \over 6} \GMab \Gmcd \theta^\beta \theta^\delta
\partial^\gamma 
\nonumber \\
X_\mu &=&  \partial_\mu + {1 \over 2} \Gmab \theta^\beta
\partial^\alpha 
\nonumber \\
Z^\mu &=&  \partial^\mu + {1 \over 2} \GMab \theta^\beta
\partial^\alpha 
\nonumber \\
Z^\alpha &=&  \partial^\alpha\qquad (\text{note \ } \partial^\mu \equiv {\partial
\over\partial\varphi_\mu}\ , \, \, \,  \partial^\alpha \equiv {\partial \over
\partial\varphi_\alpha} )\quad .
\label{ssLIvf}
\end{eqnarray}
Effecting a right group
translation in~(\ref{ssLIf}) and reexpressing the result in terms
of the $\Pi$'s, or computing the exponential in~(\ref{Texpr}), 
one finds for  $(T^t)^{-1}$ 
{\small
\begin{equation}
(T^t)^{-1} = \left( \begin{array}{cc|cc}
{\delta_\gamma}^\alpha 
& - (C \Gamma^\mu)_{\gamma \delta} \theta^\delta
& - (C \Gamma_\kappa)_{\gamma \delta} \theta^\delta \, \, 
& \, \, (C \Gamma_\lambda)_{\gamma \beta} x^\lambda
  + (C \Gamma^\nu)_{\gamma \beta} \varphi_\nu 
  + (C \Gamma_\lambda)_{\beta \delta} (C \Gamma^\lambda)_{\gamma
     \epsilon} \theta^\epsilon \theta^\delta
\\
0 
& {\delta_\nu}^\mu
& 0
& - (C \Gamma_\nu)_{\beta \delta} \theta^\delta
\\
\hline
0 & 0 
& {\delta^\rho}_\kappa 
& - (C \Gamma^\rho)_{\beta \delta} \theta^\delta
\\
0 & 0 & 0
& {\delta^\zeta}_\beta
\end{array}
\right)
\label{ssTti}
\end{equation}}
\noindent \mbox{}(the matrix indices are $\,_{\gamma}
\,_{\nu} \,^{\rho} \,^{\zeta}$ for the rows and $\,^{\alpha}
\,^{\mu} \,_{\kappa} \,_{\beta}$ for the columns).
When eqns.~(\ref{ssgcl}) are linearized in the primed variables 
(viewed as the parameters of the transformation), 
they provide a realization of the algebra~(\ref{ssLa}), 
acting on the coordinate (unprimed) variables. 
For our particular choice of parametrization~(\ref{ssge}), this 
action is rendered non-linear by the
last term in the expression for $\varphi''_\alpha$.
This term can be eliminated by modifying the two-cocycle by a
two-coboundary, the latter being generated by a suitable
spinor-valued function
$\eta_\alpha$ on ${\tilde\Sigma}(\theta^\alpha,x^\mu,\varphi_\mu)$. 
Indeed, 
with $\eta_\alpha \equiv {1 \over 6}  \Gmab x^\mu \theta^\beta 
+{1 \over 6} \GMab \varphi_\mu \theta^\beta$ we find
for the coboundary $\xi^{cob}_\alpha (g',g) \equiv \eta_\alpha(g'g)
-\eta_\alpha(g') -\eta_\alpha(g)$, the expression
\begin{eqnarray}
\xi^{cob}_\alpha &=& 
{1 \over 6} \Gmab (x'^\mu \theta^\beta + \theta'^\beta x^\mu) 
+ {1 \over 6} \GMab (\varphi'_\mu \theta^\beta +\theta'^\beta
\varphi_\mu)
\nonumber \\
 & & + {1 \over 6} \Gmab \GMcd (\theta'^\gamma
\theta'^\beta \theta^\delta -  \theta'^\delta
\theta^\gamma \theta^\beta)
\quad.
\label{xicob}
\end{eqnarray}
The new cocycle $\bar{\xi}_{\alpha}= \xi_{\alpha}+\xi^{cob}_{\alpha}$ 
modifies the last eqn. in~(\ref{ssgcl}) to read
\begin{eqnarray}
\bar{\varphi}''_\alpha &=&
  \bar{\varphi}'_\alpha + \bar{\varphi}_\alpha 
+ \Gmab ({2 \over 3} \theta'^\beta x^\mu 
- {1 \over 3} x'^\mu \theta^\beta) 
+ \GMab ({2 \over 3} \theta'^\beta \varphi_\mu 
- {1 \over 3} \varphi'_\mu \theta^\beta) 
\nonumber \\
 & & + {1 \over 3} \Gmab \GMcd \theta'^\gamma
\theta'^\beta \theta^\delta
\quad,
\label{bvpgcl}
\end{eqnarray}
which is linear in the unprimed variables
 although the definite symmetry
properties under exchange of primed and unprimed variables are
now lost. The terms in~(\ref{bvpgcl}) linear in the primed
coordinates agree (omitting $\varphi_\mu$) with the (first-order) 
results
of~\cite{Der.Gal:97}, where the (equivalent) coordinate
redefinition $\varphi_\alpha \rightarrow \bar{\varphi}_\alpha
\equiv\varphi_\alpha + \eta_\alpha$ is given. For the LI one-form 
associated with the new coordinate we find
\begin{eqnarray}
\Pi_{(\bar{\varphi}_\alpha)} &=& d \bar{\varphi}_\alpha 
- {2 \over 3} \Gmab \theta^\beta d x^\mu 
- {2 \over 3} \GMab \theta^\beta d \varphi_\mu 
\nonumber \\
 & & + {1 \over 3} \left( \Gmab x^\mu 
     +  \GMab \varphi_\mu
 +(C \Gamma_\mu)_{\alpha \delta} (C \Gamma^\mu)_{\beta \gamma} 
   \theta^\gamma \theta^\delta \right) d \theta^\beta
\quad.
\label{newPia}
\end{eqnarray}

The manifestly invariant WZ term for the  
superstring action is given by
\begin{equation}
S_{WZ} =\int_W \phi^{*}(\tilde{b}) =   
\int_W \phi^{*}(\Pi_\mu^{(\varphi)} 
 \Pi^\mu + {1 \over 2} \Pi_\alpha  \Pi^\alpha) \quad,
\label{ssWZt}
\end{equation}
which differs from the one in~\cite{Sie:94} by the term 
in $\Pi_\mu^{(\varphi)} \Pi^\mu$. It is immediately checked, 
using~(\ref{ssFDA}), (\ref{ssLIf}), that $d
\tilde{b}=db= \Gmab \Pi^\mu \Pi^\alpha \Pi^\beta$ and hence that
$\phi^*(\tilde{b})$ and the standard WZ term $\phi^*(b)$  are 
equivalent, differing only by an exact differential.
Since the string tension $T$ has dimensions [$T$]=$ML^{-1}$ and
[$\Pi_\mu^{(\varphi)}$]=$L$, [$\Pi_\alpha$]=$L^{3/2}$, the
products $T \Pi_\alpha \Pi^\alpha$,
$T\Pi_\mu^{(\varphi)}\Pi^\mu$, have the dimensions $ML$ of an action. 

To compute the conserved Noether currents, we start 
from~(\ref{JAS}) which gives the closed forms $J^A$ on
$\tilde{\Sigma}$.  With
$B^{-1}$ given by the lower right block of $(T^t)^{-1}$
(see~(\ref{ssTti}), (\ref{Tblock})) we get
\be
\addtolength{\arraycolsep}{-.6ex}
\begin{array}{rcl}
J^\mu & 
=
& {(B^{-1})^\mu}_\nu \Pi^\nu 
         + {1 \over 2} {(B^{-1})^\mu}_\beta \Pi^\beta
 \, = \, \Pi^\mu - {1 \over 2} (C \Gamma^\mu)_{\beta \gamma}
\theta^\gamma d \theta^\beta \, = \, dx^\mu
\nonumber \\[1.5ex]
J^\alpha &=& {1 \over 2} {(B^{-1})^\alpha}_\beta \Pi^\beta 
               \, = \, {1 \over 2} d \theta^\alpha
\, ,
\end{array}
\label{ssJAS}
\ee
which, after pulling back on the worldvolume, gives for 
the charge $Q^\mu$ (see~(\ref{jAexpr}))
\begin{equation}
Q^\mu = \int^{2\pi}_0 d \sigma \, {\partial x^\mu \over \partial \sigma}
\quad,
\label{sscc}
\end{equation}
$Q^\alpha$ being zero because we assume that $\theta$ is periodic
in $\sigma$. The integral~(\ref{sscc}) may lead to a non-zero
result if the topology is nontrivial~\cite{Azc.Gau.Izq.Tow:89}.
%%%%%%%%%%%%%%%%%%%%%%%%%%%%%%%%%%%%%%%%%%%%%%%%%%%%%%%%%%%%%%%%
%%%%%%%%%%%%%%%%%%%%%%%%%%%%%%%%%%%%%%%%%%%%%%%%%%%%%%%%%%%%%%%%
\subsection{$D=11$ and the case of the supermembrane}
\label{csm}

Our starting point is the FDA of Sec.~\ref{NCESA} with $p=2$.
We fix the normalisation of the forms by setting
$(a_s,a_0,a_1,a_2)=({1 \over 2},{1 \over 2},1,-{1 \over 2})$ 
so that the dual Lie algebra becomes
\begin{eqnarray}
\{ D_\alpha,D_\beta\} &=& \GMab X_\mu + 
(C \Gamma_{\mu\nu})_{\alpha\beta}
Z^{\mu\nu}
\nonumber \\
 \left[ X_\mu,D_\alpha \right] &=& -\Gmnab Z^{\nu\beta}
\nonumber \\
\left[ X_\mu,X_\nu \right] &=& \Gmnab \ZAB
\nonumber \\
\left[ X_\mu, Z^{\lambda\tau} \right] &=& {1 \over 2}
\delta^{\left[  \lambda \right.}_\mu
(C \Gamma^{\left. \tau \right]})_{\alpha\beta} \ZAB
\nonumber \\
\left[ D_\alpha,\ZMN \right] &=& (C \Gamma^{\left[
\mu \right.})_{\alpha\beta} Z^{\left.\nu \right] \beta}
\nonumber \\
\{ D_\alpha, Z^{\nu\beta} \} &=& 
({1 \over 4}(C \Gamma^\nu)_{\gamma\delta}
\delta^\beta_\alpha +2(C \Gamma^\nu)_{\gamma\alpha}
\delta^\beta_\delta) Z^{\gamma\delta}
\quad,
\label{smLa}
\end{eqnarray}
coinciding with that given in~\cite{Ber.Sez:95}.
The associated extended superspace group
manifold is parametrized by the coordinates
$(\tA,\xM,\vfmn,\vfma,\vfab)$ via\footnote{We use a
parametrization different from the one in~\cite{Ber.Sez:95}.}
\begin{equation}
g = e^{\tA D_\alpha + \xM X_\mu + \vfmn \ZMN + \vfma \ZMA + \vfab \ZAB}
\quad,
\label{smge}
\end{equation}
where $\vfmn$, $\vfab$ are antisymmetric and symmetric,
respectively, in their indices.
Application (with the help of FORM) of the BCH formula, 
where now terms of order five and higher
vanish, results in the following group law
\begin{eqnarray}
{\theta^{\alpha}}^{''} & = &  {\theta^{\alpha}}^{'} 
                              + {\theta^{\alpha}}^{}
\\
{x^\mu}''  &=&  {x^{\mu}}^{'} + {x^{\mu}}^{}
                 - 1/2\, \, {\theta^{\alpha_{2}}}^{'}\, 
                 {\theta^{\alpha_{3}}}^{}\,  
                 (C \Gamma^{\mu})_{\alpha_{2} \alpha_{3}}
\\
\varphi''_{\mu_1 \mu_2} 
&=&  \varphi^{'}_{\mu_{1} \mu_{2}}+
    \varphi^{}_{\mu_{1} \mu_{2}}
    - 1/2\, \, {\theta^{\alpha_{3}}}^{'}\, 
    {\theta^{\alpha_{4}}}\, 
    (C \Gamma_{\mu_{1} \mu_{2}})_{\alpha_{3} \alpha_{4}}
\\
\varphi''_{\mu_{1} \alpha_{2}} 
&=&  \varphi^{'}_{\mu_1 \alpha_2}- 1/2\, \,
{\theta^{\alpha_{3}}}^{'}\,
    (C \Gamma^{\mu_{4}})_{\alpha_{3} \alpha_{2}}\,
    \varphi_{\mu_{1} \mu_{4}}
\nonumber \\ 
& & - 1/2\, \, {\theta^{\alpha_{3}}}^{'}\, 
    (C \Gamma_{\mu_{1}
    \mu_{4}})_{\alpha_{3} \alpha_{2}}\,  {x^{\mu_{4}}}
\nonumber \\ 
& & + 1/12\, \, {\theta^{\alpha_{3}}}^{'}\, 
    {\theta^{\alpha_{4}}}^{'}\,  {\theta^{\alpha_{5}}}\, 
    (C \Gamma^{\mu_{6}})_{\alpha_{3} \alpha_{2}}\,  
    (C \Gamma_{\mu_{1} \mu_{6}})_{\alpha_{4} \alpha_{5}} 
\nonumber \\ 
& & + 1/12\, \, {\theta^{\alpha_{3}}}^{'}\, 
   {\theta^{\alpha_{4}}}^{'}\,  {\theta^{\alpha_{5}}}\, 
   (C \Gamma^{\mu_{6}})_{\alpha_{4} \alpha_{5}}\,  
   (C \Gamma_{\mu_{1} \mu_{6}})_{\alpha_{3} \alpha_{2}}
   \pm ((1) \leftrightarrow (2)) 
\\
%\end{eqnarray}
%\begin{eqnarray}
\varphi''_{\alpha_{1} \alpha_{2}} 
&=&      \varphi^{'}_{\alpha_{1} \alpha_{2}}
- 1/4\, \, (C \Gamma^{\mu_{3}})_{\alpha_{1} \alpha_{2}}\, \,
        {x^{\mu_{4}}}^{'}\, \, \varphi_{\mu_{3} \mu_{4}}
\nonumber \\ 
& &     + 1/2\, \, (C \Gamma_{\mu_{3} \mu_{4}})_{\alpha_{1} 
        \alpha_{2}}\, \, {x^{\mu_{3}}}^{'}\, \, {x^{\mu_{4}}}
\nonumber \\ 
& &     + 1/6\, \, {\theta^{\alpha_{3}}}^{'}\, \,
        {\theta^{\alpha_{4}}}^{'}\, \, (C \Gamma^{\mu_{5}})_{\alpha_{3} 
        \alpha_{1}}\, \,
        (C \Gamma^{\mu_{6}})_{\alpha_{4} \alpha_{2}}\, \, 
        \varphi_{\mu_{5}
        \mu_{6}}
\nonumber \\ 
& &     + 1/6\, \, {\theta^{\alpha_{3}}}^{'}\, \,
        {\theta^{\alpha_{4}}}^{'}\, \, (C \Gamma^{\mu_{5}})_{\alpha_{3} 
        \alpha_{1}}\, \,
        (C \Gamma_{\mu_{5} \mu_{6}})_{\alpha_{4} \alpha_{2}}\, \,
        {x^{\mu_{6}}}
\nonumber \\ 
& &     - 1/24\, \, {\theta^{\alpha_{3}}}^{'}\, \,
        {\theta^{\alpha_{4}}}^{'}\, \, {\theta^{\alpha_{5}}}\, \,
        {\theta^{\alpha_{6}}}\, \, (C \Gamma^{\mu_{7}})_{\alpha_{3} 
        \alpha_{1}}\, \,
        (C \Gamma^{\mu_{8}})_{\alpha_{4} \alpha_{6}}\, \, 
        (C \Gamma_{\mu_{7} \mu_{8}})_{\alpha_{5} \alpha_{2}}
\nonumber \\ 
& &     - 1/24\, \, {\theta^{\alpha_{3}}}^{'}\, \,
        {\theta^{\alpha_{4}}}^{'}\, \, {\theta^{\alpha_{5}}}\, \,
        {\theta^{\alpha_{6}}}\, \, (C \Gamma^{\mu_{7}})_{\alpha_{3} 
        \alpha_{1}}\, \,
        (C \Gamma^{\mu_{8}})_{\alpha_{5} \alpha_{2}}\, \, 
        (C \Gamma_{\mu_{7} \mu_{8}})_{\alpha_{4} \alpha_{6}}
\nonumber \\ 
& &     - 1/24\, \, {\theta^{\alpha_{3}}}^{'}\, \,
        {\theta^{\alpha_{4}}}\, \, (C \Gamma^{\mu_{5}})_{\alpha_{1} 
        \alpha_{2}}\, \,
        (C \Gamma^{\mu_{6}})_{\alpha_{3} \alpha_{4}}\, \, 
        \varphi^{'}_{\mu_{5} \mu_{6}}
\nonumber \\ 
& &     + 1/48\, \, {\theta^{\alpha_{3}}}^{'}\, \,
        {\theta^{\alpha_{4}}}\, \, (C \Gamma^{\mu_{5}})_{\alpha_{1} 
        \alpha_{2}}\, \,
        (C \Gamma_{\mu_{5} \mu_{6}})_{\alpha_{3} \alpha_{4}}\, \,
        {x^{\mu_{6}}}^{'}
\nonumber \\ 
& &     - 1/6\, \, {\theta^{\alpha_{3}}}^{'}\, \,
        {\theta^{\alpha_{4}}}\, \, (C \Gamma^{\mu_{5}})_{\alpha_{3} 
        \alpha_{1}}\, \,
        (C \Gamma^{\mu_{6}})_{\alpha_{4} \alpha_{2}}\, \, 
        \varphi^{'}_{\mu_{5} \mu_{6}}
\nonumber \\ 
& &     - 1/6\, \, {\theta^{\alpha_{3}}}^{'}\, \,
        {\theta^{\alpha_{4}}}\, \, (C \Gamma^{\mu_{5}})_{\alpha_{3} 
        \alpha_{1}}\, \,
        (C \Gamma_{\mu_{5} \mu_{6}})_{\alpha_{4} \alpha_{2}}\, \,
        {x^{\mu_{6}}}^{'}
\nonumber \\ 
& &     + 1/48\, \, {\theta^{\alpha_{3}}}^{'}\, \,
        {\theta^{\alpha_{4}}}\, \, (C \Gamma^{\mu_{5}})_{\alpha_{3} 
        \alpha_{4}}\, \,
        (C \Gamma^{\mu_{6}})_{\alpha_{1} \alpha_{2}}\, \, 
        \varphi^{'}_{\mu_{5} \mu_{6}}
\nonumber \\ 
& &     + 1/12\, \, {\theta^{\alpha_{3}}}^{'}\, \,
        {\theta^{\alpha_{4}}}\, \, (C \Gamma^{\mu_{5}})_{\alpha_{3} 
        \alpha_{4}}\, \,
        (C \Gamma_{\mu_{5} \mu_{6}})_{\alpha_{1} \alpha_{2}}\, \,
        {x^{\mu_{6}}}^{'}
\nonumber \\ 
& &     - {\theta^{\alpha_{3}}}^{'}\, \, \varphi_{\mu_{4}
        \alpha_{1}}\, \, (C \Gamma^{\mu_{4}})_{\alpha_{3} \alpha_{2}}
\nonumber \\ 
& &     - 1/8\, \, {\theta^{\alpha_{3}}}^{'}\, \, \varphi_{\mu_{4}
        \alpha_{3}}\, \, (C \Gamma^{\mu_{4}})_{\alpha_{1} \alpha_{2}}
        \pm ((1) \leftrightarrow (2))
\quad
\label{smgcl}
\end{eqnarray}
(in the expression for $\varphi_{\alpha_1 \alpha_2}$, the r.h.s. 
is assumed to be
symmetrised, with unit weight,  in $\alpha_1$, $\alpha_2$). 
The $\pm ((1) \leftrightarrow (2))$ instruction in the last two
expressions 
means that for each term displayed, one has to add (subtract) a
similar term, with the primed and unprimed variables
 exchanged (taking into account
statistics), if the order of the term is odd (even) in the
coordinates (\eg \ $\theta'^\alpha \varphi_{\mu \beta} \pm  ((1)
\leftrightarrow (2)) = \theta'^\alpha \varphi_{\mu \beta} - (-
\varphi'_{\mu \beta} \theta^\alpha)$). This symmetry property 
can be seen to hold in general: 
from the BCH formula  $e^{A'} e^A=e^{f(A',A)}$
it follows that $f(-A,-A')$=$-f(A',A)$ and hence, terms of
order $n$ in the coordinates are symmetric ($n$ odd) or
antisymmetric ($n$ even) under the above exchange of the two
spaces. The linearized (in the primed variables) form of the
previous expressions has been given in~\cite{Der.Gal:97}. 
Starting from~(\ref{smgcl}), we find for 
the LI vector fields of~(\ref{smLa}) 
\begin{eqnarray}
D_{\alpha_1} &=& 
          \partial_{\alpha_{1}}
+ 1/12\, \, {\theta^{\alpha_{2}}}^{}\, \,
{\theta^{\alpha_{3}}}^{}\, \, \left((C \Gamma^{\mu_{4}})_{\alpha_{2}
\alpha_{5}}\, \, (C \Gamma_{\mu_{6} \mu_{4}})_{\alpha_{3}
\alpha_{1}} + (C \Gamma_{\mu_{4}\mu_{6}})_{\alpha_{2}
\alpha_{5}}\, \, (C\Gamma^{\mu_{4}})_{\alpha_{3} \alpha_{1}}
\right)\, \, \partial^{\mu_{6} \alpha_{5}}
          \nonumber \\ 
 & & + 1/12\, \, {\theta^{\alpha_{2}}}^{}\, \,
\varphi^{}_{\mu_{3} \mu_{4}}\, \, \left((C
\Gamma^{\mu_{3}})_{\alpha_{2} \alpha_{5}}\, \, (C
\Gamma^{\mu_{4}})_{\alpha_{1} \alpha_{6}}+
(C\Gamma^{\mu_{3}})_{\alpha_{5} \alpha_{6}}\, \,
(C\Gamma^{\mu_{4}})_{\alpha_{2} \alpha_{1}} \right)\, \,
\partial^{\alpha_{5} \alpha_{6}}
          \nonumber \\
 & & - 1/3\, \, {\theta^{\alpha_{2}}}^{}\, \,
{x^{\mu_{3}}}^{}\, \, (C \Gamma^{\mu_{4}})_{\alpha_{2}
\alpha_{5}}\, \, (C \Gamma_{\mu_{3} \mu_{4}})_{\alpha_{6}
\alpha_{1}}\, \, \partial^{\alpha_{5} \alpha_{6}}
          \nonumber \\ & & + 1/8\, \, {\theta^{\alpha_{2}}}^{}\, \,
{x^{\mu_{3}}}^{}\, \, (C \Gamma^{\mu_{4}})_{\alpha_{5}
\alpha_{6}}\, \, (C \Gamma_{\mu_{3} \mu_{4}})_{\alpha_{2}
\alpha_{1}}\, \, \partial^{\alpha_{5} \alpha_{6}}
 \nonumber \\
 & & + 1/2\, \, {\theta^{\alpha_{2}}}^{}\, \, (C
\Gamma^{\mu_{3}})_{\alpha_{2} \alpha_{1}}\, \,  \partial_{\mu_{3}}
          \nonumber \\
 & & + {\theta^{\alpha_{2}}}^{}\, \, 
       (C \Gamma_{\mu_{3} \mu_{4}})_{\alpha_{2} \alpha_{1}}\, \, 
\partial^{\mu_{3} \mu_{4}}
          \nonumber \\
 & & - 1/2\, \, {x^{\mu_{2}}}^{}\, \, 
(C \Gamma_{\mu_{2} \mu_{3}})_{\alpha_{1} \alpha_{4}}\, \, 
\partial^{\mu_{3} \alpha_{4}}
          \nonumber \\ & & - 1/2\, \, \varphi^{}_{\mu_{2} \mu_{3}}\, \, (C
\Gamma^{\mu_{2}})_{\alpha_{1} \alpha_{4}}\, \, \partial^{\mu_{3} \alpha_{4}}
          \nonumber \\ & & + 1/8\, \, \varphi^{}_{\mu_{2} \alpha_{1}}\, \, (C
\Gamma^{\mu_{2}})_{\alpha_{3} \alpha_{4}}\, \, \partial^{\alpha_{3} \alpha_{4}}
          \nonumber \\ & & + \varphi^{}_{\mu_{2} \alpha_{3}}\, \, (C
\Gamma^{\mu_{2}})_{\alpha_{1} \alpha_{4}}\, \, \partial^{\alpha_{4} \alpha_{3}}
 \\
X_{\mu_1} &=&
           \partial_{\mu_{1}}
 - 1/6\, \, {\theta^{\alpha_{2}}}^{}\, \,
   {\theta^{\alpha_{3}}}^{}\, \, (C
\Gamma^{\mu_{4}})_{\alpha_{2} \alpha_{5}}\, \, (C \Gamma_{\mu_{1}
\mu_{4}})_{\alpha_{6} \alpha_{3}}\, \, \partial^{\alpha_{5} \alpha_{6}}
          \nonumber \\ & & + 1/2\, \, {\theta^{\alpha_{2}}}^{}\, \, (C
\Gamma_{\mu_{1} \mu_{3}})_{\alpha_{2} \alpha_{4}}\, \, \partial^{\mu_{3} \alpha_{4}}
          \nonumber \\ & & + 1/2\, \, {x^{\mu_{2}}}^{}\, \, (C \Gamma_{\mu_{2}
\mu_{1}})_{\alpha_{3} \alpha_{4}}\, \, \partial^{\alpha_{3} \alpha_{4}}
          \nonumber \\ & & - 1/4\, \, \varphi^{}_{\mu_{1} \mu_{2}}\, \, (C
\Gamma^{\mu_{2}})_{\alpha_{3} \alpha_{4}}\, \, \partial^{\alpha_{3} \alpha_{4}}
 \\
Z^{\mu_1 \mu_2} &=&
           \partial^{\mu_{1} \mu_{2}}
+ 1/12\, \, {\theta^{\alpha_{3}}}^{}\, \,
{\theta^{\alpha_{4}}}^{}\, \, (C \Gamma^{\mu_{1}})_{\alpha_{3}
\alpha_{5}}\, \, (C \Gamma^{\mu_{2}})_{\alpha_{4} \alpha_{6}}\,
\, \partial^{\alpha_{5} \alpha_{6}}
          \nonumber \\ & & - 1/12\, \, {\theta^{\alpha_{3}}}^{}\, \,
{\theta^{\alpha_{4}}}^{}\, \, (C \Gamma^{\mu_{2}})_{\alpha_{3}
\alpha_{5}}\, \, (C \Gamma^{\mu_{1}})_{\alpha_{4} \alpha_{6}}\,
\, \partial^{\alpha_{5} \alpha_{6}}
          \nonumber \\ & & + 1/4\, \, {\theta^{\alpha_{3}}}^{}\, \, (C
\Gamma^{\mu_{1}})_{\alpha_{3} \alpha_{4}}\, \, \partial^{\mu_{2} \alpha_{4}}
          \nonumber \\ & & - 1/4\, \, {\theta^{\alpha_{3}}}^{}\, \, (C
\Gamma^{\mu_{2}})_{\alpha_{3} \alpha_{4}}\, \, \partial^{\mu_{1} \alpha_{4}}
          \nonumber \\ & & + 1/4\, \, {x^{\mu_{1}}}^{}\, \, (C
\Gamma^{\mu_{2}})_{\alpha_{3} \alpha_{4}}\, \, \partial^{\alpha_{3} \alpha_{4}}
 \\
Z^{\mu_1 \alpha_2} &=&
           \partial^{\mu_{1} \alpha_{2}}
           + 1/8\, \, {\theta^{\alpha_{2}}}^{}\, \, 
          (C \Gamma^{\mu_{1}})_{\alpha_{3} \alpha_{4}}\, \, 
          \partial^{\alpha_{3} \alpha_{4}}
          + {\theta^{\alpha_{3}}}^{}\, \, 
          (C \Gamma^{\mu_{1}})_{\alpha_{3} \alpha_{4}}\, \, 
          \partial^{\alpha_{4} \alpha_{2}}
 \\
Z^{\alpha_1 \alpha_2} &=& \partial^{\alpha_{1} \alpha_{2}}
\quad
\label{smLIvf}
\end{eqnarray}
(antisymmetrisation with unit weight in $\mu_1$, $\mu_2$ is 
understood in the r.h.s. of the expression for $Z^{\mu_1 \mu_2}$).
The manifestly invariant WZ term on the extended superspace
$\tilde{\Sigma}$ is given by~\cite{Ber.Sez:95}
\begin{equation}
\tilde{b}={2 \over 3} \Pi_{\mu \nu} \Pi^\mu \Pi^\nu 
- {3 \over 5} \Pi_{\mu \alpha} \Pi^\mu \Pi^\alpha
- {2 \over 15} \Pi_{\alpha \beta} \Pi^\alpha \Pi^\beta
\label{smbt}
\end{equation}
([${\tilde b}$]=$L^3$, [$Tb$]=$ML$).
It depends on the additional $\varphi$ variables through total
differentials 
since $d \tilde{b}=db=h= (C \Gamma_{\mu
\nu})_{\alpha\beta} \Pi^\mu \Pi^\nu \Pi^\alpha \Pi^\beta$. 

The computation of the full $T$ matrix for the supermembrane is 
rather tedious. For the
Noether currents though (corresponding to the new variables) we
only need $B^{-1}$. Reading off the relevant structure constants
from~(\ref{smLa}), we find 
\begin{eqnarray}
\rho^{(f)}_{adj} \, (D_\alpha) &=& 
\left( \begin{array}{ccc}
0 & 0 & 0
\\
(C \Gamma^{[\mu_1})_{\alpha \alpha_2} {\delta_{\kappa_2}}^{\nu_1]} 
& 0 & 0 
\\
0 & {1 \over 4} (C \Gamma^{\kappa_1})_{\beta_2 \gamma_2}
{\delta_\alpha}^{\alpha_1} + 2 (C \Gamma^{\kappa_1})_{\beta'_2
\alpha} {\delta_{\gamma'_2}}^{\alpha_1} & 0
\end{array} \right)
\quad, 
\nonumber \\
\rho^{(f)}_{adj} \, (X_\rho) &=&
\left( \begin{array}{ccc}
0 & 0 & 0
\\
0 & 0 & 0
\\
{1 \over 2} {\delta_\rho}^{[\mu_1} (C \Gamma^{\nu_1]})_{\beta_2
\gamma_2} & 0 & 0
\end{array} \right)
\quad.
\label{DXadj}
\end{eqnarray}
Using these in the exponential in~(\ref{Bexpr}) we get for
$B^{-1}$
\be
B^{-1}= \left( \begin{array}{ccc}
\delta^{\mu_1 \nu_1}_{\mu_2 \nu_2}
& 
- \theta^\alpha (C \Gamma^{[\mu_1})_{\alpha \alpha_2}
  {\delta^{\nu_1]}}_{\kappa_2}
&
- {1 \over 2} x^{[\mu_1} (C \Gamma^{\nu_1]})_{\beta_2 \gamma_2}
+ \theta^\alpha \theta^\beta (C \Gamma^{[\mu_1})_{\alpha
\gamma'_2} (C \Gamma^{\nu_1]})_{\beta'_2 \beta}
\\
0 & \delta^{\kappa_1 \alpha_1}_{\kappa_2 \alpha_2} 
& -{1 \over 4} \theta^{\alpha_1} (C \Gamma^{\kappa_1})_{\beta_2
\gamma_2} -2 \theta^{\alpha} (C \Gamma^{\kappa_1})_{\beta'_2
\alpha} {\delta^{\alpha_1}}_{\gamma'_2}
\\
0 & 0 & \delta^{\beta_1 \gamma_1}_{\beta_2 \gamma_2}
\end{array} \right)
\quad
\label{smBi}
\ee
(the external indices are $\,^{\mu_1 \nu_1} \,\,^{\kappa_1
\alpha_1} \,\,^{\beta_1 \gamma_1}$ for the rows and $\,_{\mu_2
\nu_2} \,\,_{\kappa_2 \alpha_2} \,\,_{\beta_2 \gamma_2}$ for the
columns).
Substituting now in~(\ref{JAS}) we find for the forms $J^A$ on $\Sigma$ 
\begin{eqnarray}
J^{\mu_1 \nu_1} &=& {2 \over 3} dx^{\mu_1} dx^{\nu_1}
  + {1 \over 15} \theta^\alpha (C \Gamma^{[\nu_1})_{\alpha \beta}
    dx^{\mu_1]} d\theta^\beta 
  + {1 \over 15} x^{[\mu_1} (C \Gamma^{\nu_1]})_{\alpha \beta}
    d \theta^\alpha d \theta^\beta
\nonumber \\
 &=& d \left( {2 \over 3} x^{[\mu_1} dx^{\nu_1]} 
       + {1 \over 15} \theta^\alpha x^{[\mu_1} 
         (C \Gamma^{\nu_1]})_{\alpha \beta} d\theta^\beta \right)
\quad,
\nonumber \\
J^{\kappa_1 \alpha_1} &=& -{3 \over 5} dx^{\kappa_1}
   d\theta^{\alpha_1} 
 - {1 \over 30} (C \Gamma^{\kappa_1})_{\alpha_3 \alpha_2}
   \theta^{\alpha_3} d \theta^{\alpha_2} d \theta^{\alpha_1}
 + {1 \over 30} (C \Gamma^{\kappa_1})_{\alpha_3 \alpha_2}
   \theta^{\alpha_1} d \theta^{\alpha_3} d \theta^{\alpha_2}
\nonumber \\
 &=&  d \left({3 \over 5} dx^{\kappa_1}  \theta^{\alpha_1}
       - {1 \over 30} (C \Gamma^{\kappa_1})_{\alpha_3 \alpha_2}
       \theta^{\alpha_3}  \theta^{\alpha_1} d
       \theta^{\alpha_2} \right)
\quad,
\nonumber \\
J^{\beta_1 \gamma_1} &=& -{2 \over 15} d \theta^{\beta_1} 
                           d \theta^{\gamma_1} 
                      = d (-{2 \over 15}\theta^{\beta_1}
                            d\theta^{\gamma_1})
\quad.
\label{JAsm}
\end{eqnarray}
The above locally exact expressions result from rather 
non-trivial cancellations. The currents 
are obtained by pulling back to $W$ the forms (\ref{JAsm}). For 
periodic $\theta$'s the charges 
$Q^{\kappa_1 \alpha_1}$, $Q^{\beta_1 \gamma_1}$ (but not
$Q^{\mu_1\nu_1}$
for a non-trivial two-cycle \cite{Azc.Gau.Izq.Tow:89}) 
turn out to be zero. Thus, this case 
provides a realization of the algebra (\ref{smLa}) where only the 
$Q^{\mu_1\nu_1}$ term is non-zero.
%%%%%%%%%%%%%%%%%%%%%%%%%%%%%%%%%%%%%%%%%%%%%%%%%%%%%%%%%%%%%%%%
%%%%%%%%%%%%%%%%%%%%%%%%%%%%%%%%%%%%%%%%%%%%%%%%%%%%%%%%%%%%%%%%
\section{The case of D-branes}
\label{casedbr}

Let us consider first a bosonic background for which all forms of the 
R$\otimes$R sector and the dilaton vanish so that 
the two-form ${\cal F}\equiv F-B$, where $F$=$dA$ and $B$ is the
NS$\otimes$NS two-form, reduces to $F$; $A$ is the Born-Infeld 
one-form on the worldvolume $A(\xi)=A_i(\xi)d\xi^i$. 
Then the action of the D$p$-brane reduces to
\begin{equation}
 I= \int d^{p+1} \xi \sqrt{-\det(\partial_i x^\mu \partial_j x_\mu
+F_{ij})}
\quad .
\label{borninfeld}                                    
\end{equation} 
Let us look for a manifestly supersymmetric generalisation
of this action on a suitable extension of flat superspace
(we shall consider here only $D=10$, IIA D-branes). 
For the ordinary $p$-branes the supersymmetrisation is 
achieved by substituting first $\Pi^\mu_i$ for $\partial_ix^\mu$ 
and then by adding a WZ term $b$, $db=h$, with $h$ characterised 
\cite{Azc.Tow:89} by being a non-trivial CE cohomology 
$(p+2)$-cocycle on superspace $\Sigma$.
It was shown in the previous sections how to make these 
WZ terms manifestly invariant. We shall extend this 
to the D$p$-branes case
by showing first that the WZ terms
may be characterised and classified by CE-cocycles as well, and
then by finding manifestly supersymmetric potentials ${\tilde b}$ 
on the superspaces $\tilde \Sigma$ which are obtained by the
techniques of Sec. \ref{NCESA} or by dimensional reduction from 
these. We shall restrict ourselves here to the D$2$-brane case, 
and hence to its associated $\tilde \Sigma$ 
parametrised by $(\theta^\alpha,x^\mu,\varphi_\mu,\varphi_\alpha,
\varphi_{\mu\nu},\varphi_{\mu\alpha},\varphi_{\alpha\beta},
\varphi)$.

\subsection{CE-cocycle classification of D-branes}
\label{CEdB}

The new feature in the D$p$-brane case is the field $A_i(\xi)$
directly defined on the worldvolume. The one-form $A$ transforms 
under supersymmetry in such a way that $ {\cal F}\equiv dA-B$ 
is invariant, where $B$ is a two-form on superspace such 
that
\begin{equation}
    dB=-\Pi^\mu(C\Gamma_\mu\Gamma_{11})_{\alpha\beta}\Pi^\alpha\Pi^\beta
 \quad .                                           \label{6.3}
\end{equation} 
Let us now consider $A$ as an abstract form. In our approach, the 
possible WZ terms will be some non-trivial $(p+2)$-cocycles
 of the cohomology of a certain FDA
(here, of IIA type). This FDA is
generated by the supersymmetric invariant $\Pi^\alpha$, $\Pi^\mu$ and $\cal 
F$ and defined by the structure relations
\be
d\Pi^\alpha = 0 
\, , \qquad
d\Pi^\mu = \frac{1}{2}(C\Gamma^\mu)_{\alpha\beta}
                \Pi^\alpha\Pi^\beta
\, , \qquad
d{\cal F} = \Pi^\mu(C\Gamma_\mu\Gamma_{11})_{\alpha\beta}
                 \Pi^\alpha\Pi^\beta
\, .
\label{freedifferential}
\ee
Note that $dd\equiv 0$ because the identity
$(C\Gamma^\mu\Gamma_{11})_{\alpha{'}\beta{'}}
(C\Gamma_\mu)_{\delta{'}\epsilon{'}}=0$ is valid in $D=10$.
The non-trivial $(p+2)$-cocycles are given by
closed ($p+2$)-forms $h$ constructed from $\Pi^\mu$, 
$\Pi^\alpha$, ${\cal F}$ that cannot be written as the differential of a 
($p+1$)-form constructed from them, and 
with the same dimensions as the kinetic 
Lagrangian, {\it i.e.} $[h]=L^{p+1}$. This second 
requirement is necessary to avoid introducing dimensionful constants in the 
action other than the tension, [$T$]=$ML^{-p}$.

Since ${\cal F}$ is a two-form, $h$ can be expanded in powers of
${\cal F}$ as
\begin{equation}     
h=\sum^{\left[\frac{p+2}{2}\right]}_{n=0}\frac{1}{n!}h^{(p+2-2n)}(\Pi^\mu, 
\Pi^\alpha){\cal F}^n \quad ,
\label{expansionh}
\end{equation}
where $h^{(k)}$ is a form of order $k\equiv p+2-2n$ and, since $D=10$, 
$p\leq 8$ excluding the degenerate case $p+1=10$. 
 Moreover, since $[h^{(p+2-2n)}]=L^{p+1-2n}$, the $k$-forms $h^{(k)}$,
$[h^{(k)}]=L^{k-1}$, must be
\begin{equation}
          h^{(k)}=a^{(k)}\Pi^{\mu_1}\dots \Pi^{\mu_{k-2}}(C\Gamma_{\mu_1\dots
\mu_{k-2}}\Gamma)_{\alpha\beta}\Pi^\alpha\Pi^\beta\, ,\qquad k=2,\dots,p+2
\, , \label{generalform}
\end{equation}
where $\Gamma={\bf 1}$ or $\Gamma_{11}$ so that
$(C\Gamma_{\mu_1\dots \mu_{k-2}}\Gamma)_{\alpha\beta}$ 
is symmetric (\emph{i.e}, $k-2=1,2,5,6,9,10$ for ${\bf 1}$ and
$0,1,4,5,8,9$ for $\Gamma_{11}$). Since
\begin{equation}
   dh=\sum_{n=0}^{\left[\frac{p+2}{2}\right]}\frac{1}{n!}dh^{(p+2-2n)}
    {\cal F}^n
         +\sum_{n=1}^{\left[\frac{p+2}{2}\right]}\frac{1}{(n-1)!}
        (-1)^p h^{(p+2-2n)}(C\Gamma_\mu\Gamma_{11})_{\alpha\beta}\Pi^\mu
        \Pi^\alpha\Pi^\beta {\cal F}^{n-1}\quad ,             \label{difh}
\end{equation}
the required
closure of $h$ is equivalent to the following 
set of equations
\begin{equation}
\begin{array}{rclcl}
 dh^{(p-2[p/2])} &=& 0 \, , 
& \quad & 
\text{for}\; n=\left[\frac{p+2}{2}\right]
\\[1.3ex]
dh^{(p+2-2n)}+(-1)^p h^{(p-2n)}\Pi^\mu(C\Gamma_\mu\Gamma_{11})_{\alpha\beta}
\Pi^\alpha\Pi^\beta &=& 0 \, ,
& \quad &
\text{for}\; n=\left[\frac{p}{2}\right],\dots, 0 \, .  
\end{array}
\label{recgeneral}
\end{equation}
At this point it is convenient to examine separately the $p$ odd 
and $p$ even cases.
\medskip

\noindent {\it a) $p$ even}
\medskip

The first eqn. in~(\ref{recgeneral}) gives
$dh^{(0)}=0$. This means 
$h^{(0)}=0$ because $h^{(0)}\neq 0$ would imply by (\ref{expansionh})
having an additional dimensionful constant,
$[h^{(0)}]=L^{-1}$ ([${\cal F}]= L^2$).
For $n=\frac{p}{2}$ the second of~(\ref{recgeneral}) gives $dh^{(2)}=0$. Inserting
$h^{(2)}$ from eq. (\ref{generalform}), we obtain an identity, so 
$a^{(2)}$ is arbitrary.\footnote{We note in passing that in the heterotic case, for 
which $N=1$ ($\Pi^\alpha$ is MW), $a^{(2)}=0$ because 
$(C\Gamma_{11})_{\alpha\beta}\Pi^\alpha
\Pi^\beta=C_{\alpha\beta}\Pi^\alpha\Pi^\beta=0$ since
${(\Gamma_{11})^\alpha}_\beta\Pi^\beta=\Pi^\alpha$. 
So the chain of equations that follows does 
not appear and there are no non-trivial WZ terms.
This shows, as expected, that there are 
no D-branes in the heterotic case.}
 The remaining equations (for $n=\frac{p-2}{2}$, etc.)
are equivalent, by factoring 
out products of forms $\Pi^\mu$ and $\Pi^\alpha$, to
\bae
 a^{(4)}(C\Gamma^{\mu_2})_{\alpha{'}\beta{'}}
   (C\Gamma_{\mu_1\mu_2})_{\delta{'}
   \epsilon{'}}-a^{(2)}(C\Gamma_{11})_{\alpha{'}\beta{'}}
   (C\Gamma_{\mu_1}
   \Gamma_{11})_{\delta{'}\epsilon{'}} &=& 0
\nonumber \\
  2a^{(6)}(C\Gamma^{\mu_4})_{\alpha{'}\beta{'}}
  (C\Gamma_{\mu_1\dots\mu_4} \Gamma_{11})_{\delta{'}
  \epsilon{'}}-a^{(4)}(C\Gamma_{[\mu_1\mu_2})_{\alpha{'}\beta{'}}
  (C\Gamma_{\mu_3]}
  \Gamma_{11})_{\delta{'}\epsilon{'}} &=& 0
\nonumber \\
  3a ^{(8)}(C\Gamma^{\mu_6})_{\alpha{'}\beta{'}}
  (C\Gamma_{\mu_1\dots\mu_6})_{\delta{'}
  \epsilon{'}}-a^{(6)}(C\Gamma_{[\mu_1\dots \mu_4}
  \Gamma_{11})_{\alpha{'} \beta{'}}(C\Gamma_{\mu_5]}
   \Gamma_{11})_{\delta{'}\epsilon{'}} &=& 0
\nonumber \\
  4a^{(10)}(C\Gamma^{\mu_8})_{\alpha{'}\beta{'}}
  (C\Gamma_{\mu_1\dots\mu_8}\Gamma_{11})_{\delta{'}
  \epsilon{'}}-a^{(8)}(C\Gamma_{[\mu_1\dots \mu_6})_{\alpha{'}
  \beta{'}}(C\Gamma_{\mu_7]}
  \Gamma_{11})_{\delta{'}\epsilon{'}} &=& 0
\, .
\label{equationsc}
\eae
Note that the 
number of identities from (\ref{equationsc}) that are necessary 
to show that $h$ is closed depends on the values of $p$ since
$2\leq k\leq p+2$, $k$ even. Specifically, for the
D$2$-brane only the first identity is used, for the D$4$-brane the 
first two identities are relevant and the first three (four) 
identities are required for the existence of the D$6$-(D$8$-)branes.

Equations (\ref{equationsc}) have to be identically satisfied for 
certain values of the 
$a$'s to be determined. To find them, one may multiply 
the equations by $(\Gamma^\nu C^{-1})^{\beta\delta}$ (although the 
resulting system is not equivalent to the original one). 
This procedure gives some 
equalities between gamma matrices that are only satisfied if $a^{(2)}=
-a^{(4)}$, $a^{(4)}=-6a^{(6)}$, $a^{(6)}=-15a^{(8)}$ and 
$a^{(8)}=-28a^{(10)}$.
These values are the solution of eqs. (\ref{equationsc}) and determine
closed forms $h$ by (\ref{expansionh}), (\ref{generalform}) 
provided that the following  identities are satisfied 
\bae
  (C\Gamma^{\mu_2})_{\alpha{'}\beta{'}}(C\Gamma_{\mu_1\mu_2})_{\delta{'}
   \epsilon{'}}+(C\Gamma_{11})_{\alpha{'}\beta{'}}(C\Gamma_{\mu_1}
   \Gamma_{11})_{\delta{'}\epsilon{'}} &=& 0
\nonumber \\
  (C\Gamma^{\mu_4})_{\alpha{'}\beta{'}}(C\Gamma_{\mu_1\dots\mu_4}
  \Gamma_{11})_{\delta{'}
  \epsilon{'}}+3(C\Gamma_{[\mu_1\mu_2})_{\alpha{'}\beta{'}}
  (C\Gamma_{\mu_3]}
  \Gamma_{11})_{\delta{'}\epsilon{'}} &=& 0
\nonumber \\
  (C\Gamma^{\mu_6})_{\alpha{'}\beta{'}}
  (C\Gamma_{\mu_1\dots\mu_6})_{\delta{'}
  \epsilon{'}}+5(C\Gamma_{[\mu_1\dots \mu_4}
  \Gamma_{11})_{\alpha{'} \beta{'}}(C\Gamma_{\mu_5]} 
  \Gamma_{11})_{\delta{'}\epsilon{'}} &=& 0
\nonumber \\
  (C\Gamma^{\mu_8})_{\alpha{'}\beta{'}}
  (C\Gamma_{\mu_1\dots\mu_8}\Gamma_{11})_{\delta{'}
  \epsilon{'}}+7(C\Gamma_{[\mu_1\dots \mu_6})_{\alpha{'}
  \beta{'}}(C\Gamma_{\mu_7]}
  \Gamma_{11})_{\delta{'}\epsilon{'}} &=& 0 \, ,
\label{Identities}
\eae
as it is indeed the case (see Appendix B).
Therefore we have shown that, for $p$ even, there exist 
closed $(p+2)$-forms $h$ with the required dimensions for all 
even values of $p$, $p\leq 8$.

To prove that the $h$'s obtained from eq. (\ref{generalform}) for 
the appropriate values of $a^{(k)}$ are not CE-trivial, it is 
sufficient to note that if there were a potential form 
$b(\Pi^\mu,\Pi^\alpha,{\cal F})$, 
$db=h$, then this form would be a Lorentz-invariant ($p+1$)-form with physical 
dimensions $L^{p+1}$, which does not exist for $p+1<10$ since $p$ is even.

\medskip

\noindent{\it b) $p$ odd}

\medskip

In this case, the first eqn. in~(\ref{recgeneral}) gives
$dh^{(1)}=0$. Again, this means 
$h^{(1)}=0$ because obviously there are no Lorentz-scalar 
one-forms that can be constructed from $\Pi^\mu$, $\Pi^\alpha$ 
and $\cal F$. On the other hand, $k$ in  $h^{(k)}$ has 
now the range $3\leq k\leq p+2$, $k$ odd. Of these $h^{(k)}$, 
those corresponding to $k=5$ and $k=9$ vanish independently of the 
type of the matrix $\Gamma$ (see table). This leaves us with $h^{(3)}$ 
and $h^{(7)}$ and the second of (\ref{recgeneral}) leads to 
\begin{equation}
\begin{array}{rclcrcl}
  \displaystyle
   a^{(3)}(C\Gamma^{\mu_1})_{\alpha{'}\beta{'}}(C\Gamma_{\mu_1}\Gamma)_{\delta{'}
\epsilon{'}} &=& 0 \, ,
& \qquad &
  a^{(3)}(C\Gamma_{[\mu_1}\Gamma)_{\alpha{'}\beta{'}}(C\Gamma_{\mu_2]}\Gamma_{11}
    )_{\delta{'}\epsilon{'}} &=& 0 
  \\[1.3ex]
  \displaystyle
  a^{(7)}(C\Gamma^{\mu_7})_{\alpha{'}\beta{'}}(C\Gamma_{\mu_1\dots \mu_5}\Gamma
)_{\delta{'}\epsilon{'}} &= & 0 \, ,
  & \qquad &
   a^{(7)}(C\Gamma_{[\mu_1\dots\mu_5}\Gamma)_{\alpha{'}\beta{'}}(C\Gamma_{\mu_6]}
  \Gamma_{11})_{\delta{'}\epsilon{'}} &=& 0 \, .
\end{array}                             
\label{oddcase}
\end{equation}
In the $a^{(3)}$ equations,  $\Gamma$ has to be $\Gamma_{11}$ (the other 
possibility, ${\bf 1}_{32}$, may be shown to be inconsistent 
by multiplying the second expression by 
$(\Gamma^\nu C^{-1})^{\beta\delta}$). Multiplying the fourth equation
by $(\Gamma^\nu C^{-1})^{\beta\delta}$ shows that 
$a^{(7)}=0$ for both $\Gamma$=${\bf 1}$, $\Gamma_{11}$. Thus we 
have shown that the only candidate for a WZ term in the 
odd $p$ case is obtained from $h^{(3)}$ \emph{i.e.}, from
\begin{equation}
    h=\Pi^\mu(C\Gamma_\mu\Gamma_{11})_{\alpha\beta}\Pi^\alpha\Pi^\beta
  {\cal F}^{\frac{p-1}{2}}\quad ,\quad p\geq 1\quad .    \label{triviala}
\end{equation}
But then $h=d(\frac{2}{p+1}{\cal F}^{(p+1)/2})$ 
by (\ref{freedifferential}), 
and hence is a trivial CE cocycle. Therefore in the $D=10$, 
IIA theory there are no non-trivial WZ terms for the D-branes with 
$p$ odd. The other values for $p$ found in our discussion are 
precisely those for which D-branes of type IIA are known 
to exist. Thus the \emph{IIA D-branes are, as the scalar $p$-branes} 
\cite{Azc.Tow:89}, \emph{characterized by non-trivial CE cocycles}.

\bigskip

\subsection{D-branes defined on extended superspace}
\label{Dbext}

As we saw, one reason for considering superspace extensions 
associated with extended objects is that it is possible to 
find on $\tilde{\Sigma}$ manifestly invariant WZ terms since 
then $h$ may be expressed as the differential of a LI form 
${\tilde b}$. We shall now show that this is also possible 
for the D$2$-brane. The starting point is now the FDA given in 
eq. (\ref{2.15}) with the generators with more than two vector 
indices absent,
plus the equation for $d{\cal F}$ {\it i.e.}
\be
\addtolength{\arraycolsep}{-.5ex}
\begin{array}{rclcrcl}
\dis
d\Pi^\alpha&=&0
& \qquad \quad &
\dis
d\Pi^\mu &=& \frac{1}{2}(C\Gamma^\mu)_{\alpha\beta}\Pi^\alpha\Pi^\beta
\\[1.3ex]
\dis
d\Pi &=& \frac{1}{2}(C\Gamma_{11})_{\alpha\beta}\Pi^\alpha\Pi^\beta
& \qquad \quad &
\dis
d\Pi_{\mu\nu} &=& \frac{1}{2}(C\Gamma_{\mu\nu})_{\alpha\beta}\Pi^\alpha
                  \Pi^\beta
\\[1.3ex]
\dis
d\Pi_\mu^{(z)} &=& \frac{1}{2}(C\Gamma_\mu\Gamma_{11})_{\alpha\beta}
                   \Pi^\alpha\Pi^\beta 
& \qquad \quad &
\dis
d{\cal F}&=& (C\Gamma_\mu\Gamma_{11})_{\alpha\beta}\Pi^\mu\Pi^\alpha
              \Pi^\beta   \, .
 \label{freedbrane}
\end{array}
\ee
The reason one should start from (\ref{freedbrane}) 
is that the dual of the algebra defined 
by the first five equations is the one obtained 
when one computes the algebra of the Noether charges associated with 
the supertranslations in the case of the type IIA D$2$-brane (see
\cite{Ham:98}).

The next step, as was done in Sec. \ref{NCESA}, is extending this algebra 
with the generators obtained by replacing vector indices by 
spinorial ones. In the case of the D$2$-brane this is not difficult 
to do because, apart from the equation for $d{\cal F}$, 
the free differential algebra one starts with is actually the dimensional 
reduction of the $11$-dimensional one
\begin{equation}
d\Pi^{\tilde\mu}=\frac{1}{2}(C\Gamma^{\tilde\mu})_{\alpha\beta}\Pi^\alpha
      \Pi^\beta
\, ,\qquad \quad
d\Pi_{\tilde\mu\tilde\nu}=
      \frac{1}{2}(C\Gamma_{\tilde\mu\tilde\nu})_{\alpha\beta}\Pi^\alpha
  \Pi^\beta  \, ,
\label{freeeleven}
\end{equation}
in which one sets $\Pi^{\tilde\mu}\equiv(\Pi^\mu,\Pi^{10}=\Pi)$,
$\Pi_{\tilde\mu\tilde\nu}
\equiv(\Pi_{\mu\nu},\Pi_{\mu 10}=\Pi_\mu^{(z)})$. Since 
this $D=11$ algebra has already been extended
recursively in Sec. \ref{NCESA} by the new one-forms 
$\Pi_{\tilde\mu\alpha}$ and $\Pi_{\alpha\beta}$, the extended 
algebra in $D=10$ is simply its dimensional
reduction, for which $\Pi_{\tilde\mu\alpha}\equiv(\Pi_{\mu\alpha},
\Pi_\alpha^{(z)})$. The result is given by eqs. (\ref{freedbrane}) plus
\begin{eqnarray}
   d\Pi_{\mu\alpha}&=&(C\Gamma_{\nu\mu})_{\alpha\beta}\Pi^\nu\Pi^\beta+
    (C\Gamma_{11}\Gamma_\mu)_{\alpha\beta}\Pi\Pi^\beta+
(C\Gamma^\nu)_{\alpha\beta}\Pi_{\nu\mu}\Pi^\beta-(C\Gamma_{11})_{\alpha\beta}
    \Pi_\mu^{(z)}\Pi^\beta\quad ,
    \nonumber
   \\
d\Pi_\alpha^{(z)}&=&(C\Gamma_\nu\Gamma_{11})_{\alpha\beta}\Pi^\nu\Pi^\beta
   +(C\Gamma^\nu)_{\alpha\beta}\Pi_\nu^{(z)}\Pi^\beta\quad ,
    \nonumber
    \\
    d\Pi_{\alpha\beta}&=&-\frac{1}{2}(C\Gamma_{\mu\nu})_{\alpha\beta}
     \Pi^\mu\Pi^\nu-
   (C\Gamma_\mu\Gamma_{11})_{\alpha\beta}\Pi^\mu\Pi-\frac{1}{2}
   (C\Gamma^\mu)_{\alpha\beta} \Pi_{\mu\nu}\Pi^\nu
   \nonumber
\\
    & &  \mbox{} +\frac{1}{2}(C\Gamma_{11})_{\alpha\beta}
   \Pi_{\mu}^{(z)}\Pi^\mu-
    \frac{1}{2}(C\Gamma^\mu)_{\alpha\beta}
     \Pi_\mu^{(z)}\Pi
   +\frac{1}{4}(C\Gamma^\mu)_{\alpha\beta}\Pi_{\mu\delta}\Pi^\delta
\nonumber
\\
 & & \mbox{}+ \frac{1}{4}(C\Gamma_{11})_{\alpha\beta}\Pi_\delta^{(z)}\Pi^\delta
   +2(C\Gamma^\mu)_{\delta\alpha{'}}\Pi_{\mu\beta{'}}\Pi^\delta+
    2(C\Gamma_{11})_{\delta\alpha{'}}\Pi_{\beta{'}}^{(z)}\Pi^\delta
  \quad ,
                                    \label{freedbraneext}
\end{eqnarray}
which, apart from the $d{\cal F}$ equation, corresponds to
the dimensional reduction of (\ref{smLa}) (with $Z^\mu=2Z^{\mu 10}$) 
\begin{eqnarray}
\{ D_\alpha,D_\beta\} &=& (C\Gamma^\mu)_{\alpha\beta} X_\mu + 
(C \Gamma_{\mu\nu})_{\alpha\beta}
Z^{\mu\nu}+(C\Gamma_{11})_{\alpha\beta}Z+(C\Gamma_\mu
\Gamma_{11})_{\alpha\beta}Z^\mu
\nonumber \\
 \left[ X_\mu,D_\alpha \right] &=& - 
(C\Gamma^{\mu\nu})_{\alpha\beta} Z^{\nu\beta}-
(C\Gamma_\mu
\Gamma_{11})_{\alpha\beta} Z^\beta
\nonumber \\
\left [Z,D_\alpha\right] &=& (C\Gamma_{11}\Gamma_\mu)_{\alpha\beta}Z^{\beta\mu}
\nonumber \\
\left[ X_\mu,X_\nu \right] &=& (C\Gamma^{\mu\nu})_{\alpha\beta} Z^{\alpha\beta}
\nonumber \\
\left[Z,X_\mu\right] &=& (C\Gamma_{11}\Gamma_\mu)_{\alpha\beta}Z^{\alpha\beta}
\nonumber \\
\left[ X_\mu, Z^{\lambda\tau} \right] &=& {1 \over 2}
\delta^{\left[  \lambda \right.}_\mu
(C \Gamma^{\left. \tau \right]})_{\alpha\beta}Z^{\alpha\beta}
\nonumber \\
\left[ X_\mu,Z^\nu\right] &=& \frac{1}{2}\delta^\nu_\mu(C\Gamma_{11})_{\alpha\beta}
Z^{\alpha\beta}
\nonumber \\
\left[ Z,Z^\mu\right] &=& -\frac{1}{2}(C\Gamma^\mu)_{\alpha\beta}Z^{\alpha\beta}
\nonumber \\
\left[ D_\alpha,Z^{\mu\nu} \right] &=& (C \Gamma^{\left[
\mu \right.})_{\alpha\beta} Z^{\left.\nu \right] \beta}
\nonumber \\
\left[ D_\alpha,Z^\mu\right] &=& -(C\Gamma_{11})_{\alpha\beta}Z^{\mu\beta}+
(C\Gamma^\mu)_{\alpha\beta}Z^\beta
\nonumber \\
\{ D_\alpha, Z^{\nu\beta}\} &=& 
({1 \over 4}(C \Gamma^\nu)_{\gamma\delta}
\delta^\beta_\alpha +2(C \Gamma^\nu)_{\gamma\alpha}
\delta^\beta_\delta) Z^{\gamma\delta}
\nonumber \\
\{ D_\alpha, Z^\beta\} &=& ({1 \over 4}(C 
\Gamma_{11})_{\gamma\delta}\delta^\beta_\alpha+2
(C
\Gamma_{11})_{\gamma\alpha}\delta^\beta_\delta) Z^{\gamma\delta}
\quad .
\label{smLaext}
\end{eqnarray}
 We can now show, using the new forms in (\ref{freedbraneext}), that 
it is possible to find an invariant
WZ term  ${\tilde b}$, $h=d{\tilde b}$, on 
the extended superspace. In our case $h$ is given by (eqs. (\ref{expansionh}), 
(\ref{generalform}); $k=2,4$)
\begin{equation}
      h=(C\Gamma_{\mu\nu})_{\alpha\beta}\Pi^\mu\Pi^\nu\Pi^\alpha\Pi^\beta
    -(C\Gamma_{11})_{\alpha\beta}\Pi^\alpha\Pi^\beta {\cal F}
                         \quad .             \label{hagain}
\end{equation}
Again, it is possible to expand ${\tilde b}$ as 
${\tilde b}=b^{(3)}+b^{(1)}{\cal F}$.
Using this expression in $h=d{\tilde b}$, and identifying the result with $h$, yields
$db^{(1)}=-(C\Gamma_{11})_{\alpha\beta}\Pi^\alpha\Pi^\beta$, from which follows 
that $b^{(1)}=-2\Pi$. Similarly,
\begin{equation}
    db^{(3)}=-2(C\Gamma_\mu\Gamma_{11})_{\alpha\beta}\Pi\Pi^\mu\Pi^\alpha
  \Pi^\beta +(C\Gamma_{\mu\nu})_{\alpha\beta}\Pi^\mu\Pi^\nu\Pi^\alpha
  \Pi^\beta=(C\Gamma_{\tilde\mu\tilde\nu})_{\alpha\beta}
  \Pi^{\tilde\mu}\Pi^{\tilde\nu}\Pi^\alpha\Pi^\beta
                                         \label{wztrivial}
\end{equation}
where in the last equality we have rewritten the expression using the
eleven-dimensional notation. This has the advantage that the expression
for $b^{(3)}$ in $D=11$ was given in (\ref{smbt}), 

\begin{equation}
    b^{(3)}=\frac{2}{3}\Pi_{\tilde\mu\tilde\nu}\Pi^{\tilde\mu}\Pi^{\tilde
     \nu}-\frac{2}{15}\Pi_{\alpha\beta}\Pi^\alpha\Pi^\beta-
     \frac{3}{5}\Pi_{\tilde\mu\alpha}\Pi^{\tilde\mu}\Pi^\alpha \quad.
                                             \label{bersez}
\end{equation}
Reducing (\ref{bersez}) to $D=10$ and adding $b^{(1)}{\cal F}=
-2\Pi {\cal F}$ we find the invariant WZ term,
\begin{equation}
  {\tilde b}=
 \frac{2}{3}\Pi_{\mu\nu}\Pi^\mu\Pi^\nu+\frac{4}{3}\Pi_\mu^{(z)}\Pi^\mu
    \Pi-\frac{2}{15}\Pi_{\alpha\beta}\Pi^\alpha\Pi^\beta
   -\frac{3}{5}\Pi_{\mu\alpha}\Pi^\mu\Pi^\alpha-\frac{3}{5}\Pi_\alpha^{(z)}
   \Pi\Pi^\alpha-2\Pi {\cal F}\quad .
                                 \label{bdbrane}
\end{equation}
This shows that on the extended superspace corresponding to
eqs.  (\ref{freedbrane}) and (\ref{freedbraneext}), the WZ term 
of the type IIA D$2$-brane can be made invariant, 
as was the case for the ordinary $p$-branes.
We expect that this result holds for the other values of $p$. 

In contrast with the case of ordinary $p$-branes, the extended
free differential algebra is not the dual of a Lie algebra 
because of the equation for the two-form $d{\cal F}$.
However,  it is easy to check 
that 
\begin{equation}
      d(\frac{1}{2}\Pi^\alpha\Pi_\alpha^{(z)}-\Pi^\mu\Pi_\mu^{(z)})=
    (C\Gamma_\mu\Gamma_{11})_{\alpha\beta}\Pi^\mu\Pi^\alpha
      \Pi^\beta                                \label{6.22}
\end{equation}
so that, on the extended superspace, 
we may set
\begin{equation}
{\cal F}=
\frac{1}{2}\Pi^\alpha\Pi_\alpha^{(z)}-\Pi^\mu\Pi_\mu^{(z)}\quad,
\label{fonsuper}
\end{equation}
in accordance with (\ref{freedbrane}), (\ref{freedbraneext}).
This is not a surprising fact since from (\ref{freedbrane}) 
we see that $d{\cal F}$ is equal to the $h$ corresponding 
to the WZ term of the type IIA superstring on a flat background. 
So it has to be possible to write it as the differential of an 
invariant form ${\tilde b}=
{\tilde b}(\Pi^\mu,\Pi^\alpha,\Pi_\mu^{(z)},\Pi_{\alpha}^{(z)})$ 
on the fully extended superspace ${\tilde \Sigma}$ of the IIA 
superstring. Since ${\cal F}=dA-B$ and $B$ is defined on $\Sigma$, 
$dA$ may be written on ${\tilde \Sigma}$.
Making use of its LI forms \ie,  of
\be
\addtolength{\arraycolsep}{-0.8ex}
\begin{array}{rclcrcl}
\dis
\Pi^\alpha &=& \dis d\theta^\alpha
\, , & \quad &
\dis
\Pi^\mu&=& \dis dx^\mu+\frac{1}{2}(C\Gamma^\mu)_{\alpha\beta}\theta^\alpha
     d\theta^\beta
\\[1.3ex]
\dis 
\Pi_\mu^{(z)}&=& \dis d\varphi_\mu+\frac{1}{2}
(C\Gamma_\mu\Gamma_{11})_{\alpha\beta}
   \theta^\alpha d\theta^\beta
\, , & \quad &
\displaystyle
\Pi_\alpha^{(z)}&=& \dis d\varphi_\alpha-(C\Gamma_\mu\Gamma_{11})_{\alpha\beta}
   dx^\mu\theta^\beta-(C\Gamma^\mu)_{\alpha\beta}d\varphi_\mu\theta^\beta
\\[1.3ex]
\displaystyle
 & & & & & & \dis -\frac{1}{6}\left[ (C\Gamma_\mu\Gamma_{11})_{\alpha\beta}(C\Gamma^\mu)_{\delta
   \epsilon} \! + \! (C\Gamma_\mu\Gamma_{11})_{\delta\epsilon}
(C\Gamma^\mu)_{\alpha
   \beta} \right]\theta^\beta \theta^\delta d\theta^\epsilon
\, ,
\end{array}
\ee
it is easy to identify $A$ as the one-form on $\tilde\Sigma$
\begin{equation}
    A=\varphi_\mu dx^\mu+\frac{1}{2}\varphi_\alpha d\theta^\alpha \quad .
                                                  \label{newA}
\end{equation}
In this way, {\it the customary Born-Infeld worldvolume field $A_i(\xi)$  
becomes here $\phi^*(A)$}, with $A$ on $\tilde\Sigma$
given by (\ref{newA}), and its existence may be looked at as
a consequence of extended supersymmetry. 

The previous discussion shows that it is natural
to rewrite the action of the D-branes on a flat background 
by using \emph{only} objects that are initially defined on the
appropriately extended superspace. We show now that 
the Euler Lagrange (EL) equations are still
the same (provided a rather natural condition is met) and that the gauge 
transformations of $A_i(\xi)$ can be reinterpreted in the new language. So at this 
point it seems that the geometric difference between the ordinary $p$-branes and the 
D$p$-branes is that while the action of the former may be defined
from forms on ordinary superspace $\Sigma$, the action of the latter
requires the extended superspace of the IIA superstring if one 
whishes to avoid objects that only have a meaning on 
the worldvolume. In the IIA superstring case 
the extended superspace was also considered, but the 
new variables appeared only in the WZ term and as total derivatives 
(Sec. \ref{css}) and thus had trivial EL equations. In the D-brane 
case, in contrast, these variables have nontrivial EL equations.

Let us now see how the EL equations change by making the substitution
$A\rightarrow \varphi_\mu dx^\mu+\frac{1}{2}\varphi_\alpha d\theta^\alpha$.
Let $I[x^\mu,\theta^\alpha,A_i]$ be the action before making the 
subtitution, where $A_i$ are the coordinates  of the form  
$A$ ($A=A_i d\xi^i$). The EL equations are
\begin{equation}
     \frac{\delta I}{\delta x^\mu}=0
\, ,\qquad 
\frac{\delta I}{\delta \theta^\alpha}=0
\, ,\qquad
\frac{\delta I}{\delta A_j}=0 \, .
\label{ELbefore}
\end{equation}  
When the substitution is made, there are new terms in the equations, and they 
read
\begin{equation}
\label{elr}
 \begin{array}{rclcrcll}
\displaystyle
\int d\xi{'}^{p+1}\frac{\delta I}{\delta A_j(\xi{'})}
\frac{\delta A_j(\xi{'})}{\delta x^\mu(\xi)}+\frac{\delta I}{\delta 
x^\mu} & = & 0
\, , & \qquad &
\displaystyle
\frac{\delta I}{\delta\varphi_\mu} =
\frac{\delta I}{\delta A_j}\partial_j x^\mu & = & 0
\\[2.6ex]
\displaystyle
\int d\xi{'}^{p+1}\frac{\delta I}{\delta A_j(\xi{'})}
\frac{\delta A_j(\xi{'})}{\delta \theta^\alpha(\xi)}+
\frac{\delta I}{\delta \theta^\alpha} &= & 0
\, , & \qquad  &
\displaystyle
   \frac{\delta I}{\delta\varphi_\alpha}=
     \frac{1}{2} \frac{\delta I}{\delta A_j}\partial_j
\theta^\alpha & = & 0 \, ,
  \end{array}
\end{equation}
where the additional contributions come from the partial functional 
derivative terms. If 
$\frac{\delta I}{\delta A_j}\partial_j x^\mu$ and 
$\frac{\delta I}{\delta A_j}\partial_j \theta^\alpha$ were zero
without $\frac{\delta I}{\delta A_j}$ being zero, this 
would imply the collapse of one worldvolume dimension. Thus, we 
must have $\frac{\delta I}{\delta A_j}=0$ which
in the first equation of each set in (\ref{elr})
implies eqs. (\ref{ELbefore}). Hence both actions are equivalent.
In fact, it may be shown that there is an additional
gauge freedom which accounts for the difference of
degrees of freedom between $A$ (eqn. (\ref{newA})) and $A_i(\xi)$, 
but we shall not discuss this here\footnote{We thank P. Townsend
for discussions on this point.}. 
This seems to indicate that, when
the action on $W$ is obtained from entities on 
${\tilde \Sigma}$, there is an additional gauge freedom 
which in our formulation plays a r\^ole complementing 
that of $\kappa$-symmetry.

Finally, the $U(1)$ gauge field $A_i(\xi)$ on $W$ 
 has a gauge transformation $\delta A_i(\xi)=\partial_i 
\Lambda(\xi)$. The question now is what is the gauge transformation 
of the component fields  if one writes $A_id\xi^i$ as $\phi^*(A)$. 
In other words, for a 
given $\Lambda(\xi)$, there should be 
a transformation of $\varphi_\mu$ and $\varphi_\alpha$ in (\ref{newA})
reproducing $\delta_i\Lambda$. This may be obtained by 
means of a superfield $\lambda$ such that
$\phi^*\lambda(x^\mu,
\theta^\alpha)=\Lambda(\xi)$. Then if under a 
gauge transformation one defines
$\delta \varphi_\mu= \partial_\mu \lambda$ and $\delta \varphi_\alpha=
2\partial_\alpha \lambda$, $\phi^*(A)$ behaves as expected since then
$\delta\phi^* [\varphi_\mu dx^\mu+
\frac{1}{2}\varphi_\alpha d\theta^\alpha]=\partial_i \Lambda$.
The fact that when the
supersymmetry transformations of a field close only
modulus a gauge transformation one obtains
an extension of a FDA is not restricted to D$p$-branes. 
In fact, one may achieve manifest invariance by 
introducing an electromagnetic potential on
the worldsheet in the Green-Schwarz superstring action,
in which case the string tension is the circulation of
the potential around the string \cite{Tow:92}
(see also \cite{Azc.Izq.Tow:92})
and a similar result applies to the other scalar 
$p$-branes ~\cite{Ber.Lon.Tow:92}. Clearly, the
worldvolume fields introduced there could be defined
on our appropriate extended superspaces as well. As for the 
IIB D$p$-brane, an analysis similar to that in this
section for the IIA case would classify them by first 
showing that WZ terms exist for odd $p$. In a second stage, 
the worldvolume gauge field $A$ may be expressed as the pull-back 
of a IIB superspace one-form. In fact, this last point for the $A$ 
in the $p$=1 IIB D-string case has been discussed very 
recently in \cite{Sak:98}
by introducing an appropriate extended group manifold.
We may conclude, then, that the different worldvolume fields
may be expressed in terms of forms defined on suitably extended
superspaces.

\section{Noether charges and D-brane actions}
\label{2bn}

It follows from the discussion of Sec. \ref{casedbr} that the 
worldvolume field $A(\xi)$ that appears in the D$2$-brane action 
may be written on the superstring extended superspace 
parametrized by $(x^\mu,\theta^\alpha,\varphi_\mu, \varphi_\alpha)$. 
On the other hand, the D2-WZ term, which is quasi-invariant in 
these coordinates, can be made strictly invariant by further extending the 
previous superspace to
${\tilde \Sigma}=(\theta^\alpha,x^\mu,\varphi_\mu,
\varphi_\alpha$, $\varphi_{\mu\nu},\varphi_{\mu\alpha},\varphi_{\alpha\beta},
\varphi)$. In this way, the whole action is invariant.

If one now computes the canonical commutators (or Poisson brackets)
of the charges 
corresponding to 
the symmetries of the action, the resulting algebra is exactly 
the RI version of the Lie algebra dual to (\ref{freedbrane}) 
(removing its last line) plus (\ref{freedbraneext}), 
given in (\ref{smLaext}). 
The RI generator algebra ($\{Q,Q\}$, etc.) is 
the same as (\ref{smLaext}) with an 
additional minus sign on the r.h.s.
Let us concentrate on the $\{ Q_\alpha, Q_\beta\}$ commutator,
\begin{equation}
 \{ Q_\alpha,Q_\beta\}= (C\Gamma^\mu)_{\alpha\beta}P_\mu +
  (C\Gamma_\mu\Gamma_{11})_{\alpha\beta}{\hat Z}^\mu +
  (C\Gamma_{\mu\nu})_{\alpha\beta}{\hat Z}^{\mu\nu}
  + (C\Gamma_{11})_{\alpha\beta}{\hat Z}\quad   
  \label{QQ}
\end{equation}
(there has been a redefinition of the generators so that $\{ Q , Q \}
= + C\Gamma^\mu P_\mu$ etc).
 Let us assume that we had written the action, as it is
customary, in terms of $(x^\mu,\theta^\alpha,A)$ alone, 
with $A=A(\xi)$ directly defined on $W$. In this case, 
the $C\Gamma_{\mu\nu}$ and $C\Gamma_{11}$ contributions
would come from the quasi-invariance of the WZ Lagrangian, while
$C\Gamma_{11}\Gamma_\mu$ would be the result of the contribution of the
$A(\xi)$ field to the Noether current \cite{Ham:98} (see also 
\cite{Ber.Tow:98}).
This follows easily from the appropriate definition of the 
conserved Noether currents and charges (see, \eg, \cite{Azc.Izq:95})
which include an additional piece when the Lagrangian is 
quasi-invariant, a common feature of the conventional actions
for $p$-branes \cite{Azc.Gau.Izq.Tow:89}. In the present 
D-branes case, there is an additional contribution due to the
\emph{worldvolume} field $A(\xi)$ since its transformation 
properties, $\delta A=\Delta$, are \emph{postulated} to 
compensate for those of the composite object $B$, 
$\delta B= d\Delta$,
so that ${\cal F}= dA-B$ is invariant. As a result, the 
supersymmetry transformations {\it do not close on $A$}, and this
produces an additional term by a mechanism similar to the one in
the standard quasi-invariance case.
 
     These modifications become evident in our context \ie,
by formulating the action on the extended superspace. Let us 
consider the D2-brane Lagrangian with the quasi-invariant WZ term 
$b=b(x^\mu,\theta^\alpha,\varphi_\mu,\varphi_\alpha)$. The conserved 
Noether currents then have to include the quasi-invariance
piece. If we wrongly ignored this additional term, the 
(canonical formalism) algebra of the corresponding (non-conserved, 
non-Noether) charges would be the algebra of the symmetries
$x^\mu,\theta^\alpha,\varphi_\mu,\varphi_\alpha$ of the 
Lagrangian, \emph{i.e.} 

\begin{equation}
\{ Q_\alpha,Q_\beta\}= (C\Gamma^\mu)_{\alpha\beta}P_\mu +
  (C\Gamma_\mu\Gamma_{11})_{\alpha\beta}{\hat Z}^\mu \quad .    
\label{2Db}
\end{equation} 
The algebra of the conserved Noether charges is not (\ref{2Db}),
however, because these must include the quasi-invariance
contribution. We may find the correct algebra immediatley 
by replacing the quasi-invariant WZ term $b$, by 
${\tilde b}={\tilde b}(x^\mu,\theta^\alpha,\varphi_\mu,\varphi_\alpha,
\varphi_{\mu\nu},\varphi_{\mu\alpha},\varphi_{\alpha\beta},\varphi)$,  
which is manifestly invariant since the transformation properties 
of the additional variables $(\varphi_{\mu\nu},
\varphi_{\mu\alpha},\varphi_{\alpha\beta},\varphi)$ remove  the 
quasi-invariance of $b$. By definition, the transformation 
properties of all the coordinates obviously close into the group 
law or algebra. Hence, it follows that the algebra of charges
computed using the canonical formalism reproduces (\ref{QQ}), and 
that the contributions to ${\hat Z}^{\mu\nu}$ and $\hat Z$ are 
entirely due to the WZ term $\tilde b$ (or to the quasi-invariance 
of $b(x^\mu,\theta^\alpha,\varphi_\mu,\varphi_\alpha)$ if we
used $b$ instead). 

%%%%%%%%%%%%%%%%%%%%%%%%%%%%%%%%%%%%%%%%%%%%%%%%%%%%%%%%%%%%%%%%%%%%%%%
%%%%%%%%%%%%%%%%%%%%%%%%%%%%%%%%%%%%%%%%%%%%%%%%%%%%%%%%%%%%%%%%%%%%%%%
%%%%%%%%%%%%%%%%%%%%%%%%%%%%%%%%%%%%%%%%%%%%%%%%%%%%%%%%%%%%%%%%%%%%%%%
\section{Branes with higher order tensors: the case of the 
M$5$-brane}
\label{m5b}
%%%%%%%%%%%%%%%%%%%%%%%%%%%%%%%%%%%%%%%%%%%%%%%%%%%%%%%%%%%%%%%%%%%%%%%

We shall now show that the previous analysis can be applied also
to other extended objects that are neither ordinary $p$-branes 
nor D-branes. We shall consider here the case of the $D=11$
$M$5-brane, which contains a worldvolume two-form field $A$
in the action (see \cite{Ban.Lec:97}). The action in a flat bosonic 
background depends on $A$ trough $H=dA-C$, where $C$ is a 
background three-form. We shall take as our starting point the case with 
$C=0$ and with all other forms of rank higher than one in that action 
vanishing. We do not need to worry about the (generalized)
self-duality condition for A on the worldvolume, since this 
condition may arise as a field equation for an auxiliary field 
(see \cite{Ban.Lec:97, Aga.Par:97}). The supersymmetric 
action of the M$5$-brane is obtained in two steps. First, 
one substitutes $H=dA-C$ for $dA$, where $C$ is a form on 
ordinary flat superspace such that
\begin{equation}
    dC=-(C\Gamma_{\mu\nu})_{\alpha\beta}\Pi^\mu\Pi^\nu\Pi^\alpha\Pi^\beta
 \quad ,                                                      \label{11a}
\end{equation}
and the transformation properties of the worldvolume field $A$ are fixed so 
that $H$ is invariant. Secondly, a WZ term is added to
obtain $\kappa$ symmetry.

   Let us now find the WZ term in our framework. It should be 
obtained from the FDA generated by the abstract invariant forms 
$\Pi^\alpha$, $\Pi^\mu$, $H$,
\begin{equation}
d\Pi^\alpha=0
\, ,
\qquad
d\Pi^\mu=\frac{1}{2}(C\Gamma^\mu)_{\alpha\beta}\Pi^\alpha\Pi^\beta\, ,
\qquad
dH=(C\Gamma_{\mu\nu})_{\alpha\beta}\Pi^\mu\Pi^\nu\Pi^\alpha\Pi^\beta
\, ,
\label{11b}
\end{equation}
and be given by a CE-non-trivial potential $b$ of a closed form 
$h(\Pi^\alpha,\Pi^\mu,H)$. Thus, we have 
to solve the problem of finding nontrivial $(p+2)$-cocycles of the FDA 
(\ref{11b}). We shall find that there is no solution unless $p=5$. 

A general $(p+2)$-form on (\ref{11b}) can be written as
\begin{equation}
     h=h^{(p+2)}+h^{(p-1)}H\quad ;                       \label{11c}
\end{equation}
there are no further powers of $H$ since $H^2\equiv H\wedge
H=0$. The closure of  $h$ gives
\begin{equation}
    dh^{(p+2)}+dh^{(p-1)}H-(-1)^ph^{(p-1)}
  (C\Gamma_{\mu\nu})_{\alpha\beta}\Pi^\mu\Pi^\nu\Pi^\alpha\Pi^\beta=0\quad ,
                                                          \label{11ca}
\end{equation}
which is equivalent to 
\begin{equation}
\addtolength{\arraycolsep}{.3ex}
\begin{array}{l}
   \displaystyle
   dh^{(p-1)}=0
 \\[1.1ex]
  \displaystyle
dh^{(p+2)}=(-1)^ph^{(p-1)}(C\Gamma_{\mu\nu})_{\alpha\beta}
\Pi^\mu\Pi^\nu\Pi^\alpha\Pi^\beta \, .               
\end{array}                                              \label{11d}
\end{equation}
Now, since  $[h]=L^{p+1}$ and $[H]=L^3$,
\begin{equation}
\addtolength{\arraycolsep}{.3ex}
\begin{array}{l}
    \displaystyle
        h^{(p+2)}=a^{(p+2)}(C\Gamma_{\mu_1\dots\mu_p})_{\alpha\beta}
         \Pi^{\mu_1}\dots\Pi^{\mu_p}\Pi^\alpha\Pi^\beta 
  \\[1.1ex]
   \displaystyle
       h^{(p-1)}=a^{(p-1)}(C\Gamma_{\mu_1\dots\mu_{p-3}})_{\alpha\beta}
         \Pi^{\mu_1}\dots\Pi^{\mu_{p-3}}\Pi^\alpha\Pi^\beta\, ,
\end{array}                                               \label{11e}
\end{equation}
for some constants $a^{(p+2)}$ and $a^{(p-1)}$. The first equation in 
(\ref{11d}) requires $a^{(p-1)}=0$ unless $p-3=2$, in which case
$a^{(p-1)}$ is arbitrary due to the identity 
$(C\Gamma_{\mu\nu})_{\alpha{'}\beta{'}}(C\Gamma^\nu)_{\delta{'}\epsilon{'}}=0$,
valid in $D$=$4,5,7,11$. 
If $p\neq 5$, $a^{(p-1)}=0$ and $h^{(p-1)}=0$, and the second equation of 
(\ref{11d}) gives  $dh^{(p+2)}=0$, which again implies 
$a^{(p+2)}=0$ unless $p=2$. But if $p=2$, we have $h=h^{(2+2)}\propto 
(C\Gamma_{\mu\nu})_{\alpha\beta}\Pi^{\mu}\Pi^{\nu}\Pi^\alpha\Pi^\beta=dH$,
in which case $h$ is the differential of a  LI form
and hence CE-trivial. Thus we are just left with the case $p=5$, 
$a^{(5-1)}$ arbitrary. Inserting (\ref{11e}) in (\ref{11d}) gives,
factoring out the $\Pi^\alpha$'s and $\Pi^\mu$'s,
\begin{equation}
      \frac{5}{2}a^{(5+2)}(C\Gamma^{\mu_5})_{\alpha{'}\beta{'}}
      (C\Gamma_{\mu_1\dots\mu_5})_{\delta{'}\epsilon{'}}+
    a^{(5-1)}(C\Gamma_{[\mu_1\mu_2})_{\alpha{'}\beta{'}}
(C\Gamma_{\mu_3\mu_4]})_{\delta{'}\epsilon{'}}=0\quad .    \label{11f}
\end{equation}
The second identity in (\ref{11identities}) gives 
$a^{(7)}=-\frac{2}{15} a^{(4)}$. The resulting closed form
\begin{equation}
      h\propto (C\Gamma_{\mu_1\dots\mu_5})_{\alpha\beta}
         \Pi^{\mu_1}\dots\Pi^{\mu_5}\Pi^\alpha\Pi^\beta -\frac{15}{2}
         (C\Gamma_{\mu_1\mu_2})_{\alpha\beta}
         \Pi^{\mu_1}\Pi^{\mu_2}\Pi^\alpha\Pi^\beta H        \label{11g}
\end{equation}
is not CE-exact, as may be seen by an argument analogous to that used in 
the IIA D-branes case: a LI potential form $b$ would have to be a 
scalar six-form depending on 
$\Pi^\alpha$, $\Pi^\mu$ and $H$ with dimensions $L^6$, which does not 
exist. 

It is possible to see, however, that a LI expression for $H$ exists 
on the appropriate extended superspace. Since $H$ is a three-form, 
it has formally the same properties as the invariant WZ term $\tilde b$ 
of the M$2$-brane, the extended superspace of which is the 
one corresponding to the Lie FDA obtained by the methods of 
Sec. \ref{NCESA}. Eqs. (\ref{2.2}), (\ref{dPimu}), 
 (\ref{dPim1a1}) and 
the first of (\ref{dPim1a24})  with $(a_s,a_0,
a_1,a_2)=(\frac{1}{2},\frac{1}{2},1,-\frac{1}{2})$ give,
respectively,
\be
\addtolength{\arraycolsep}{-.6ex}
\begin{array}{rclcrcl}
\dis
d\Pi^\alpha &=& \dis 0 
\, , & \qquad &
\dis
d\Pi^\mu & = & \dis \frac{1}{2}(C\Gamma^\mu)_{\alpha\beta}\Pi^\alpha\Pi^\beta
\\[1.3ex]
\dis
d\Pi_{\mu\nu} &=& \dis \frac{1}{2}(C\Gamma_{\mu\nu})_{\alpha\beta}\Pi^\alpha
\Pi^\beta
\, , & \qquad &
\dis
d\Pi_{\alpha\beta} &=& 
\dis \mbox{}-\frac{1}{2}(C\Gamma_{\mu\nu})_{\alpha\beta}
\Pi^\mu 
\Pi^\nu-\frac{1}{2}(C\Gamma^\mu)_{\alpha\beta}
\Pi_{\mu\nu}\Pi^\nu
\\[1.3ex]
\dis
d\Pi_{\mu\alpha} & =  &
\dis (C\Gamma_{\nu\mu})_{\alpha\beta}\Pi^\nu\Pi^\beta+
     (C\Gamma^\nu)_{\alpha\beta}\Pi_{\nu\mu}\Pi^\beta 
\, , & \qquad & 
 & & 
\dis
\mbox{} + \frac{1}{4}(C\Gamma^\mu)_{\alpha\beta}
\Pi_{\mu\delta}\Pi^\delta+ 2(C\Gamma^\mu)_{\delta\alpha{'}}
\Pi_{\mu\beta{'}}\Pi^\delta 
\, ,
\label{11h}
\end{array}
\ee
\ie, the dual of (\ref{smLa})). Using again (\ref{smbt}) we may then write 
\begin{equation}
     H=\frac{2}{3}\Pi^\mu\Pi^\nu\Pi_{\mu\nu}+\frac{3}{5}\Pi^\mu\Pi^\alpha
   \Pi_{\mu\alpha}-
      \frac{2}{15}\Pi_{\alpha\beta}\Pi^\alpha \Pi^\beta      
                 \quad .                                         \label{11i}  
\end{equation}

We might now go on and show that there exists a LI $\tilde b$ such that 
$h=d{\tilde b}$ on a suitably extended superspace; 
we shall omit its expression~\cite{Sez:97} since it is not needed 
below. What we wish to show is that now we may 
use (\ref{11i}) to replace the worldvolume two-form $A(\xi)$ by 
 the pull-back of the two-form $A$ on extended superspace given by
\begin{eqnarray}
     A &=& \frac{2}{3}\varphi_{\mu\nu}dx^\mu dx^\nu-\frac{3}{5}\varphi_{\mu\alpha}
   dx^\mu d\theta^\alpha-
 \frac{2}{15}\varphi_{\alpha\beta}d\theta^\alpha d\theta^\beta
\nonumber \\
  & & \mbox{} +\frac{1}{30}\varphi_{\mu\nu}x^\mu(C\Gamma^\nu)_{\alpha\beta}
  d\theta^\alpha d\theta^\beta+ 
  \frac{11}{30}\varphi_{\mu\nu}dx^\mu(C\Gamma^\nu)_{\alpha\beta}
  \theta^\alpha d\theta^\beta-\frac{13}{180}\varphi_{\mu\nu}
  (C\Gamma^\mu)_{\alpha\beta}(C\Gamma^\nu)_{\delta\epsilon}
\theta^\alpha d\theta^\beta \theta^\delta d\theta^\epsilon
\nonumber \\
 & & \mbox{} + \frac{1}{10}\varphi_{\mu\alpha}(C\Gamma^\mu)_{\delta\epsilon}
\theta^\delta d\theta^\epsilon d\theta^\alpha 
  + \frac{1}{20}\varphi_{\mu\alpha}(C\Gamma^\mu)_{\delta\epsilon}
d\theta^\delta d\theta^\epsilon \theta^\alpha \quad .
 \label{11j}
\end{eqnarray}
Again, this expression may also be used to find the gauge 
transformation $\delta A(\xi)=d\Lambda(\xi)$. This is achieved by the 
 one-form on 
superspace $\lambda=\lambda_\mu dx^\mu+\lambda_\alpha d\theta^\alpha$, 
$\phi^*(\lambda)=\Lambda(\xi)$. Then, if 
\begin{eqnarray}
\delta\varphi_{\mu\nu}&=&\frac{3}{2} 
\partial_{[\mu}\lambda_{\nu]}\quad ,
\nonumber \\
 \delta\varphi_{\mu\alpha}&=&\mbox{} -\frac{5}{3}
(\partial_\mu\lambda_\alpha+\partial_\alpha\lambda_\mu)+\frac{11}{12}
  \partial_{[\mu}\lambda_{\nu]}(C\Gamma^\nu)_{\alpha\beta}\theta^\beta\quad ,
  \nonumber \\
\delta\varphi_{\alpha\beta}&=&\mbox{} -\frac{15}{2}\partial_{\alpha{'}}
\lambda_{\beta{'}}+\frac{15}{21}(C\Gamma^\mu)_{\alpha\beta}\theta^\delta
(\partial_\mu\lambda_\delta+\partial_\delta\lambda_\mu)
 +\frac{15}{12}(C\Gamma^\mu)_{\delta\alpha{'}}\theta^\delta
(\partial_\mu\lambda_{\beta{'}}+\partial_{\beta{'}}\lambda_\mu)
\nonumber \\
& & \mbox{}- \frac{139}{240}(C\Gamma^\mu)_{\delta\beta}(C\Gamma^\nu)_{\alpha\epsilon}
\theta^\delta\theta^\epsilon \partial_{[\mu}\lambda_{\nu]}+
\frac{1}{20}x^\mu(C\Gamma^\nu)_{\alpha\beta}\partial_{[\mu}\lambda_{\nu]}
\label{gaugemem}
\end{eqnarray}
one obtains $\delta\phi^*(A)=d\Lambda(\xi)$.

As in the previous D-brane case, the EL equations derived from the 
action $I[x^\mu,\theta^\alpha,A_{ij}]$,
\begin{equation}
 \frac{\delta I}{\delta x^\mu}=0\quad ,
\quad \quad \quad
    \frac{\delta I}{\delta \theta^\alpha}=0\quad ,
\quad \quad \quad
    \frac{\delta I}{\delta A_{ij}}=0\, ,           \label{11k}
\end{equation}
are equivalent to the ones corresponding to the new action in which 
$A(\xi)$ is the pull-back of (\ref{11j}).  
Indeed, the equation for $\varphi_{\alpha\beta}$
gives $\frac{\delta I}{\delta A_{ij}}\partial_i
    \theta^\alpha\partial_j\theta^\beta=0$, and substituting it into that of 
$\varphi_{\mu\alpha}$,
\be
\frac{\delta I}{\delta A_{ij}}
\left(-\frac{3}{5}\partial_ix^\mu\partial_j
\theta^\alpha+\frac{1}{10}(C\Gamma^\mu)_{\delta\epsilon}\theta^\delta
\partial_i\theta^\epsilon\partial_j \theta^\alpha+ \frac{1}{20}
(C\Gamma^\mu)_{\delta\epsilon}\theta^\alpha \partial_i\theta^\delta
\partial_j\theta^\epsilon\right)=0
\quad,
\label{onemoreeqn}
\ee
 gives $\frac{\delta I}{\delta A_{ij}}\partial_i x^\mu\partial_j 
\theta^\alpha=0$ and so on. Therefore one obtains  
\[
\addtolength{\arraycolsep}{-.5ex}
\begin{array}{rclcrclcrcl}
\displaystyle
\int d\xi{'}^{p+1}\frac{1}{2}\frac{\delta I}{\delta A_{ij}(\xi{'})}
   \frac{\delta A_{ij}(\xi{'})}{\delta x^\mu(\xi)}+\frac{\delta I}{\delta
   x^\mu} 
& = & 
\dis
0
\, , & \qquad & 
\dis
\frac{\delta I}{\delta A_{ij}} \partial_i x^\mu \partial_j x^\nu
& = & 
0
\, , & \qquad &
\dis
 & & 
\\[3ex]
\dis
\int d\xi{'}^{p+1}\frac{1}{2}\frac{\delta I}{\delta A_{ij}(\xi{'})}
   \frac{\delta A_{ij}(\xi{'})}{\delta \theta^\alpha(\xi)}+    
   \frac{\delta I}{\delta \theta^\alpha}
& = & 
0
\, , & \qquad &
\dis
\frac{\delta I}{\delta A_{ij}}\partial_i x^\mu\partial_j \theta^\alpha
& = & 
0
\, , & \qquad &
\dis
\frac{\delta I}{\delta A_{ij}}\partial_i \theta^\alpha\partial_j\theta^\beta
&=&
0
\, .
\end{array} 
\label{11l}
\]
 The second equation 
implies $\frac{\delta I}{\delta A_{ij}} \partial_i x^\mu=0$ 
for all $\mu$ if one wants to avoid the possibility of one 
dimension of the object collapsing. This in turn implies 
$\frac{\delta I}{\delta A_{ij}}=0$ for the same reason, and 
inserting this equation into (\ref{11l}) gives (\ref{11k}).

%%%%%%%%%%%%%%%%%%%%%%%%%%%%%%%%%%%%%%%%%%%%%%%%%%%%%%%%%%%%%%%%%%%%%
%%%%%%%%%%%%%%%%%%%%%%%%%%%%%%%%%%%%%%%%%%%%%%%%%%%%%%%%%%%%%%%%%%%%

\section{Conclusions}

We have provided in this paper a unified 
approach to the study of various $p$-branes by defining 
them on suitably extended superspaces. 
All of these are supergroup manifolds, extensions of the basic odd abelian 
groups ${\text{sTr}}_D$ determined by the spinors of the 
specific theory considered. The extension algorithms in 
Secs. ~\ref{Mces} and ~\ref{NCESA} show how they depend,
when they do, on specific identities for $\Gamma$-matrices. 
The central extensions do not need any
$\Gamma$-identities, but the non-central ones require the identities
(\ref{Gammaprop}), precisely the ones needed to define the WZ
terms of the old branescan. 

The centrally extended superspaces are associated with (topological) 
charges, but the introduction of manifestly supersymmetric WZ terms 
requires the addition of non-central variables, already 
for the branes of the old branescan. When the procedure is 
applied to D$p$-branes, it is seen that all the fields in their 
action may also be defined by pullbacks of entities on the 
previously introduced superspaces. In the language of FDA's, our
results show that all the FDA's involved in the formulation 
of the $p$-branes considered here become {\it Lie} FDA's on
suitably extended superspaces. We conjecture, in view of the
previous discussion, that this is the case in general and that 
there exists an extended superspace definition for all fields 
appearing in the action of the various $p$-branes. In other
words, there exists a kind of field/extended superspaces democracy
by which all brane worldvolume fields are pullbacks from some target
superspace $\tilde\Sigma$. 
The appropriate ${\tilde \Sigma}$ of the theory is given by an 
extension of a certain $\text{sTr}_D$ and, 
using ${\tilde \Sigma}$, the action can be defined in a manifestly 
invariant form. In fact, in this field/extended superspace 
democracy context, the invariance properties 
seem to characterize essentially the superbrane actions.
It should not come then as a suprise 
that $\kappa$-symmetric actions may also be introduced
for D$p$-branes, as in \cite{Aga.Pop.Sch:97} for D-branes 
with rigid IIA  and IIB superPoincar\'e symmetry. As 
is the case for ordinary $p$-branes, $\kappa$-symmetry is
achieved when the relative coefficient of the 
kinetic and WZ-like part is such that the Bogomol'nyi 
bound is saturated.

   Our extensions provide at the same time a connection between
the CE cocycles and the mechanism of partial breaking of
supersymmetry. The CE ($p$+2)-cocycles lead to 
(extended) loop-type or worldvolume current algebras 
(see, {\it e.g.} \cite{Azc.Gau.Izq.Tow:89,Azc.Izq.Tow:91, 
Ber.Del.Sok:91,Ber.How.Pop.Sez.Sok:91,Ber.Per.Sez.Ste.Tow:93})
and the two-cocycle to the corresponding 
algebra of charges defining the extended superspace algebra. The 
new variables in the extended superspaces are also essential
to define (invariant) actions. They may also be relevant 
in the problem of quantisation, the formulation of dualities 
(see Sec.~\ref{sectene1}) and in the formulation of 
the additional gauge freedom hidden in the definition 
of some superbrane fields, the worldvolume definition of which
reflects an election of gauge. We suspect that the mathematical 
existence of the extensions considered here has a deeper 
meaning beyond the aspects discussed in this paper.
%%%%%%%%%%%%%%%%%%%%%%%%%%%%%%%%%%%%%%%%%%%%%%%%%%%%%%%%%%%%%%%%%%
\vskip .5cm
\noindent
\emph{Note added}. After completion of this paper an 
article \cite{Abe.Hat.Kam.Tok:99} has appeared in hep-th, where 
an approach similar to that in Sec. \ref{Dbext} for the field 
$A$ is given for the IIB D-brane case.

\vskip .5cm
\noindent
{\bf Acknowledgements}:
This research has been partially supported by the research grant 
PB96-0756 from the DGICYT, Spain. C.C. wishes to thank the Spanish
Ministry of Education and Culture for a post-doctoral fellowship
and J.M.I acknowledges a grant from the Junta de Castilla y
Le\'on. J.C.P.B. thanks the CSIC and the Ministerio 
de Educaci\'on for an FPI grant. Finally, the authors 
wish to thank Paul Townsend for helpful discussions.

%%%%%%%%%%%%%%%%%%%%%%%%%%%%%%%%%%%%%%%%%%%%%%%%%%%%%%%%%%%%%%%%%
%%%%%%%%%%%%%%%%%%%%%%%%%%%%%%%%%%%%%%%%%%%%%%%%%%%%%%%%%%%%%%%%%
%%%%%%%%%%%%%%%%%%%%%%%%%%%%%%%%%%%%%%%%%%%%%%%%%%%%%%%%%%%%%%%%%
\vskip .8cm
\noindent \appendix{\Large \bf Appendix}
\vskip .5cm
%%%%%%%%%%%%%%%%%%%%%%%%%%%%%%%%%%%%%%%%%%%%%%%%%%%%%%%%%%%%%%%%%%%%
%%%%%%%%%%%%%%%%%%%%%%%%%%%%%%%%%%%%%%%%%%%%%%%%%%%%%%%%%%%%%%%%%%%%
%%%%%%%%%%%%%%%%%%%%%%%%%%%%%%%%%%%%%%%%%%%%%%%%%%%%%%%%%%%%%%%%%%%%
\section{Non-central extensions}
We give here some details of the derivation of~(\ref{dPim1a24}).
After introducing $\Pi_{\mu_1 \dots \mu_{p-1} \alpha_1}$
satisfying~(\ref{dPim1a1}), we look for non-trivial CE 2-cocycles
with external indices ($\mu_1 \dots \mu_{p-2} \alpha_1
\alpha_2$). There are four available LI two-forms with these
indices,
\bae
\rho^{(1)} &=& (C \Gamma_{\nu \rho \mu_1 \dots \mu_{p-2}})_{
\alpha_1\alpha_2} \Pi^\nu \Pi^\rho
\nonumber \\
\rho^{(2)} &=& (C \Gamma^\nu)_{\alpha_1 \alpha_2}\Pi_{\nu \rho
\mu_1 \dots \mu_{p-2}} \Pi^\rho 
\nonumber \\
\rho^{(3)} &=& (C \Gamma^\nu)_{\alpha_1 \alpha_2} \Pi_{\nu \mu_1
\dots \mu_{p-2} \beta} \Pi^\beta
\nonumber \\
\rho^{(4)} &=& (C \Gamma^\nu)_{\alpha'_1 \beta} \Pi_{\nu \mu_1
\dots \mu_{p-2} \alpha'_2} \Pi^\beta
\quad,
\label{rho14}
\eae
none of which is closed. Looking for a linear combination 
$\rho \equiv \rho^{(1)} + \lambda_2 \rho^{(2)} + \lambda_3
\rho^{(3)} + \lambda_4 \rho^{(4)}$
that is closed, we compute (making use of the MC equations
for the available $\Pi$'s)
\bae
d\rho &=& 
\left\{ 
 2 a_s (C \Gamma_{\mu \nu \mu_1 \dots \mu_{p-2}})_{\alpha_1
 \alpha_2} (C \Gamma^\mu)_{\gamma \delta} 
 + \lambda_2 a_0 (C \Gamma^\mu)_{\alpha_1 \alpha_2} (C
\Gamma_{\mu \nu \mu_1 \dots \mu_{p-2}})_{\gamma \delta}
\right.
\nonumber \\
 & & \left. +\lambda_3 a_1 (C \Gamma^\nu)_{\alpha_1 \alpha_2} (C
\Gamma_{\mu \nu \mu_1 \dots \mu_{p-2}})_{\gamma \delta}
 - \lambda_4 a_1 (C \Gamma^\mu)_{\alpha'_1 \gamma} (C
   \Gamma_{\mu \nu \mu_1 \dots \mu_{p-2}})_{\alpha'_2 \delta}
\right\}
\Pi^\nu \Pi^\gamma \Pi^\delta
\nonumber \\
& & + \left\{ -\lambda_2 \alpha_1 (C \Gamma^\mu)_{\alpha_1
\alpha_2} (C \Gamma^\nu)_{\gamma \delta} + \lambda_3 {a_s a_1
\over a_0} (C \Gamma^\nu)_{\alpha_1 \alpha_2} (C
\Gamma^\mu)_{\gamma \delta} \right.
\nonumber \\
 & & + \left. \lambda_4 {a_s a_1 \over a_0}
(C \Gamma^\nu)_{\alpha'_1 \gamma} (C \Gamma^\mu)_{\alpha'_2
\delta} 
\right\} 
\Pi_{\mu \nu \mu_1 \dots \mu_{p-2}} \Pi_\gamma \Pi_\delta
\quad.
\label{dPiv1}
\eae
Inside the first curly brackets above, one can combine the third term
with the second, changing at the same time its sign
(the $\Gamma$'s are antisymmetric in the vector indices). In the
fourth term, one can also symmetrise over $\gamma$, $\delta$
(since $\Pi_\gamma$ and $\Pi_\delta$ commute). Effecting
explicitly this symmetrisation, as well as the indicated one (by
the primes) over
$\alpha_1$, $\alpha_2$, one gets four terms, which, together with
the other two, give exactly the six permutations
of~(\ref{Gammaprop}) (the $\Gamma$'s are symmetric in the
spinorial indices, so that the twenty-four permutations
of~(\ref{Gammaprop}) reduce to six). The sum of all six terms
will be zero (due to~(\ref{Gammaprop})) if their coefficients are
equal -- this gives the equations
\be
a_0 \lambda_2 - a_1 \lambda_3 = 2 a_s \quad, 
\quad \quad \quad
-{1 \over 4} a_1 \lambda_4 = 2 a_s
\quad.
\label{linsys1}
\ee
Inside the second curly brackets in~(\ref{dPiv1}), the last term is
zero because of antisymmetry in $\mu$, $\nu$. The sum of the
first two will be zero (for the same reason) if their
coefficients are equal, \ie \ if
\be
a_s \lambda_2 + {a_s a_1 \over a_0} \lambda_3 = 0
\quad.
\label{linsys2}
\ee
Solving the linear system of~(\ref{linsys1}), (\ref{linsys2}) one
gets
\be
\lambda_2= {a_s \over a_0}
\quad,
\quad \quad \quad
\lambda_3= -{a_s \over a_1}
\quad,
\quad \quad \quad
\lambda_4= -8 {a_s \over a_1}
\quad,
\label{lambdas}
\ee
which leads to the first of~(\ref{dPim1a24}).

For the next extension, looking for LI two-forms with indices
($\mu_1 \dots \mu_{p-3} \alpha_1 \alpha_2 \alpha_3$) we find
\bae
\rho^{(1)} &=& (C \Gamma^\nu)_{\alpha'_1 \alpha'_2} \Pi_{\nu \rho
\mu_1 \dots \mu_{p-3} \alpha'_3} \Pi^\rho
\nonumber \\
\rho^{(2)} &=& (C \Gamma^\nu)_{\alpha'_1 \beta} \Pi_{\nu 
\mu_1 \dots \mu_{p-3} \alpha'_2 \alpha'_3} \Pi^\beta
\nonumber \\
\rho^{(3)} &=& (C \Gamma^\nu)_{\alpha'_1 \alpha'_2} \Pi_{\nu 
\mu_1 \dots \mu_{p-3} \beta \alpha'_3} \Pi^\beta
\quad,
\label{rho13}
\eae
none of which is closed (we use, for simplicity, the same
symbols $\rho$, $\lambda$ as in the previous extension). For 
their linear
combination $\rho \equiv \rho^{(1)} + \lambda_2 \rho^{(2)} +
\lambda_3 \rho^{(3)}$ we compute
\bae
d \rho  &=& (C \Gamma^\nu)_{\alpha'_1 \alpha'_2} a_1 
\left\{ (C \Gamma_{\lambda \nu \rho \mu_1 \dots \mu_{p-3}})_{\beta
\alpha'_3} \Pi^\lambda \Pi^\beta
+ {a_s \over a_0} (C \Gamma^\sigma)_{\beta \alpha'_3} \Pi_{\sigma
\nu \rho \mu_1 \dots \mu_{p-3}}) \Pi^\beta 
\right\} \Pi^\rho
\nonumber \\
& &
 -a_s (C \Gamma^\nu)_{\alpha'_1 \alpha'_2} (C
\Gamma^\nu)_{\gamma \delta}
\Pi_{\nu \rho \mu_1 \dots \mu_{p-3} \alpha'_3} \Pi^\gamma
\Pi^\delta
\nonumber \\
& & 
 + \lambda_2 a_4 (C \Gamma^\nu)_{\alpha'_1 \beta} 
\left\{ (C \Gamma_{\sigma \rho \nu \mu_1 \dots 
\mu_{p-3}})_{\alpha'_2 \alpha'_3} \Pi^\sigma \Pi^\rho
+ {a_s \over a_0} (C \Gamma^\sigma)_{\alpha'_2 \alpha'_3}
\Pi_{\sigma \rho \nu \mu_1 \dots \mu_{p-3}} \Pi^\rho
\right.
\nonumber \\
 & & \left. -{a_s \over a_1} (C \Gamma^\sigma)_{\alpha'_2
\alpha'_3} \Pi_{\sigma \nu \mu_1 \dots \mu_{p-3} \gamma}
\Pi^\gamma 
- 8 {a_s \over a_1} (C \Gamma^\sigma)_{\alpha'_2 \gamma}
  \Pi_{\sigma \nu \mu_1 \dots \mu_{p-3} \alpha'_3} \Pi^\gamma 
\right\} \Pi^\beta
\nonumber \\
& & 
+ \lambda_3 a_2 (C \Gamma^\nu)_{\alpha'_1 \alpha'_2}
\left\{
(C \Gamma_{\sigma \rho \nu \mu_1 \dots  \mu_{p-3}})_{\alpha'_2
\alpha'_3} \Pi^\sigma \Pi^\rho
+ {a_s \over a_0} (C \Gamma^\sigma)_{\beta \alpha'_3}
 \Pi_{\sigma \rho \nu \mu_1 \dots \mu_{p-3}} \Pi^\rho 
\right.
\nonumber \\
 & & 
\left.  -{a_s \over a_1} (C \Gamma^\sigma)_{\beta \alpha'_3}
\Pi_{\sigma \nu \mu_1 \dots \mu_{p-3} \gamma} \Pi^\gamma
- 8 {a_s \over a_1} (C \Gamma^\sigma)_{\bar{\beta} \gamma}
\Pi_{\sigma \nu \mu_1 \dots \mu_{p-3} \bar{\alpha'}_3} \Pi^\gamma 
\right\} \Pi^\beta
\label{drho2}
\eae
(the barred indices in the last term denote a second
symmetrisation, besides the one over the primed indices). There is a 
novelty here compared with the
previous extension: there are four different types of terms in
the $\Pi$'s, the coefficients of which must separately vanish,
giving rise to four linear equations for the two unknowns
$\lambda_2$, $\lambda_3$ (care must be taken of the fact that
when the second symmetrisation 
in the last term above is effected, corresponding to
the barred indices, one obtains contributions to two different
types of terms in the $\Pi$'s). Making use of~(\ref{Gammaprop})
and of the symmetry properties of the $\Gamma$'s, as in the
previous extension, one arrives at the (overdetermined) linear system
\be
\lambda_2 -\lambda_3 = {a_1 \over a_2}
\quad,
\quad \quad \quad
4 {a_s a_2 \over a_1} \lambda_3 = a_s
\quad,
\quad \quad \quad
\lambda_2 -5 \lambda_3 =0
\ee
(the first equation appears twice) which nevertheless admits the 
solution
\be
\lambda_2 = {5 a_1 \over 4 a_2}
 \quad,
\quad \quad \quad
\lambda_3 = {a_1 \over 4 a_2} 
\quad,
\ee
leading to the second of~(\ref{dPim1a24}). 
The last
of~(\ref{dPim1a24}), as well as~(\ref{dPim1ak2}), are proved
similarly.
%%%%%%%%%%%%%%%%%%%%%%%%%%%%%%%%%%%%%%%%%%%%%%%%%%%%%%%%%%%%%%%%
%%%%%%%%%%%%%%%%%%%%%%%%%%%%%%%%%%%%%%%%%%%%%%%%%%%%%%%%%%%%%%%%
%%%%%%%%%%%%%%%%%%%%%%%%%%%%%%%%%%%%%%%%%%%%%%%%%%%%%%%%%%%%%%%%
\section{$D=10$ $\Gamma$-matrix identities}
%%%%%%%%%%%%%%%%%%%%%%%%%%%%%%%%%%%%%%%%%%%%%%%%%%%%%%%%%%%%%%%%
We prove here the $\Gamma$-identities needed in 
Sec.~\ref{CEdB}\footnote{In the search for further possible
$\Gamma$-identities the results in \cite{Nai.Osa.Fuk:86} may be useful.}. 
The first two identities in (\ref{Identities}) 
follow by dimensional reduction 
from the known $D=11$ relations
\begin{eqnarray}
  (C\Gamma^{\tilde \mu_2})_{\alpha{'}\beta{'}}(C\Gamma_{\tilde \mu_1
  \tilde \mu_2})_{\delta{'}\epsilon{'}} &=& 0
\quad, \nonumber \\
  (C\Gamma^{\tilde \mu_5})_{\alpha{'}\beta{'}}
  (C\Gamma_{\tilde \mu_1\dots
  \tilde \mu_5})_{\delta{'}\epsilon{'}}-3
  (C\Gamma_{[\tilde\mu_1\tilde\mu_2})_{\alpha{'}\beta{'}}
  (C\Gamma_{\tilde\mu_3\tilde\mu_4]})_{\delta{'}\epsilon{'}} &=& 0
\quad.
\label{11identities}
\end{eqnarray}
where the tilded indices $\tilde \mu=0,1,\dots,10$.

The third identity can be proved as follows. First, using that
\begin{equation}
    \Gamma_{\mu_1\dots\mu_6}=\Gamma_{\mu_1\dots\mu_5}\Gamma_{\mu_6}-
  5 \Gamma_{[\mu_1\dots\mu_4}\eta_{\mu_5]\mu_6}    \label{dirac}
\end{equation}
 and the fact that $\Gamma_{11}^2=1$, we see that
\begin{equation}
(C\Gamma^{\mu_6})_{\alpha{'}\beta{'}}
(C\Gamma_{\mu_1\dots\mu_6})_{\delta{'}\epsilon{'}}=
(C\Gamma^{\mu_6})_{\alpha{'}\beta{'}}(C\Gamma_{\mu_1\dots\mu_5}
\Gamma_{11})_{\delta{'}\lambda}{(\Gamma_{11}\Gamma_{\mu_6})^\lambda}_{
\epsilon{'}}
\end{equation}
since the second term in (\ref{dirac}) does not contribute because
$(C\Gamma_{\mu_1\dots\mu_4})_{\delta\epsilon}$ is antisymmetric
(primed indices are symmetrised).
Now, due to the identity 
$(C\Gamma_\mu)_{\alpha{'}\beta{'}}(C\Gamma_{11}
\Gamma^\mu)_{\delta{'}\epsilon{'}}=0$, we have \\
$(C\Gamma_\mu)_{\alpha{'}\beta{'}}
{(\Gamma_{11}\Gamma^\mu)^\lambda}_{\epsilon{'}}=-
{(\Gamma_\mu)^\lambda}_{\alpha{'}}(\Gamma_{11}\Gamma^\mu)_{\beta{'}
\epsilon{'}}$ so that
\begin{eqnarray}
 (C\Gamma^{\mu_6})_{\alpha{'}\beta{'}}
(C\Gamma_{\mu_1\dots\mu_6})_{\delta{'}\epsilon{'}} &=&
-(C\Gamma_{\mu_1\dots\mu_5}\Gamma_{11}\Gamma^{\mu_6})_{\delta{'}\alpha{'}}
(\Gamma_{11}\Gamma_{\mu_6})_{\beta{'}\epsilon{'}}
\nonumber \\
&=&5(C\Gamma_{[\mu_1\dots\mu_4}\Gamma_{11})_{\delta{'}\alpha{'}}
(C\Gamma_{11}\Gamma_{\mu_5]})_{\beta{'}\epsilon{'}} \quad,
\end{eqnarray}
where in the second equality use has been made of (\ref{dirac}) and the 
fact that $(C\Gamma_{\mu_1\dots\mu_6}\Gamma_{11})_{\delta\alpha}$
is antisymmetric. 

Finally, the fourth equation 
in (\ref{Identities})
may be shown to be equivalent to the second. 
Indeed, by multiplying the fourth identity by $\epsilon^{\mu_1\dots\mu_7
\nu_1\nu_2\nu_3}$ and using that $\Gamma_{\mu_1\dots\mu_q}\Gamma_{11}
\propto \frac{1}{(10-q)!}
\epsilon_{\mu_1\dots\mu_q \rho_{1}\dots \rho_{10-q}}
\Gamma^{\rho_{1}\dots \rho_{10-q}}$, one obtains
\begin{equation}
      \frac{7!3!}{2}(C\Gamma^{[\nu_1})_{\alpha{'}\beta{'}}
   (C\Gamma^{\nu_2\nu_3]})_{\delta{'}\epsilon{'}}+
 \frac{6!4!7}{4!}(C\Gamma^{\mu_7\nu_1\nu_2\nu_3}\Gamma_{11})_{\alpha{'}\beta{'}}
 (C\Gamma_{\mu_7}\Gamma_{11})_{\delta{'}\epsilon{'}}=0
\quad,
\end{equation}
which is equivalent to the second equation in (\ref{11identities})
due to the fact that $i\Gamma_\mu\Gamma_{11}$ realize the same 
Clifford algebra as $\Gamma_\mu$.
%%%%%%%%%%%%%%%%%%%%%%%%%%%%%%%%%%%%%%%%%%%%%%%%%%%%%%%%%%%%%%%%%%
%%%%%%%%%%%%%%%%%%%%%%%%%%%%%%%%%%%%%%%%%%%%%%%%%%%%%%%%%%%%%%%%%%
%%%%%%%%%%%%%%%%%%%%%%%%%%%%%%%%%%%%%%%%%%%%%%%%%%%%%%%%%%%%%%%%%%

%\bibliographystyle{article}
%\bibliography{biblio-1,1}

\begin{thebibliography}{10}

\bibitem{Ach.Eva.Tow.Wil:87}
A.~Ach\'ucarro, J.~M. Evans, P.~K. Townsend and D.~L. Wiltshire, 
\emph{Super $p$-branes}, Phys. Lett. \textbf{198B}, 441--446 (1987).

\bibitem{Azc.Tow:89}
J.~A. de~Azc\'arraga and P.~K. Townsend, \emph{Superspace geometry and
  classification of supersymmetric extended objects}, Phys. Rev. Lett.
  \textbf{62}, 2579--2512 (1989).

\bibitem{Che.Eil:48}
C.~Chevalley and S.~Eilenberg, \emph{Cohomology theory of Lie groups 
and Lie algebras}, Trans. Am. Math. Soc. \textbf{63}, 85--124 (1948).

\bibitem{Duf.Khu.Lu:95}
M.~J. Duff, R.~R. Khuri and J.~X. Lu, \emph{String solitons}, Phys. Rep.
  \textbf{259}, 213--326 (1995).

\bibitem{Duf:96b}
M.~J. Duff, \emph{Supermembranes}, CTP-TAMU-61/96, hep-th/9611203.

\bibitem{Pol:95}
J. Polchinski, \emph{Dirichlet branes and Ramond-Ramond charges}, 
Phys. Rev. Lett. {\bf 75}, 4724-4727 (1995)

\bibitem{Pol.Cha.Joh:96}
J.~Polchinski, S.~Chaudhuri and C.~V. Johnson, \emph{Notes on $D$-branes},
  hep-th/9602052.

\bibitem{Aga.Pop.Sch:97}
M.~Aganagic, C.~Popescu and J.~H. Schwarz, \emph{$D$-brane actions 
with local $\kappa$-symmetry}, Phys. Lett. \textbf{B393}, 311--315 (1997).

\bibitem{How.Rae.Rud.Sez:98}
P.~S. Howe, O.~Raetzel, I.~Rudychev and E.~Sezgin, \emph{L-branes},
  KCL-MTH-98-39 / CPT-TAMU-3498, hep-th/9810081.

\bibitem{Sch:97b}
J.~H. Schwarz, \emph{Lectures on superstring and $M$-theory dualities}, 
Nucl. Phys. Proc. Suppl. \textbf{55B}, 1 (1997).

\bibitem{Ban.Fis.She.Sus:97}
T.~Banks, W.~Fischler, S.~H. Shenker and L.~Susskind, \emph{M theory as a
  matrix model: a conjecture}, Phys. Rev. \textbf{D55}, 5112--5128 (1997).

\bibitem{Azc.Gau.Izq.Tow:89}
J.~A. de~Azc\'arraga, J.~P. Gauntlett, J.~M. Izquierdo and P.~K. Townsend,
  \emph{Topological extensions of the supersymmetry algebra for extended
  objects}, Phys. Rev. Lett. \textbf{63}, 2443--2446 (1989).

\bibitem{Gre:89}
M.~B. Green, \emph{Super-translations, superstrings and Chern-Simons forms},
  Phys. Lett. \textbf{B223}, 157--164 (1989).

\bibitem{Ber.Sez:89}
E.~Bergshoeff and E.~Sezgin, \emph{New spacetime superalgebras and their
  Kac-Moody extensions}, Phys. Lett. \textbf{B232}, 96--103 (1989).

\bibitem{Azc.Izq.Tow:91}
J.~A. de~Azc\'arraga, J.~M. Izquierdo and P.~K. Townsend, \emph{Classical
anomalies of supersymmetric extended objects}, Phys. Lett. \textbf{B267},
366--373 (1991).

\bibitem{Ber.Sez:95}
E.~Bergshoeff and E.~Sezgin, \emph{Super $p$-brane theories and new 
spacetime superalgebras}, Phys. Lett. \textbf{B354}, 256--263 (1995).


\bibitem{Bar:96}
I.~Bars, \emph{Supersymmetry, p-brane duality and hidden spacetime
     dimensions}, Phys. Rev. {\bf D54}, 5202--5210 (1996).

\bibitem{Sor.Tow:97}
D.~Sorokin and P.~K. Townsend, \emph{$M$-theory superalgebra from the
  $M$-five-brane}, Phys. Lett. \textbf{B412}, 265--273 (1997).


\bibitem{Sez:97}
E.~Sezgin, \emph{The $M$-algebra}, 
Phys. Lett. \textbf{B392}, 323--331 (1997).


\bibitem{Der.Gal:97}
A.~Deriglazov and A.~Galajinsky, \emph{A linear realization for the new
  spacetime superalgebras in 10 and 11 dimensions}, Mod. Phys. Lett.
  \textbf{A12}, 1517--1529 (1997).

\bibitem{Bar:97}
I.~Bars, \emph{$S$--theory}, Phys. Rev. \textbf{D55}, 2373--2381 (1997).

\bibitem{Sak:98}
M.~Sakaguchi, \emph{Type II superstrings and new spacetime superalgebras},
  hep-th/9809113.

\bibitem{Ban.Pas.Sor.Ton.Vol:95}
I. Bandos, P. Pasti, D. Sorokin, M. Tonin and D. Volkov,
\emph{Superstrings and supermembranes in the doubly supersymmetrical
approach}, Nucl. Phys. {\bf B446}, 79-119 (1995)

\bibitem{How.Sez:97}
P.S. Howe and E. Sezgin, \emph{Superbranes}, Phys. Lett. {\bf B390}, 
133-142 (1997)


\bibitem{Ald.Azc:85}
V.~Aldaya and J.~A. de~Azc\'arraga, 
\emph{A note on the meaning of covariant derivatives in supersymmetry}, 
J. Math. Phys. \textbf{26}, 1818--1821 (1985).


\bibitem{Azc.Izq:95}
J.~A. de~Azc\'arraga and J.~M. Izquierdo, \emph{Lie groups, Lie algebras,
cohomology and some applications in physics} (Camb. Univ. Press, 1995).

\bibitem{Haa.Lop.Soh:75}
R.~Haag, J.~T. Lopusza\'nski and M.~Sohnius, 
\emph{All possible generators of supersymmetry of the $S$--matrix}, 
Nucl. Phys. \textbf{B88}, 257--274 (1975).

\bibitem{Hol.Pro:82}
J.~W. van Holten and A.~Van Proeyen, \emph{N=1 supersymmetry algebras 
in $d$=2, 3, 4 mod 8}, J. Phys. \textbf{A15}, 3763--3783 (1982).

\bibitem{DAu.Fre:82}
R. D'Auria and P. Fr\'e,
\emph{Geometric supergravity in $D$=11 and its hidden supergroup},
Nucl. Phys. {\bf B201}, 101-140 (1982) 
(E.: {\it ibid.} {\bf B206}, 496 (1982))


\bibitem{Ziz:84b}
P.~A. Zizzi, \emph{An extesion of the Kaluza-Klein picture for the $N=4$
supersymmetric Yang-Mills theory}, Phys. Lett. \textbf{137B}, 57--61 (1984).

\bibitem{Ziz:84}
P.~A. Zizzi, \emph{Antisymmetric tensors in supersymmetric algebras and
spontaneous compactification}, Phys. Lett. \textbf{149B}, 333--336 (1984).

\bibitem{Azc:92}
J.~A. de~Azc\'arraga, \emph{Wess-Zumino terms, extended algebras and 
anomalies in classical physics}, Contemp. Math. \textbf{132}, 
75--98 (1992).

\bibitem{Sie:94}
W.~Siegel, \emph{Randomizing the superstring}, Phys. Rev. \textbf{D50},
2799--2805 (1994).


\bibitem{Ber:79}
F. Berezin, \emph{The mathematical basis of supersymmetric theories},
Sov. J. Nucl. Phys. {\bf 29}, 857--866 (1979) [Yad. Fiz. {\bf 29}, 1670]
(Section 5).

\bibitem{Sch:80}
A. S. Schwarz, \emph{Supergravity and field space democracy},
Nucl. Phys. {\bf B 171}, 154--156 (1980)

\bibitem{Gay.Rom.Sch:81}
A. V. Gayduk, V. N. Romanov and A. S. Schwarz, {\it Supergravity and
field space democracy}, Commun. Math. Phys. {\bf 79}, 507--528 (1981).

\bibitem{Ald.Azc:83}
V. Aldaya and J. A. de Azc\'arraga, {\it Geometric quantisation in 
the presence of an electromagnetic field}, Int. J. Theor. Phys.
{\bf 22}, 1--18 (1983).

\bibitem{Ham:98}
H. Hammer, \emph{Topological extensions of Noether charge algebras
carried by D$p$-branes}, Nucl. Phys. {\bf B521}, 503-546 (1998).

\bibitem{Sul:77}
D.~Sullivan, \emph{Infinitesimal computations in topology}, Ins. des Haut.
  \'Etud. Sci., Pub. Math. \textbf{47}, 269--331 (1977).

\bibitem{DAu.Fre.Reg:80}
R. D'Auria, P. Fr\'e and T. Regge \emph{Graded Lie algebra, cohomology and
supergravity}, Riv. Nuov. Cim. {\bf 3}, fasc. 12 (1980).

\bibitem{PvN:83}
P. van Nieuwenhuizen, \emph{Free graded differential superalgebras}, 
in Lect. Notes in Phys. {\bf 180}, 228-245, Springer-Verlag (1983).

\bibitem{Nie:84}
P.~Van Nieuwenhuizen, \emph{An introduction to simple gravity and the
  Kaluza--Klein program}, in \emph{Relativit\'e, groupes et topologie II},
  B.~S.~De Witt and R.~Stora (eds.), pp. 823--932 (Elsevier, 1984).

\bibitem{Wes:98}
P.~C. West, \emph{Supergravity, brane dynamics and string duality}, 
1997 Isaac Newton lectures, KCL-MTH-98-55, hep-th/9811101.

\bibitem{Tow:96}
P.~K. Townsend, \emph{$p$--brane democracy}, in \emph{Particles, 
strings and cosmology},  J.~Bagger, G.~Domokos, A.~Falk and
A.~Kovesi-Domokos (eds.), pp. 271--285 (World Sci., 1996), hep-th/9507048.

\bibitem{Tow:97}
P.~K. Townsend, \emph{$M$ theory from its superalgebra}, 1997 Carg\`ese
  lectures, hep-th/9712004.

\bibitem{Giv.Por.Rab:94}
A.~Giveon, M.~Porrati and E.~Rabinovici, 
\emph{Target space duality in string theory}, 
Phys. Rep. \textbf{244}, 77--202 (1994).

\bibitem{Eva:88}
J.~M. Evans, \emph{Super $p$-brane Wess-Zumino terms}, Class. Quant. Grav.
  \textbf{5}, L87--L90 (1988).

\bibitem{Ber.Tow:98}
E. Bergshoeff and P.K. Townsend, \emph{Super D-branes revisited},
Nucl. Phys. {\bf B531}, 226-238 (1998).

\bibitem{Tow:92}
P. K. Townsend, \emph{Worldsheet electromagnetism and the 
superstring tension}, Phys. Lett. {\bf B272}, 285-288 (1992).

\bibitem{Azc.Izq.Tow:92}
J.~A. de~Azc\'arraga, J.~M. Izquierdo and P.~K. Townsend,
\emph{Kaluza-Klein origin for the superstring tension}, Phys. Rev.
 {\bf D45}, R3321-R3325 (1992).

\bibitem{Ber.Lon.Tow:92}
E. Bergshoeff, L.A.J. London and P.K. Townsend, \emph{Spacetime 
  invariance and the super $p$-brane}, Class. Quantum Grav. {\bf 9},
  2545-2556 (1992).

\bibitem{Ban.Lec:97}
I. Bandos, K. Lechner, A.Nurmagambetov, P. Pasti, D.Sorokin and M. Tonin,
\emph{Covariant action for the super-five-brane of M-theory},
Phys. Rev. Lett. {\bf 78}, 4332-4434 (1997).

\bibitem{Aga.Par:97}
M. Aganagic, J. Park, C. Popescu and J. H. Schwarz, 
\emph{Worldvolume action of the M-theory five-brane}, 
Nucl. Phys. {\bf B496}, 191--214 (1997)

\bibitem{Ber.Del.Sok:91}
E. Bergshoeff, F. Delduc and E. Sokatchev,
\emph{Light-like integrability in loop superspace, Kac-Moody central
charges and Chern-Simons terms},
Phys. Lett. {\bf B262}, 444--450, (1991)

\bibitem{Ber.How.Pop.Sez.Sok:91}
E. Bergshoeff, P.S. Howe, C.N. Pope, E. Sezgin and E. Sokatchev,
\emph{Ten-dimensional supergravity from light-like integrability
in loop superspace}, Nucl. Phys. {\bf B354}, 113-128 (1991)

\bibitem{Ber.Per.Sez.Ste.Tow:93}
E. Bergshoeff, R. Percacci, E. Sezgin, K.S. Stelle and P.K. Townsend,
\emph{$U(1)$-Extended gauge algebras in $p$-loop space},
Nucl. Phys. {\bf B398}, 343-358 (1993)

\bibitem{Abe.Hat.Kam.Tok:99}
M. Abe, M. Hatsuda, K. Kamimura and T. Tokunaga, 
\emph{$SO(2,1)$ covariant IIB superalgebra}, hep-th/9903234.

\bibitem{Nai.Osa.Fuk:86}
S. Naito, K. Osada and T. Fukui, \emph{Fierz identities and invariance of
$11$-dimensional supergravity action}, Phys. Rev. {\bf D34}, 536-552 (1986)
\end{thebibliography}

\end{document}